\documentclass[conference,compsoc]{IEEEtran}

\usepackage{graphicx} 
\usepackage{amsmath,amsfonts,paralist,algorithm,algorithmic,multicol,paralist}
\usepackage{hyperref}
\usepackage{cleveref}
\usepackage{booktabs}
\usepackage{multirow}
\usepackage{colortbl}
\usepackage[table,xcdraw]{xcolor}
\usepackage{pdflscape}
\usepackage{rotating}
\usepackage{subcaption}
\usepackage{array}
\usepackage{filecontents}
\usepackage{tikz}
\usepackage{stfloats}

\newcommand\Tstrut{\rule{0pt}{2.6ex}}         
\newcommand\Bstrut{\rule[-0.9ex]{0pt}{0pt}}   
\newcommand{\fig}[1]{\hyperref[#1]{Figure~\ref*{#1}}}
\newcommand{\tbl}[1]{\hyperref[#1]{Table~\ref*{#1}}}
\newcommand{\figs}[2]{\hyperref[#1]{Figures~\ref*{#1}--\ref*{#2}}}

\newcommand{\karim}[1]{{\color{teal} Karim: #1}}
\newcommand{\diff}[1]{{\color{black} #1}}

\newcommand{\ignore}[1]{}

\begin{document}
\date{}

\title{Private Data Imputation}


\author{
\IEEEauthorblockN{Abdelkarim Kati}
\IEEEauthorblockA{University of Waterloo\\
Email: akati@uwaterloo.ca}

\and
\IEEEauthorblockN{Florian Kerschbaum}
\IEEEauthorblockA{University of Waterloo\\
Email: florian.kerschbaum@uwaterloo.ca}

\and
\IEEEauthorblockN{Marina Blanton}
\IEEEauthorblockA{University at Buffalo\\
Email: mblanton@buffalo.edu}
}


\maketitle

\begin{abstract}
Data imputation is an important data preparation task where the data analyst replaces missing or erroneous values to increase the expected accuracy of downstream analyses.
The accuracy improvement of data imputation extends to private data analyses across distributed databases.
However, existing data imputation methods violate the privacy of the data rendering the privacy protection in the downstream analyses obsolete.
We conclude that private data analysis requires private data imputation.

In this paper, we present the first optimized protocols for private data imputation.
We consider the case of horizontally and vertically split data sets.
Our optimization aims to reduce most of the computation to private set intersection (or at least oblivious programmable pseudo-random function) protocols which can be very efficiently computed.
We show that private data imputation has -- on average across all evaluated datasets -- an accuracy advantage of 20\% in case of vertically split data and 5\% in case of horizontally split data over imputing data locally. 
In case of the worst data split we observed that imputing using our method resulted in an increase of up to 32.7 times in the quality of imputation over the vertically split data and 3.4 times in case of horizontally split data. 
Our protocols are very efficient and run in 2.4 seconds in case of vertically split data and 8.4 seconds in case of horizontally split data for 100,000 records evaluated in the 10 Gbps network setting, performing one data imputation.
\end{abstract}

\section{Introduction}

\diff{Consider two hospitals seeking to improve patient diagnosis through collaborative analytics. Hospital A has 50,000 patient records
; Hospital B has 30,000. When Hospital B conducts research, incomplete records would benefit from imputation using similar cases from Hospital A's larger dataset. Yet privacy regulations like HIPAA prohibit sharing raw patient records. This creates a dilemma: imputation accuracy improves with more data, yet privacy requirements prevent data sharing. This challenge extends beyond healthcare to banks, research consortia, and any domain where distributed parties hold similar information about different entities.}

Data cleaning \cite{ilyasbook} is a necessary step before any data analysis.
This applies even to private analyses such as those supported by secure multi-party computation or private set intersection (PSI)~\cite{google}.
However, existing data cleaning methods operate on a single database, which prevents their use in privacy-preserving data analysis over databases distributed across mutually distrusting parties.
Blass and Kerschbaum~\cite{bla23} have studied the problem of detecting errors across distributed databases.
Their solution reduces to PSI which makes it very efficient.
However, correcting errors is still an open problem.

Data imputation helps to correct errors.
It is the task of replacing non-existent data entries with the most likely data value.
\diff{It is well known among statisticians that analysis over an imputed dataset leads to better models than ignoring records with missing entries~\cite{Schafer2002}.}
However, existing data imputation algorithms operate on a single database.
Operating over multiple databases increases accuracy of the imputation and hence accuracy of any downstream analysis task.
\diff{In our evaluation we show that, on average, imputing over horizontally split databases (where parties hold different records with the same attributes) improves accuracy by 20\%, while vertically split databases (where parties hold different attributes for the same records) yield a 5\% improvement for the most commonly evaluated type of missing data (MCAR -- missing completely at random), even when considering all our performance-enhancing modifications. In worst-case data splits, our methods achieve improvements of up to 32.7× for vertical splits and 3.4× for horizontal splits. These substantial accuracy gains demonstrate that distributed imputation is essential for high-quality data analysis.}

However, privately imputing over multiple databases is non-trivial.
While there exist privacy-preserving training algorithms over distributed data using secure multi-party computation, e.g., \cite{Knott21,Zhou24} or federated learning, e.g., \cite{Kairouz21}, these require already imputed data.
\diff{Yet, when exchanging the data in the clear for the purpose of data imputation, privacy of the entire training dataset gets violated, undermining the privacy guarantees of downstream privacy-preserving analyses.}
Hence, we need private data imputation to complete the data science process from start to end before we can engage in privacy-preserving training of machine learning models.
We study this problem in this paper.

Secure multi-party computation~\cite{goldreich2004} can certainly be used to implement private data imputation.
However, in current data analysis problems we often deal with millions, if not billions, of data records.
Generic multi-party computation is not likely to scale to such problem sizes and hence we need to employ optimizations.
One optimization we share with Blass and Kerschbaum~\cite{bla23} is that we make extensive use of PSI which is a very fast privacy-enhancing technology.
However, data imputation is necessarily more complicated than just detecting errors as in~\cite{bla23} and we need to design more complicated protocols while preserving privacy.

Jagannathan and Wright~\cite{jag06,jag08} have already studied data imputation over horizontally partitioned databases with privacy.
They infer the missing value using a decision tree.
However, 
a recent benchmark study~\cite{benchmark} on data imputation recommends other data imputation algorithms with higher accuracy.
\diff{The benchmark demonstrates that simple methods like $k$-nearest neighbor ($k$-NN) and random forest consistently outperform complex deep learning approaches%
. Critically, when trained on incomplete data---the scenario our protocols address---ML methods' advantages largely disappear while $k$-NN maintains robust performance. Furthermore, federated learning approaches would require multiple training rounds with secure aggregation plus expensive privacy-preserving inference over encrypted or secret-shared data for each imputation query. The benchmark shows ML training times often exceed simple methods by 10--100×, while adding privacy-preserving neural network inference would increase costs by another 100--1000× compared to our PSI-based approach. Even with these substantial computational costs, federated ML approaches would deliver inferior imputation quality according to comprehensive benchmark evidence (detailed discussion in Section~\ref{sec:prelim}).}

In this paper, we study the problem of {\em private data imputation with high accuracy}.
We assume two parties, Alice and Bob, each with their own private database.
Bob has a missing data value and it should be imputed.
We consider both horizontally and vertically split data.

\diff{We solve the problem using a secure multi-party computation which uses PSI to perform the most time-consuming operations and hence our protocol is very efficient. A recent benchmark study~\cite{benchmark} recommends nearest-neighbor approaches as top performers with low computational complexity. Unlike traditional $k$-NN which selects exactly $k$ neighbors regardless of their distance, we use a radius-based nearest-neighbor algorithm that considers all tuples within a learned distance threshold $r_j$ for each attribute $j$.}

\medskip \noindent \textbf{Our Contributions.}
In summary, our private data imputation protocol provides the first practical solution to private data imputation across distributed datasets.
Our contributions are:
\begin{itemize}
\item We present the first private data imputation protocols for horizontally and vertically split data without prerequisite learning.
\item We show that our private data imputation protocols have higher imputation accuracy than local data imputation on a single database {\em even when using performance-enhancing modifications such as quantization}.
\item We show that our private data imputation protocols are cryptographically secure in the semi-honest model with clearly specified small leakage, such as which attribute is being imputed.
\item We show that our private data imputation 
is efficient by imputing over $100,000$ records with $10$ attributes in less than $2.4$ seconds in case of vertically split data, and $8.4$ seconds in case of horizontally split data evaluated in the 10 Gbps network setting.
\end{itemize}


\section{Related Work}
\label{sec:rel-work}

\noindent \textbf{Conventional data imputation and cleaning.} Data imputation can be realized by a variety of different mechanisms from simple such as the mean (average) or mode (most frequent value) computation to more complex techniques based on deep learning (DL) such as discriminative DL (e.g., \cite{jin19}) or generative DL (e.g., \cite{kin14,yoo18}) imputation. However, according to \cite{benchmark}, the $k$-nearest neighbor ($k$-NN) approach that samples a random value from the nearest neighbors provides superior accuracy compared to other solutions.

\medskip \noindent \textbf{Privacy-preserving data imputation and cleaning.}
Privacy-preserving data imputation was first considered by Jagannathan and Wright~\cite{jag06,jag08}. The work considered horizontally partitioned data between two parties and used a lazy decision tree imputation algorithm. Clifton et al.~\cite{cli22} follow a different approach and proposes a differentially private mechanism for multiple item data imputation.
There are other publications (e.g., \cite{ome17,jen24}), but they invoke standard tools, do not show accuracy improvement, or do not provide sufficient treatment of security.

Privacy-preserving data cleaning is a related problem of identifying malformed or misclassified items or outliers. Publications in this space, e.g., \cite{bla23} generally consider different techniques and are not suitable for data imputation. But more fundamentally, data cleaning is the first step of the process, while data correction, as we do in this work, is to be performed afterwards.

\diff{Sublinear lookup methods such as Private Information Retrieval (PIR)~\cite{chor1998private} and Oblivious RAM (ORAM)~\cite{goldreich96} offer asymptotically better amortized complexity for database queries when invoked a superlinear number of times than our linear approach of using PSI. Recent work on secure approximate nearest-neighbor search~\cite{chen20} achieves sublinear complexity using locality-sensitive hashing and ORAM. However, these approaches incur significant computational overhead in practice due to large constant factors even when ignoring the linear setup cost, and they are designed for high-dimensional vector spaces. Our setting differs fundamentally: we perform multi-attribute comparisons with quantized values across distributed datasets. PSI-based approaches, while linear in dataset size, benefit from highly optimized implementations~\cite{volepsi} with small constant factors that remain practical for datasets up to several millions of records, particularly when invoked a constant or sublinear number of times. Furthermore, sublinear methods would require complex composition of multiple private queries for our multi-attribute neighbor computation adding additional complexity to the protocol.}

\section{Preliminaries}
\label{sec:prelim}

We first precisely define the problem we are addressing in this work. We then follow with defining the building blocks that our constructions utilize.

\subsection{Problem Formulation}

Let $T$ be a relational table consisting of $n$ tuples $t_i = \langle x_{i,1}, \ldots, x_{i,m} \rangle$ for $i \in [1, n]$.
Some items are missing, i.e., $x_{i,j} = \perp$, and the goal is to fill, or impute, the missing items with their approximate value. 

We consider different splits of table $T$.
When the table $T$ is \emph{horizontally split} between Alice and Bob, Alice has $n_A$ tuples $t_1$ through $t_{n_A}$ with all $m$ attributes and Bob has $n_B$ tuples $t_{n_A + 1}$ through $t_{n_A + n_B}$ with all $m$ attributes, where $n = n_A + n_B$.
When the table $T$ is \emph{vertically split} between Alice and Bob, Alice has $m_A$ attributes $A_1$ through $A_{m_A}$ and Bob has $m_B$ attributes $A_{m_A + 1}$ through $A_{m_A + m_B}$ for each of $n$ tuples, where $m = m_A + m_B$.
Alice and Bob can link tuples using a common identifier, e.g., the index $i$ with vertically split data.

Let $x_{\alpha, \beta}$ denote the element to be imputed, i.e., it is the $\beta$th attribute of tuple $t_\alpha$. Without loss of generality, we let $x_{\alpha,\beta}$ reside with Bob. That is, in the case of horizontally split data, $t_\alpha$ is among Bob's tuples, while in the case of vertically split data, $\beta$th attribute is in Bob's dataset.

\diff{The horizontal and vertical data splitting scenarios we consider arise naturally in collaborative settings. Horizontal splits occur when organizations independently collect records about different entities (e.g., hospitals in different regions with identical medical attributes). Vertical splits emerge when parties collect complementary attributes about the same entities (e.g., a hospital with clinical measurements and an insurer with claims data). Combining information from multiple parties substantially improves imputation accuracy. Our protocols enable such collaboration without revealing sensitive data, providing formal privacy guarantees throughout the computation.
}

\diff{
We base our approach on Jäger et al.~\cite{benchmark}, which evaluated six imputation methods across 69 datasets with three missingness patterns (missing completely at random MCAR, missing at random MAR, and missing not at random MNAR) and missingness fractions from the range 1--50\%. For numerical columns, $k$-NN ranks in the top three methods in 75\% of cases. For MCAR and MAR patterns with 30--50\% missingness, ML-based methods provide less than 1\% improvement for classification and 0.5\% for regression when training incomplete data~\cite{benchmark}, while $k$-NN maintains robust performance. While ML models offer reusability, this is outweighed by $k$-NN's superior accuracy on incomplete data and compatibility with efficient cryptographic protocols. Our radius-based variant (Section~\ref{sec:nn-comp}) requires no training and naturally integrates with privacy-preserving tools like PSI and OPPRF.
}

In order to replace $x_{\alpha,\beta}$ with its approximate value, we start with the approach that was determined to be among the best in~\cite{benchmark}:
Determine the set $I$ of nearest neighbors of $t_\alpha$.
This was accomplished in~\cite{benchmark} using a $k$-nearest neighbor algorithm that computes $k$ closest tuples to $t_\alpha$.

The value to replace the missing item $x_{\alpha,\beta}$ with is determined from the neighboring records as follows: 

\begin{itemize}
\item For {\em categorical} data: Uniformly select one of the tuples from $I$ and use its $\beta^{\text{th}}$ attribute to impute $x_{\alpha,\beta}$.  Note that the expected value of the selection is the mode, i.e., the most frequent element.

\item For {\em numerical} data: Compute the mean of the $\beta$th attribute values in $I$ and use it to impute $x_{\alpha,\beta}$.
\end{itemize}

\smallskip \noindent 
As far as data privacy goes, in the ideal case, Alice does not learn any information as a result of the computation and Bob learns only the imputed value $x_{\alpha, \beta}$. Note that the output itself discloses some information about Alice's dataset to Bob. For example, when the data is horizontally split, if the computed value of $x_{\alpha,\beta}$ does not appear as the $\beta$th attribute among Bob's tuples, Bob learns one of the Alice's values for that attribute.

We consider standard definitions of secure two-party computation that require that information about private inputs is not disclosed during the computation. We treat the parties as semi-honest adversaries, i.e., those that do not deviate from the computation but might try to deduce unauthorized information from their view during protocol execution. Security is normally shown via a simulation paradigm that allows one to simulate the adversary's view without access to the other party's private information and show that the simulated view is indistinguishable from the real view during the protocol execution.


\subsection{Building Blocks}
\label{sec:bb}
\noindent
\textbf{Oblivious Programmable Pseudo-Random Functions (OPPRF).}
An OPPRF is a secure, cryptographic protocol between two participants, Alice and Bob.
Alice has as input a value $x$ and Bob has as input a set of pairs of values $\mathbb{Y} = \{ (y_i, r_i) \}$ and a key $k$ for a pseudo-random function.
As output of the protocol, Alice learns a value $r$ and Bob learns no new information, i.e., has no output.
It holds that if $(x, s) \in \mathbb{Y}$, then $r = s$ and if $(x, \cdot) \notin \mathbb{Y}$, then $r$ is pseudo-random, i.e., computationally indistinguishable from a uniformly drawn random number.
In a practical protocol instance, Alice's output is usually implemented as a keyed (punctured) pseudo-random function on Alice's value $x$ with Bob's key $k$.
Note that Alice does not learn whether $(x, \cdot) \in \mathbb{Y}$, i.e., Bob should choose its values $r_i$ also randomly.

In our implementation, we use VOLE-OPPRF by Rindal and Schoppmann~\cite{volepsi}.  Each $r_i$ is from the space $\{0, 1\}^\kappa$, where $\kappa$ is the security parameter. While $x$ and $y_i$s can come from a different space, in the implementation on which we rely these values are also cast to be from the same space. 
As the bitlength of the numbers is a function of a security parameter, we use $O(\kappa)$ to denote the cost of a single cryptographic operation. 
Then this construction has a communication and computation cost of $O(n \kappa)$ for building the OPPRF on a set of size $n$ and performing $n$ OPPRF evaluations.

\medskip \noindent \textbf{Hashing and Cuckoo Hashing.} We use a regular cryptographic hash function $H$ that hashes inputs of arbitrary size into fixed-length digests $H: \{0, 1\}^* \rightarrow \{0, 1\}^\kappa$ or into another suitable output space such as $H: \{0, 1\}^* \rightarrow \mathbb{G}$ for a group $\mathbb{G}$ defined above. 

In addition, we rely on Cuckoo hashing that uses $h$ ordinary hash functions $H_1, \ldots, H_h$ with $\mu$ output bins, i.e., each $H_i: \{0, 1\}^* \rightarrow [\mu]$, where notation $[x]$ denotes the set $\{1, \ldots, x\}$. To insert an element $x$ into a Cuckoo hash, we compute $H_1(x), \ldots, H_h(x)$, which are the indices of the candidate bins. If there is an empty bin with index $H_i(x)$ for some $i \in [h]$, we place $x$ into that bin. Otherwise, we evict an element $y$ from one of the candidate bins, place $x$ in that bin, and re-insert $y$ using the same process. 

Given the size of the set $|\mathbb{X}|$ we intend to insert and statistical security parameter $\lambda$, one can compute parameters $h$ and $\mu$ so that the number of bins is linear in the set size $\mu = O(|\mathbb{X}|)$ and with an overwhelming probability $1 - 2^{-\lambda}$ there is an allocation with every bin containing at most one element.

\medskip \noindent \textbf{Private Set Intersection (PSI).} PSI is a well-studied functionality (see, e.g., \cite{fre04,kis05,dec10,jar10,bla16,volepsi}) that allows two participants, Alice and Bob, to compute the intersection of their private input sets without disclosing any information about their sets besides the set cardinality (or an upper bound on the set cardinality). That is, on input $\mathbb{X} = \{x_i\}$ from Alice and $\mathbb{Y} = \{y_i\}$ from Bob, one or both parties learn $\mathbb{X} \cap \mathbb{Y}$.

%

\medskip \noindent \textbf{VOLE-PSI.} VOLE-PSI \cite{volepsi} is a state-of-the-art PSI protocol.
It uses an OPPRF based on Vector-Oblivious Linear Evaluation (VOLE), an extension of the PaXoS data structure \cite{pinkas20}.
VOLE-PSI scales optimally with a constant number of public-key operations and linearly many symmetric key operations in the number of set elements.

VOLE-PSI can be used as a one-sided PSI protocol, where Bob computes the OPRF of his elements and Alice sends the PRFs of her elements for computing the intersection.
Moreover, VOLE-PSI can also be used as a circuit-PSI protocol, where neither party learns the intersection and the parties can proceed with further computation on the private intersection.
In that case, Bob learns the OPPRF of his values and Alice retains one corresponding PRF which matches Bob's if and only if she has the value in his set.

This construction of circuit-PSI is based on previous work~\cite{pinkas18,kerschbaum23} where the parties first place their elements in the bins, but we use a different bin placement strategy per party.
In particular, Bob hashes his elements into the bins using Cuckoo hashing and pads all empty bins to one element.
Alice hashes her elements using all Cuckoo hashing functions, i.e., she replicates the elements if necessary, and pads the size of each bin to a maximum that is only exceeded with negligible probability.
Pinkas et al.~\cite{pinkas18} show how to compute this maximum number from the number of elements.
Alice programs the VOLE-OPPRF to the same random value for each of the elements in her bin, such that if Bob's element is in Alice's bin, they share the same random value.
After such a circuit-PSI, Alice and Bob can run a two-party computation in which they compare the PRFs and then perform any computation over the set intersection.

\medskip \noindent \textbf{Secure Two-Party Computation.} We rely on general-purpose secure computation techniques for components of our solutions. There are a variety of techniques that differ in the number of supported computation participants, participant collusion threshold, and the threat model~\cite{goldreich2004}. Of interest to this work is secure two-party computation with semi-honest participants. The security expectations are such that as a result of computing on private inputs the parties do not learn any information about the private data they handle other than the agreed-upon computation outcome. 

We choose ABY~\cite{aby} for our implementation due to its versatility and the ability to combine different data representations and techniques. The operations we execute using ABY include equality, less-than comparisons, joint sampling of a random value, addition, multiplication, and multiplexor operations to realize branching with private conditions.

\section{MPC-friendly Nearest Neighbor Computation}
\label{sec:nn-comp}

Let $t_\alpha \sim t_\omega$ denote a neighboring relation between tuples $t_\alpha$ and $t_\omega$, where we seek to impute the $\beta$th attribute of $t_\alpha$, $x_{\alpha, \beta}$.
We say tuples $t_\alpha$ and $t_\omega$ are neighbors if all of their non-empty attributes are within a pre-defined distance $r_j$ from each other. That is, 
\begin{align*}
t_\alpha \sim t_\omega \mbox{ iff } & \forall j \in \{1, \dots , m\}\backslash \{\beta\}, \\
& x_{\alpha, j} = \bot \lor x_{\omega ,j} = \bot \lor | x_{\alpha, j} - x_{\omega ,j} | \leq r_j.
\end{align*}
%
Because our goal is to perform imputation using sampling from neighboring records, only tuples with a non-empty $\beta$ attribute are useful for our purpose. For that reason, we redefine the neighboring relation between $t_\alpha$ and any other tuple $t_\omega$ to require that $x_{\omega, \beta} \not= \perp$. That is,
\begin{align*}
t_\alpha \sim t_\omega \mbox{ iff } (x_{\omega,\beta} \not= \perp) \wedge (\forall j \in \{1, \dots , m\}\backslash \{\beta\}, \\
\: x_{\alpha, j} = \bot \lor x_{\omega ,j} = \bot \lor | x_{\alpha, j} - x_{\omega ,j} | \leq r_j).
\end{align*}
In the rest of this work, unless mentioned otherwise, we will use this modified relation to denote neighboring relation. The optimal values for the radius parameters $r_j$ are determined through an adaptive search process during the training phase, which we describe in detail in Section \ref{sec:radii_search}.

Evaluating the above constraints without leaking intermediate results on all tuples requires $O(nm)$ costly less-than comparisons, which we aim to avoid. We therefore reduce less-than comparisons to equality comparisons using quantization.
We quantize the value of the $j$th attribute in each tuple as follows: 
The first quantization places values into ranges $[0, r_j-1], [r_j, 2r_j-1], [2r_j, 3r_j-1], \ldots$ and the second quantization places values into ranges $[0, r_j/2 - 1], [r_j/2, 3/2 r_j - 1], [3/2 r_j, 5/2 r_j - 1], \ldots$.
We then represent each $x_{\alpha, j}$ (and $x_{\omega, j}$, respectively) as its two ranges which we denote as $q_1(x_{\alpha, j})$ and $q_2(x_{\alpha, j})$.
Now, the following relations hold:
\begin{multline*}
| x_{\alpha, j} - x_{\omega, j} | \leq \frac{r_j}{2} \\
\Longrightarrow q_1(x_{\alpha, j}) = q_1(x_{\omega, j}) \lor q_2(x_{\alpha, j}) = q_2(x_{\omega, j})
\end{multline*}
\begin{multline*}
q_1(x_{\alpha, j}) = q_1(x_{\omega, j}) \lor q_2(x_{\alpha, j}) = q_2(x_{\omega, j}) \\
\Longrightarrow | x_{\alpha, j} - x_{\omega, j} | \leq \frac{3}{2}r_j
\end{multline*}

We therefore replace testing for $| x_{\alpha, j} - x_{\omega, j} | \leq r_j$ with $q_1(x_{\alpha, j}) = q_1(x_{\omega, j}) \lor q_2(x_{\alpha, j}) = q_2(x_{\omega, j})$.

Our distance metric not only uses only equality comparisons instead of less-than comparisons, it is also composable across attributes.
This means, in the vertical setting, that Alice can compute the distance to her attributes and Bob's to his.
In this paper, we will show how to efficiently implement this algorithm as a protocol over distributed parties with privacy while preserving the accuracy of the imputation.

\section{Solution Structure}
\label{sec:structure}

We are interested in developing solutions for different types of data partitioning (such as horizontally and vertically split data) and different types of attributes (such as containing categorical and numerical values) that after optimizations result in significantly different protocols. However, we put forward a general computation structure that accommodates different types of data partitioning as well as different ways of determining the imputation value. 

Recall that the goal is to compute the imputation value for the item $x_{\alpha,\beta}$, where $\alpha$ defines the tuple $t_{\alpha}$ and $\beta$ is the attribute of the tuple.
Conceptually, the computation consists of two main components:
(i) privately determining the neighbors of the tuple $t_\alpha$ in the partitioned dataset and (ii) privately analyzing the $\beta$th attribute of the neighbor records to determining the imputation value for $x_{\alpha,\beta}$. Without loss of generality, let the item $x_{\alpha,\beta}$ be located among Bob's records. 

\begin{figure}[t]
    \framebox{
	\begin{minipage}{0.95\linewidth}

    {\sc Parameters:} There are $n$ tuples consisting of $m$ attributes horizontally or vertically partitioned between Alice and Bob.
    \medskip

    \noindent{\sc Functionality:}
    \begin{itemize}
        \item Receive private records from Alice.
        \item Receive private records as well indices $\alpha$ and $\beta$ from Bob.
        \item Determine the neighbors of $t_{\alpha}$ and create an indicator bit vector of size $n-1$, where the value in the position $\omega$($\not=\alpha$) is set to 1 if and only if $t_\omega$ is a neighbor of $t_\alpha$ and $x_{\omega,\beta}$ is not missing.
        \item Create two shares of each bit of the indicator vector and give one share to Alice and the other share to Bob.
    \end{itemize}
    
    \end{minipage}
    }
    \caption{Ideal functionality for neighbor computation.}
    \label{fig:ideal-func-1}
\end{figure}
Figure~\ref{fig:ideal-func-1} describes the ideal functionality of the first component, where Alice and Bob privately enter their data and Bob also supplies the indices $\alpha$ and $\beta$. As a result of this phase of the computation, Alice and Bob obtain a secret-shared indicator vector, where each tuple $t_\omega$ (except tuple $t_\alpha$) is marked with a bit indicating whether $t_\omega$ is a neighbor of $t_\alpha$ according to the neighbor computation described in Section~\ref{sec:nn-comp} and the value $x_{\omega,\beta}$ is not missing. By including only neighbor tuples with a non-empty value in the desired attribute, we are able to directly use those records in the computation, without having to additionally check if the desired attribute is present.

\begin{figure}[t]
    \framebox{
	\begin{minipage}{0.95\linewidth}

    {\sc Parameters:} There are $n$ tuples consisting of $m$ attributes horizontally or vertically partitioned between Alice and Bob.
    \medskip

    \noindent{\sc Functionality:}
    \begin{itemize}
        \item Receive private records and a share of the indicator vector $v$ from Alice.
        \item Receive private records, indices $\alpha$ and $\beta$, and a share of the indicator vector $v$ from Bob.
        \item Extract the $\beta$th attribute from all tuples $t_{\omega}$ where $\omega \not= \alpha$.
        \item Conditionally use each extracted value $x_{\omega,\beta}$ in the computation of the imputed value using the corresponding flag $v_\omega$ from the indicator vector $v$:
        \begin{itemize}
        \item for numerical data: $x_{\alpha,\beta} = (\sum_{\omega,v_{\omega}=1} x_{\omega,\beta}) / (\sum_\omega v_\omega)$ 
        \item for categorical data: $x_{\alpha,\beta} \stackrel{R}{\leftarrow} \bigcup_\omega \{x_{\omega,\beta} \mid v_\omega = 1\}$
        \end{itemize}
        \item Give the computed value $x_{\alpha,\beta}$ to Bob.
    \end{itemize}    
    \end{minipage}
    }
    \caption{Ideal functionality for imputation value computation.}
    \label{fig:ideal-func-2}
\end{figure}
The ideal functionality for the second component is given in Figure~\ref{fig:ideal-func-2}. It uses the indicator vector to conditionally include the $\beta$th attribute of a tuple into the computation of the imputation values. Note that the computation for determining the imputation value significantly differs for numerical data (mean computation) and categorical data (random sampling). Notation $x \stackrel{R}{\leftarrow} X$ means drawing an element uniformly at random from the set $X$.

Our solutions introduce optimizations that fit within this general structure. For example, in the case of horizontally split data Bob can locally determine the neighbors of $t_\alpha$ among his tuples and interactive computation is only needed to compare Alice's tuples to Bob's $t_\alpha$. In the case of vertically partitioned data, Alice and Bob can determine local neighbors among all records using partial attributes known to each if Alice gets the value of $\alpha$ from Bob. This reduces the size of the joint computation and consequently the indicator vector that the parties need to maintain.

\ignore{
\subsection{Batch Imputation}
\label{sec:batch-imp}

\diff{
While we focus on imputing a single value $x_{\alpha,\beta}$, our protocols naturally extend to batch imputation of multiple missing values. When Bob has $\ell$ missing values across potentially different tuples and attributes, the protocols can be trivially executed $\ell$ times. An important optimization is possible when multiple values are missing within the same tuple $t_\alpha$: the neighbor computation phase (Figure~\ref{fig:ideal-func-1}) is performed once for $t_\alpha$, and only the imputation value computation (Figure~\ref{fig:ideal-func-2}) is repeated for each missing attribute $\beta_1, \ldots, \beta_\ell$. This reduces computational overhead by eliminating $(\ell-1)$ repetitions of the expensive neighbor computation (OPPRF evaluation for horizontal splits or PSI for vertical splits), while requiring only $\ell$ lightweight aggregation or sampling operations over the shared neighbor set. Likely further optimizations are possible, but we leave them to future work.
}
 \karim{A detailed cost analysis can be provided in the Appendix.}
}

\section{Horizontally Split Data}
\label{sec:h}

In the horizontally split case, Bob has a tuple $t_\alpha$ with a missing value for attribute $\beta$ -- i.e., $x_{\alpha,\beta}$ has no value -- which he would like to impute.
Bob can find neighbors of $t_\alpha$ among his tuples using local computation, while Alice needs to identify $t_\alpha$'s neighbors among her tuples using secure comparisons. Note that in the context of the ideal functionality in Figure~\ref{fig:ideal-func-1}, the parties need to determine the indicator vector only for Alice's tuples, as Bob can locally determine the neighbor status of his tuples (and enter them directly into the computaion that follows).

\subsection{Determining Neighbor Tuples}
\label{sec:h-initial}

Let $I_A$ and $I_B$ be the set of neighbors among Alice's and Bob's tuples, respectively. As the first component of the solution, the parties need to jointly determine $I_A$.

We can determine which records among Alice's tuples have identical quantized ranges using OPPRF as follows:
Bob locates a tuple $t_\alpha$ which has a missing value for attribute $\beta$, i.e., $x_{\alpha,\beta}$ has no value.
We could compare Bob's $t_\alpha$ to each $t_\omega$ in Alice's dataset by comparing their quantized values for each attribute.
However, in that case we would still need to perform a logical-OR operation between compared pairs to obtain the final result.
Instead, we reduce the computation to one invocation of an OPPRF as described next.

For each attribute $j$, let $q'_1(x_{i,j})$ be the closest to item $x_{i,j}$'s quantized range in the first quantization method that is not the true range $q_1(x_{i,j})$. Similarly, let $q'_2(x_{i,j})$ be the closest range $\not=q_2(x_{i,j})$ using the second quantization method. 
For each attribute $j$ and each tuple index $\omega \in [n_A]$, Bob draws a random value $\sigma_{\omega, j}$ and programs the OPPRF as follows:
\begin{itemize}
\item ${\sf OPPRF}(q_1(x_{\alpha,j})||q_2(x_{\alpha,j})) = \sigma_{\omega, j}$
\item ${\sf OPPRF}(q_1(x_{\alpha,j})||q'_2(x_{\alpha,j})) = \sigma_{\omega, j}$
\item ${\sf OPPRF}(q'_1(x_{\alpha,j})||q_2(x_{\alpha,j})) = \sigma_{\omega, j}$
\item ${\sf OPPRF}(\bot) = \sigma_{\omega, j}$
\end{itemize}
where $||$ denotes concatenation.

For each tuple $t_\omega$ in Alice's dataset, Alice  uses both quantizations of each of her $x_{\omega, j}$ (for $j \not= \beta$) or $\bot$ (if $x_{\omega, j} = \bot$) to query the OPPRF, i.e., Alice privately computes with Bob:
\begin{itemize}
\item $\tau_{\omega, j} = {\sf OPPRF}(\bot)$ if $x_{\omega, j} = \bot$
\item $\tau_{\omega, j} = {\sf OPPRF}(q_1(x_{\omega, j})||q_2(x_{\omega, j}))$ otherwise.
\end{itemize}

\smallskip \noindent 
At this point, it holds that
$$
t_\alpha \sim t_\omega \Longleftrightarrow \sum_{j \in \{1, \dots , m\}\backslash \{\beta\} } \sigma_{\omega, j} = \sum_{j \in \{1, \dots , m\}\backslash \{\beta\} } \tau_{\omega, j}
$$
%
%
Note that Bob independently chooses a uniformly random $\sigma_{\omega, j}$ for each of the $n_A \cdot (m-1)$ invocations of the OPPRF, so that Alice cannot infer any information from observing multiple invocations.
That is, even if Alice queries the OPPRF on the same inputs, the outputs she observes are independent of each other.


Also note that the solution, as described above, can handle missing values in Alice's dataset, but not in Bob's tuple $t_\alpha$. 
In practice, in addition to missing $x_{\alpha, \beta}$ in $t_\alpha$, Bob may have a missing value, $x_{\alpha, \gamma}$, for another attribute $\gamma$ in tuple $t_\alpha$.
Thus, we need to address this before we can proceed. Our solution to mitigating the issue is to have Bob program the OPPRF at $\gamma$ to the same $\sigma_{\omega,\gamma}$ for each possible value of $x_{\omega,\gamma}$ in Alice's dataset. This means that OPPRF evaluation on Alice's value $x_{\omega,\gamma}$ will always produce the expected $\sigma_{\omega,\gamma}$ regardless of its value $x_{\omega,\gamma}$.  


After evaluating the OPPRF on Alice's tuples, the next step is to compute which of Alice's tuples are in $I_A$. A simple solution is for Bob to send her the expected $\sigma_{\omega} = \sum_{j\in[m]\setminus \{\beta\}} \sigma_{\omega, j}$ for each $\omega \in [n_A]$. This allows Alice to determine whether any given $t_\omega$ is a neighbor of Bob's tuple $t_\alpha$ and thus compute $I_A$. As we show below, this variant leaks an undesirable amount of information; however, we leave it for comparison purposes and denote it as \emph{plain neighbor access.} 

A secure way of computing $I_A$ consists of Alice and Bob privately comparing the computed $\tau_{\omega} = \sum_{j\in[m]\setminus \{\beta\}} \tau_{\omega, j}$ and the expected $\sigma_{\omega} = \sum_{j\in[m]\setminus \{\beta\}} \sigma_{\omega, j}$ for equality for each of Alice's tuples $\omega \in [n_A]$. This variant results in a secret-shared indicator vector, and we call it \emph{blind neighbor access}.

The next step is to use the results of neighbor computation to determine the imputation value.

\subsection{Imputation Value for Numerical Data}

In the case of numerical data, the imputation value is computed as the mean of the $\beta$th attribute among the neighboring records in $I_A \cup I_B$. 

\subsubsection{Plain Neighbor Access}
\label{sec:h-mean-plain}

Computing the imputation value from the set of neighbors $I_A \cup I_B$ is much easier if Alice and Bob know their respective sets of neighbors $I_A$ and $I_B$. After Alice receives the expected $\sigma_\omega$s, she determines $I_A$ and Bob knows $I_B$.

Next, Alice and Bob will need to compute the sum of $\beta$th attribute of their tuples in $I_A$ and $I_B$ and divide the sum by $|I_A \cup I_B|$. To achieve that, Bob computes a partial sum $s_B$ of the values in $I_B$, while Alice, respectively, computes $s_A$ using her tuples $I_A$. The remaining computation that we need to perform privately is $(s_A + s_B) / (|I_A| + |I_B|)$, with Alice possessing $s_A$ and $|I_A|$ and Bob possessing $s_B$ and $|I_B|$. This can be very efficiently evaluated using existing secure two-party computation tools such as ABY.

\subsubsection{Blind Neighbor Access}
\label{sec:h-mean-blind}

The above solution with plain neighbor access reveals the set $I_A$ of neighbors to Alice.
This allows Alice to conduct a simple inference attack on Bob's $t_\alpha$ by simply averaging her tuples in $I_A$ and treating the result as an approximation of $t_\alpha$.
To prevent this attack, we should not reveal $I_A$ to Alice. 

We thus can represent each tuple $t_\omega$ in Alice's set by two values: a private flag which is set iff $t_\omega \in I_A$ (i.e., the indicator vector) and the value of the $\beta$th attribute $x_{\omega,\beta}$. This will allow us to compute with Alice's dataset by entering all $n_A$ tuples into the computation and adding only the values from the tuples where the flag is set. 

This conditional addition can easily be implemented using existing secure computation tools using a single pass through Alice's tuples. That is, after the OPPRF evaluation in Section \ref{sec:h-initial}, Alice and Bob securely compare the expected $\sigma_\omega$ and computed $\tau_\omega$ values for equality and add the item $x_{\omega,\beta}$ to Alice's sum $s_A$ if the values are equal. The remaining computation to determine the mean proceeds as specified in Section \ref{sec:h-mean-plain}.

\subsection{Imputation Value for Categorical Data}

For categorical data, we randomly sample an element from $I_A \cup I_B$ and extract its $\beta$th element.

\subsubsection{Plain Neighbor Access}
\label{sec:h-random-plain}

In this variant, Alice knows the set of neighbors $I_A$ among her records and Bob has neighbors $I_B$ in this records. Next, Alice and Bob will need to securely sample one element from the union $I_A \cup I_B$ of their neighbors to determine the imputation value.


Let Alice randomly sample one of the records in $I_A$ and extract its $\beta$th attribute, which we denote by $v_A$. (Note that by the definition of neighbor relationship, Alice will only consider the tuples with non-empty $\beta$th attribute and any tuple in $I_A$ can be chosen as her sample.)
Similarly, Bob randomly samples one of the records in $I_B$ and extracts its $\beta$th attribute, denoted by $v_B$ (similarly, it is guaranteed to be non-empty). The parties now need to select $v_A$ or $v_B$ in such a way that the sample is drawn uniformly at random from the union $I_A \cup I_B$.

To accomplish that, the parties could draw a random integer $R \in [0, |I_A \cup I_B|-1]$ and securely compare it to $|I_A|$. 
If $R \ge |I_A|$, Bob replaces $x_{\alpha,\beta}$ with his value $v_B$; otherwise, he learns Alice's $v_A$ and replaces $x_{\alpha,\beta}$ with $v_A$. More precisely, if $c$ denotes the result of the private comparison $R \ge |I_A|$, the parties privately compute $c \cdot v_B + (1-c) \cdot v_A = c \cdot (v_B-v_A) + v_A$ using a single multiplication.

A complication is that the parties do not know the sizes of each other's set $I_A$ or $I_B$ to draw $R$ from the desired range. Our solution is to draw a value from the fixed range $[0, 1)$ and scale it by a factor $|I_A|+|I_B|$ using a single multiplication. Individual sizes $|I_A|$ and $|I_B|$ can serve the purpose of additive shares of $|I_A|+|I_B|$, making it very efficient.

In our implementation, sampling random $R$ from $[0, 1)$ is realized using $p$ bits of precision, i.e., the parties draw from the set $[0, 2^{p}-1]$. This is accomplished by Alice and Bob each entering $p$ private bits into the secure computation and XORing them prior to scaling the result by $|I_A| + |I_B|$.

Next, notice that with this solution Bob learns the result of imputation $v$, which is either $v_A$ or $v_B$ and Bob knows $v_B$. This means that if the imputed value is $v = v_A$ while $v_B$ differs, Bob knows that the sampled value came from Alice and thus Alice has $v_A$ in her $I_A$ records. If $v_A$ also appears in Bob's records $I_B$, this is the knowledge he would not be able to deduce from the output alone. That is, if Bob learns that $v = v_A$ and $v_A$ appears in his records, he should not know whether the value came from his or Alice's records.

To eliminate this leakage, we modify the solution so that Bob does not learn $v_B$ sampled from his records. In the scenario above, this would prevent him from knowing that $v_A$ originated from Alice's records. To achieve this, we have Bob enter all of the $\beta$ attributes from the records in $I_B$ into secure computation, one of which is sampled privately. Specifically, we draw a random value in the range $[0, |I_B|-1]$ inside secure two-party computation and scan through Bob's input to obliviously select the element at that index. The remaining computation proceeds as before.

\subsubsection{Blind Neighbor Access}
\label{sec:h-random-blind}

This time, to prevent inference attacks, Alice does not learn which of her tuples are in $I_A$. Rather, Alice and Bob securely compare their $\sigma_\omega$ and $\tau_\omega$ sums for each of Alice's tuples $t_\omega$ and secret share a private flag that indicates whether the tuple is in $I_A$.

Let there be $\ell$ neighbors in a set of $n_A$ tuples.
If $\ell \ll n_A$ and Alice samples blindly from her set of tuples, the probability that she samples a neighbor is low. This means that many samples are needed to ensure that she samples a neighbor at least once.
To overcome this, our solution is to combine $c$ tuples together into a single value by adding their neighbor flags and  the $\beta$th attribute's values.
If the combined neighbor count is exactly 1, then we can use the corresponding value $v_A$ that contains the sum of the attribute values in the sampling procedure above.

We can determine the optimal value of $c$ as follows:
If the probability that each tuple is a neighbor is $\frac{\ell}{n_A}$ and the probabilities are independent of each other, the probability that exactly one flag will be set among $c$ tuples is $$q(c) = c \cdot \frac{\ell}{n_A} \left(1 - \frac{\ell}{n_A}\right)^{c-1}.$$ 
That is, there are $c$ disjoint combinations for one of the draws to be a neighbor (with probability $\frac{\ell}{n_A}$) and $c-1$ other draws to be not a neighbor (with probability $1 - \frac{\ell}{n_A}$ each). The value of $c$ that maximizes the probability function $q(c)$ is $c =\frac{-1}{\ln{(1 - \frac{\ell}{n_A} ) }}$ and we set it based on the expected number of neighbors among Alice's tuples.

%

The next step is to pack the tuples into enough slots~-- each with $c$ tuples~-- so that the probability that at least one of the slots will have exactly one neighboring tuple is above a certain threshold $1 - \varepsilon$. Given the desired bound on the probability of failure $\varepsilon$, the number of sufficient slots $d$ is determined as the lowest $d$ that satisfies $$\left(1-q(c) \right)^d \le \varepsilon.$$ When $d \cdot c \ge n_A$, a single tuple can be packed more than once, but the work to compute their neighbor status is bounded by ${n_A}$. When $d \cdot c < n_A$, on the other hand, we only need to evaluate the OPPRF and compute the neighbor flags for the $d \cdot c$ tuples chosen to be packed.

That is, Bob and Alice first compute $c$ and $d$ based on an estimated number of neighbors in Alice's dataset. If $d \cdot c \ge n_A$, Alice and Bob compute OPPRF values $\sigma_\omega$ and $\tau_\omega$ for $1 \leq \omega \leq n_A$. Otherwise, they compute these values only for the $d \cdot c$ tuples selected for this purpose. Comparison of the OPPRF evaluations and all consecutive steps are performed using secure computation. 
The parties then compute packed representation of $d$ samples with $c$ tuples each. To achieve random selection, tuples need to be selected and packed in a randomized order. Each of the $d$ slots now contains the sum of $c$ neighbor flags (the count) together with the sum of the $\beta$th attributes of the corresponding tuples.

%
Sampling can be performed by scanning the slots and obliviously grabbing the value in the first slot with its count being 1.
The chosen value will be a valid neighbor with probability at least $1 - \varepsilon$. That is, there is no valid sample across all slots with probability at most $\varepsilon$. (Sample validity can easily be checked and the computation aborted otherwise). The rest proceeds as described in Section \ref{sec:h-random-plain} by drawing a random sample from Bob's neighbors and selecting either Alice's or Bob's sample. 

Choosing between Alice's and Bob's sample requires using $|I_A|$ in the computation. However, if $d \cdot c < n_A$ and we do not perform OPPRF evaluation for all $n_A$ Alice's tuples, we cannot compute $|I_A|$. In that case $|I_A|$ is approximated as $\frac{n_A}{d \cdot c} |I'_A|$, where $|I'_A|$ is the number of neighbors among the evaluated $d \cdot c$ tuples.



\section{Vertically Split Data}
\label{sec:v}

In vertically split data, Bob also locates a tuple $t_\alpha$ which has a missing value for attribute $\beta$.
This time, each party has a partial set of attributes and neighbor computation has a completely different structure.


\subsection{Determining Neighbor Tuples}

This time, Bob can determine $t_\alpha$'s neighbors in this dataset by local comparisons using the attributes he possesses.
This, however, is a superset of the actual neighbors, since he does not have access to Alice's attributes.
Similarly, Alice can determine her neighbors (assuming she knows $\alpha$) which are also a superset of the actual neighbors. Let $I_A$ and $I_B$ be the sets of row indices of Alice's and Bob's local neighbors, respectively. 

Note that because the neighbor tuples are intended for imputation, the parties need to select only those tuples as neighbors where the $\beta$th attribute is not empty. With vertically split data, this attribute resides with Bob and thus he selects only those neighbors where the $\beta$th attribute is not missing.

In the context of the ideal functionality of Figure~\ref{fig:ideal-func-1}, we do not need to maintain the indicator vector for all tuples, as the parties can locally filter out some of the tuples as not being relevant. In particular, the tuples of interest are only those in $I_A$ and $I_B$ as potential neighbors, while the remaining tuples are automatically removed from consideration as non-matching.

The set of global neighbors can be determined as $I_A \cap I_B$. We therefore have Alice and Bob engage in a private set intersection (PSI). To protect the number of local neighbors, the parties can pad their inputs by a constant fraction of the dataset size to remove any leakage about the number of local neighbors.





A simpler version involves the parties engaging in a conventional PSI, with the result being revealed to Bob. This undeniably discloses additional information to Bob beyond the indented output and we refer to it as plain neighbor access. A version that properly protects the data, blind neighbor access, involves the parties engaging in circuit-PSI that produces a secret-shared vector of bits, i.e., the indicator vector. That version will allow Alice and Bob to securely compute an imputation value using global neighbors. 

\subsection{Imputation Value for Numerical Data}
\label{sec:v-mean}

\subsubsection{Plain Neighbor Access}
\label{sec:v-mean-plain}

Computing the imputation value is easy with access to the neighboring tuples. 
That is, after Bob obtains the set intersection $I_A \cap I_B$, he can simply compute the mean of the attribute values in the set intersection since he has access to all values for attribute $\beta$.

\subsubsection{Blind Neighbor Access}
\label{sec:v-mean-blind}



To reduce the information leakage above, the parties utilize circuit-PSI to obtain secret-shared output.
We can use VOLE-PSI as our circuit-PSI protocol, which relies on Cuckoo and standard hashing as was described in Section \ref{sec:bb}. 
In that case, Alice and Bob each obtain a vector of random values as the PSI output.
These random values are equal if Bob's element is in Alice's set and otherwise they are different.
After comparing those random values using a two-party computation (ABY in our implementation), Alice and Bob hold secret shares of the indicator vector of neighboring elements.

What remains is to securely scan through the vector adding the $\beta$th attributes of the tuples in the intersection, after which the sum is divided by the size of the intersection. This is performed using secure two-party computation.

\subsection{Imputation Value for Categorical Data}

This time instead of computing the mean we sample one value from the intersection at random. The bulk of the computation remains the same as in Section \ref{sec:v-mean} and we need to modify only the last step.

\subsubsection{Plain Neighbor Access}
\label{sec:v-random-plain}

As before, this variant is easy: after Bob learns the intersection, he simply samples one value from the $\beta$th attribute of the neighbor tuples.

\subsubsection{Blind Neighbor Access}
\label{sec:v-random-blind}

Similar to the solution in Section \ref{sec:v-mean-blind}, Alice and Bob engage in circuit-PSI and obtain a vector of secret-shared bits that indicate which tuples are global neighbors. 
The parties now randomly select one of the global neighbors using secure two-party computation as follows. 
They compute the size $\eta$ of the set intersection from the indicator vector and jointly sample a random value $\mu \in [0, \eta-1]$  with $p$ bits of precision.
That is, Alice and Bob enter randomly chosen $p$-bit $r_A$ and $r_B$, respectively, add them using secure computation $r = r_A + r_B \bmod 2^p$, scale the result by multiplying it by $\eta$, and truncate the resulting $\eta \cdot r$ by $p$ bits.
Finally, they scan through the items and privately select the $\mu$th element from the intersection using the indicator vector and output its value to Bob.


\section{Analysis}
\label{sec:analysis}

We next analyze performance of our solutions as a function of the input size and security parameters, as well as show that they achieve security under the standard security definitions in the presence of semi-honest participants.

\subsection{Complexity Analysis}

\begin{table*}[t]
\centering
\caption{Complexity of proposed solutions.}
\label{tab:complexity}
\begin{tabular}{l|l|c|c} \hline
Split method & \hspace{0.32in} Sampling type & Computation & Communication \\ \hline \hline

\multirow{4}{*}{Horizontal} & Plain, Mean (Sec.~\ref{sec:h-mean-plain}) & \multirow{2}{*}{$O(\kappa(m \cdot n_A + |I_B|))$} & \multirow{2}{*}{$O(\kappa(m \cdot n_A + |I_B|))$} \\

& Plain, Random (Sec.~\ref{sec:h-random-plain}) & & \\ \cline{2-4}

& Blind, Mean(Sec.~\ref{sec:h-mean-blind}) & $O(\kappa ((m + \kappa) n_A + |I_B|))$ & $O(\kappa ((m + \kappa) n_A +  |I_B|))$ \\ 

& Blind, Random (Sec.~\ref{sec:h-random-blind}) & $O(\kappa ((m + \kappa) \min(n_A, c \cdot d) + |I_B|))$ & $O(\kappa ((m + \kappa) \min(n_A, c \cdot d) +  |I_B|))$ \\ \hline

\multirow{4}{*}{Vertical} & Plain, Mean (Sec.~\ref{sec:v-mean-plain}) & \multirow{2}{*}{$O(\kappa(|I_A| + |I_B|))$} & \multirow{2}{*}{$O(\kappa(|I_A| + |I_B|))$} \\ 

& Plain, Random (Sec.~\ref{sec:v-random-plain}) & & \\ \cline{2-4}

& Blind, Mean (Sec.~\ref{sec:v-mean-blind}) & \multirow{2}{*}{$O(\kappa(|I_A| + |I_B|))$} & \multirow{2}{*}{$O(\kappa(|I_A| + |I_B|))$} \\

& Blind, Random (Sec.~\ref{sec:v-random-blind}) & & \\ \hline
\end{tabular}
\end{table*}

\tbl{tab:complexity} summarizes the computation and communication complexity of the developed solutions. We discuss horizontally and vertically partitioned constructions in turn.

\medskip \noindent \textbf{Horizontally split data.} The core of our solutions for horizontal partitioning is OPPRF evaluation of quantized tuple values for the purposes of determining nearest neighbors among Alice's records. This involves building and evaluating an OPPRF on $O(m \cdot n_A)$ inputs. Because each evaluation involves cryptographic operations (although inexpensive) that depend on a security parameter $\kappa$, we express the overall cost as $O(\kappa \cdot m \cdot n_A)$. This is also the amount of communication associated with OPPRF evaluation.

The imputation value computation contributes additional cost. With plain neighbor access as described in Sections \ref{sec:h-mean-plain} and~\ref{sec:h-random-plain}, computing the imputation value from the nearest neighbors among Alice's records is simple. Doing so for Bob's neighbors, on other other hand, involves work linear in the number of neighbors, $|I_B|$, performed using secure two-party computation, which contributes $O(\kappa \cdot |I_B|)$ to both computation and communication.

Blind neighbor access, as described in Sections \ref{sec:h-mean-blind} and~\ref{sec:h-random-blind}, has two significant differences. First, determining the nearest neighbors among Alice's records and sampling one of them is now more complex. For the mean computation, the largest component involves performing $n_A$ equality comparisons on $\kappa$-bit inputs. While it is possible to perform this computation on 1-bit shares, e.g., using GMW~\cite{gmw} combined with oblivious transfer for multiplication triple generation~\cite{aby}, this will result in increasing the round complexity of the construction. In our implementation, we choose to proceed with garbled circuit evaluation~\cite{garbled-circuits} from ABY, which represents each bit as a $\kappa$-bit label. This means that the work and communication asymptotically rise to $O(\kappa^2 \cdot n_A)$ in our implementation.
Second, for random sampling, Alice's items are combined, which impacts the cost of OPPRF evaluation. OPPRF construction and evaluation in Section \ref{sec:h-random-blind} involves $O(\kappa \cdot m \cdot \min(n_A, c \cdot d))$ computation and communication.


\medskip \noindent \textbf{Vertically split data.} Our solutions for vertical partitioning are fundamentally different and involve evaluating PSI as the first step. This can be implemented using $O(\kappa (|I_A| + |I_B|))$ computation and communication. This time, $I_A$ and $I_B$ correspond to the nearest neighbors computed using partial matching and thus are larger in size than the set of true neighbors (and consequently larger than the corresponding sets $I_A$ and $I_B$ from the horizontally split solutions).

Determining an imputation value with plain neighbor access, as described in Sections \ref{sec:v-mean-plain} and~\ref{sec:v-random-plain} does not have additional costs, while blind neighbor access from Sections \ref{sec:v-mean-blind} and~\ref{sec:v-random-blind} requires us to protect the result of set intersection and sample from the intersection using secure two-party computation. This component is also linear in the size of the local sets of nearest neighbors, where we invoke circuit-PSI that uses a hash table of linear size and scan through the result to determine the imputation value. 

\subsection{Security Analysis}

We use a standard simulation-based formulation of security for two-party computation to demonstrate that the proposed protocols provide the expected security guarantees. Because the information disclosure with plain sampling or mean computation allows for inferences, our focus is on the blind variants of the protocols. Due to space limitations, detailed security analysis is given in Appendix~\ref{sec:security}, while \tbl{tab:disclosure} summarizes information disclosure of different constructions.
 
\begin{table}[t]
\caption{Information disclosure summary.}
\label{tab:disclosure}
\centering
\begin{tabular}{l|c|c|c} \hline
\multirow{2}{*}{Splitting} & \multirow{2}{*}{Sampling} & \multicolumn{2}{c}{Information disclosure} \\ \cline{3-4}
& & to Bob & to Alice \\ \hline
\multirow{2}{*}{Horizontal} & Plain & \cellcolor{lightgray} & $\beta$, $I_A$\\
& Blind & \cellcolor{lightgray} & $\beta$ \\ \hline
\multirow{2}{*}{Vertical} & Plain & $I_A \cap I_B$ & $\alpha$ \\
& Blind & \cellcolor{lightgray} & $\alpha$ \\ \hline
\end{tabular}
\end{table}

In Appendix~\ref{sec:ext}, we propose modifications to our protocols which eliminate all leakage, but we leave whose evaluation to future work due to their large expected run time.

\section{Evaluation}
\label{sec:eval}

In this section, we present a thorough evaluation of our private data imputation protocols, analyzing their performance across varying datasets and configurations. Our evaluation is centered on two main criteria: (i) accuracy, demonstrating the quality of imputed data when leveraging private inputs and (ii) efficiency, examining computational and communication costs. We show the effectiveness of combining private data in enhancing imputation accuracy and demonstrate that our protocols are computationally efficient, completing imputation tasks on datasets with up to 100,000 records within seconds, even in constrained network settings.

Before presenting our experimental results, we first detail our methodology for selecting the radius parameter referenced in Section \ref{sec:nn-comp}, which is a critical component that distinguishes our approach from traditional $k$-NN imputation and directly impacts both the accuracy and privacy properties of our protocol.

\subsection{Adaptive Radii Search}
\label{sec:radii_search}

For each attribute $j$ in our protocol, we determine an optimal radius $r_j$ through a systematic optimization process during the training phase. This section details our empirical methodology for radius selection, which balances imputation accuracy with computational efficiency requirements.

\medskip \noindent \textbf{Radius Optimization.}
Our radius selection process follows a data-driven approach:
\begin{itemize}
\item \textbf{Initialization:} We begin with an initial radius of $0.35$ times the standard deviation of each attribute, providing a starting point that accounts for the natural variability of each feature.
\item \textbf{Adaptive search:} We employ a convex optimization procedure that systematically explores the radius parameter space. The search begins with small radii values and incrementally increases them while monitoring imputation quality through RMSE on a validation dataset.
\item \textbf{Stopping criteria:} The search terminates after observing three consecutive performance declines, indicating we have likely passed the optimal radius value. This approach ensures we find a near-optimal radius without exhaustively searching the entire parameter space. 
\item \textbf{Parameter adaptation:} We employ an adaptive step size strategy with a decay rate of 0.9. It progressively reduces the step size after performance declines to enable finer-grained exploration around promising values.
\end{itemize}
This optimization process occurs separately for each attribute, resulting in attribute-specific radii that account for their unique statistical properties and contribution to imputation accuracy. The final selected radii are then used during the testing phase to evaluate imputation performance.

\medskip \noindent \textbf{Statistical Properties and Privacy Implications.}
Our radius-based approach offers several important statistical and privacy advantages:
\begin{enumerate}
\item \textbf{Attribute-specific calibration:} By optimizing each attribute's radius independently, our method adapts to different scales and distributions across features, avoiding the one-size-fits-all limitation of traditional $k$-NN. This is particularly important in heterogeneous datasets with attributes of vastly different characteristics.
\item \textbf{Density adaptation:} Unlike $k$-NN which always selects exactly $k$ neighbors regardless of data density, our approach naturally adapts to varying data densities by considering all records within the optimized distance threshold. This prevents forced inclusion of dissimilar records in sparse regions.
\item \textbf{Outlier robustness:} The radius approach inherently provides protection against outlier influence by excluding distant points regardless of the number of available neighbors, maintaining relevance of the imputation basis.
\item \textbf{Privacy-utility tradeoff management:} The radius directly influences the privacy-utility tradeoff by controlling both the neighbor set size and the quantization granularity. Smaller radii create finer-grained partitions that yield better imputation accuracy but potentially reveal more information about the underlying data distribution. Larger radii increase the anonymity set size for each comparison, enhancing privacy protection but potentially reducing imputation accuracy. Our optimization process finds an effective balance point on this privacy-utility curve.
\item \textbf{Compatibility with secure computation:} The quantization approach enabled by our fixed-radius method is particularly well-suited for secure multi-party computation environments, as it transforms expensive range comparisons into efficient equality checks that can be implemented using protocols like PSI and OPPRF as described in Section \ref{sec:bb}.
\end{enumerate}
The empirical results that follow demonstrate that our method consistently outperforms standard imputation techniques across different missingness types while maintaining practical computational efficiency, validating our approach to radius selection and optimization.

\begin{table}[t]
\caption{Overview of the datasets used. Imp. is the dataset's column we used to evaluate the imputation methods, Att. is the number of attributes, and Feat. is the number of numerical columns.} 
\label{table:overview}
\centering
\setlength{\tabcolsep}{0.25ex}
\begin{tabular}{c c c c c} 
 \hline
 ID & Name & Imp. & Att. & Feat. \\ 
 \hline\hline
 
 42636  & Long                & x18                 & 4477    & 20 \Tstrut\\
 287    & wine\_equality      & sulphates           & 6497    & 12 \\
 189    & kin8nm              & theta8              & 8192    & 9 \\
 197    & cpu\_act            & pgin                & 8192    & 22 \\
 198    & delta\_elevators    & curRoll             & 9517    & 7 \\
 23515  & sulfur              & a3                  & 10081   & 7 \\
 42183  & dataset\_sales      & day                 & 10738   & 15 \\
 216    & elevators           & climbRate           & 16599   & 19 \\
 218    & house\_8L           & P3                  & 22784   & 9 \\
 215    & 2dplanes            & x6                  & 40768   & 11 \\
 1200   & BNG(stock)          & company1            & 59049   & 10 \\
 23395  & COMET\_MC\_SAMPLE   & wire\_id            & 89640   & 6 \Bstrut\\
 \hline
\end{tabular}
\end{table}

\subsection{Accuracy}
\label{sec:accuracy}

For our accuracy experiments, we used twelve current real-world datasets from OpenML~\cite{openml}, each containing numeric columns. The datasets are summarized in \tbl{table:overview}. Missing values are artificially introduced in both training and testing data, allowing us to assess imputation accuracy in a scenario where models are trained and evaluated on corrupted data.

Our accuracy results focus on demonstrating how leveraging additional private data from Alice can improve the imputation quality of Bob’s data. Since imputation generally benefits from increased data availability, our secure multi-party protocols allow Bob to impute missing values by incorporating information about neighbors in Alice’s dataset. We observed that the imputation quality is significantly improved when using Alice's data in comparison to imputation using only Bob’s local data. This improvement is attributed to the inclusion of additional relevant data from Alice’s dataset, which provides a more robust basis for predicting missing values, particularly in vertically split settings where Alice holds attributes that complement those in Bob’s records. 

We conducted experiments using three missingness patterns: missing completely at random (MCAR), missing at random (MAR), and missing not at random (MNAR). For each pattern, we introduced 10\% missing values to the specified “to-be-imputed” column. The imputation quality was then evaluated by comparing the discarded values in this column, used as ground truth, with predictions generated by the imputation model.
    
MCAR, the most widely used missingness pattern, involves values missing in an entirely random fashion. In this setup, missingness is independent of any observed (i.e., in the dataset) or unobserved (i.e., missing) data. In other words, the likelihood of a data point being missing is not influenced by the values of other variables in the dataset or by the actual (but not observed) values in the column with missing data. MCAR is typically implemented by drawing a random number from a uniform distribution for each data entry, discarding a value if the generated number falls below a predefined missingness threshold.

In contrast, MAR and MNAR patterns are more complex and less commonly addressed in the literature \cite{biessmann, schelter20, schelter21}. MAR implies that missingness is related to observed data but not to the missing data itself, while MNAR, the most challenging case, involves missingness that depends on unobserved or missing data values. 
    
To model these patterns realistically, we follow approaches inspired by observations in large-scale, real-world datasets. Using the method proposed by Schelter et al.~\cite{schelter20,schelter21}, we select two random percentiles in the value range of a column -- defining upper and lower bounds.
For MAR, we discard values in a target column when an entry in a different, randomly chosen column falls within the specified percentile range. This method controls the missingness by linking it to observed data in a related column. For MNAR, values are discarded directly if they fall within the selected percentile range of the target column itself, making the missingness dependent on the values in the column being imputed.
    
To ensure statistical robustness, each experiment was conducted per dataset and missingness types, where we re-sampled the missing data for each evaluation and repeated the data splits between Alice and Bob 25 times each. This setup helped us account for the variability inherent in missingness patterns and data partitioning.
To evaluate the accuracy of our imputation, we used root mean square error $\text{RMSE} = \sqrt{\frac{1}{n} \sum_{i=1}^{n} (y_i - \hat{y}_i)^2}$, where $n$ is the total number of observations, $y_i$ is the actual (observed) value, and $\hat{y}_i$ is the predicted values. 
    
\begin{figure*}[t]
    \centering
    \captionsetup[subfigure]{}
    \includegraphics[width=0.6\textwidth]{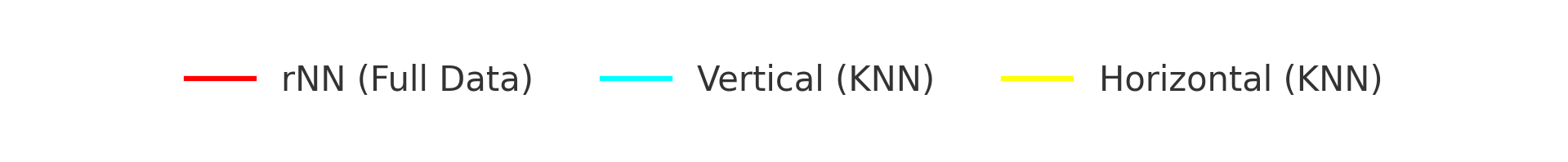}
    
    \begin{subfigure}[t]{\textwidth}
        \centering
        \includegraphics[width=0.32\textwidth]{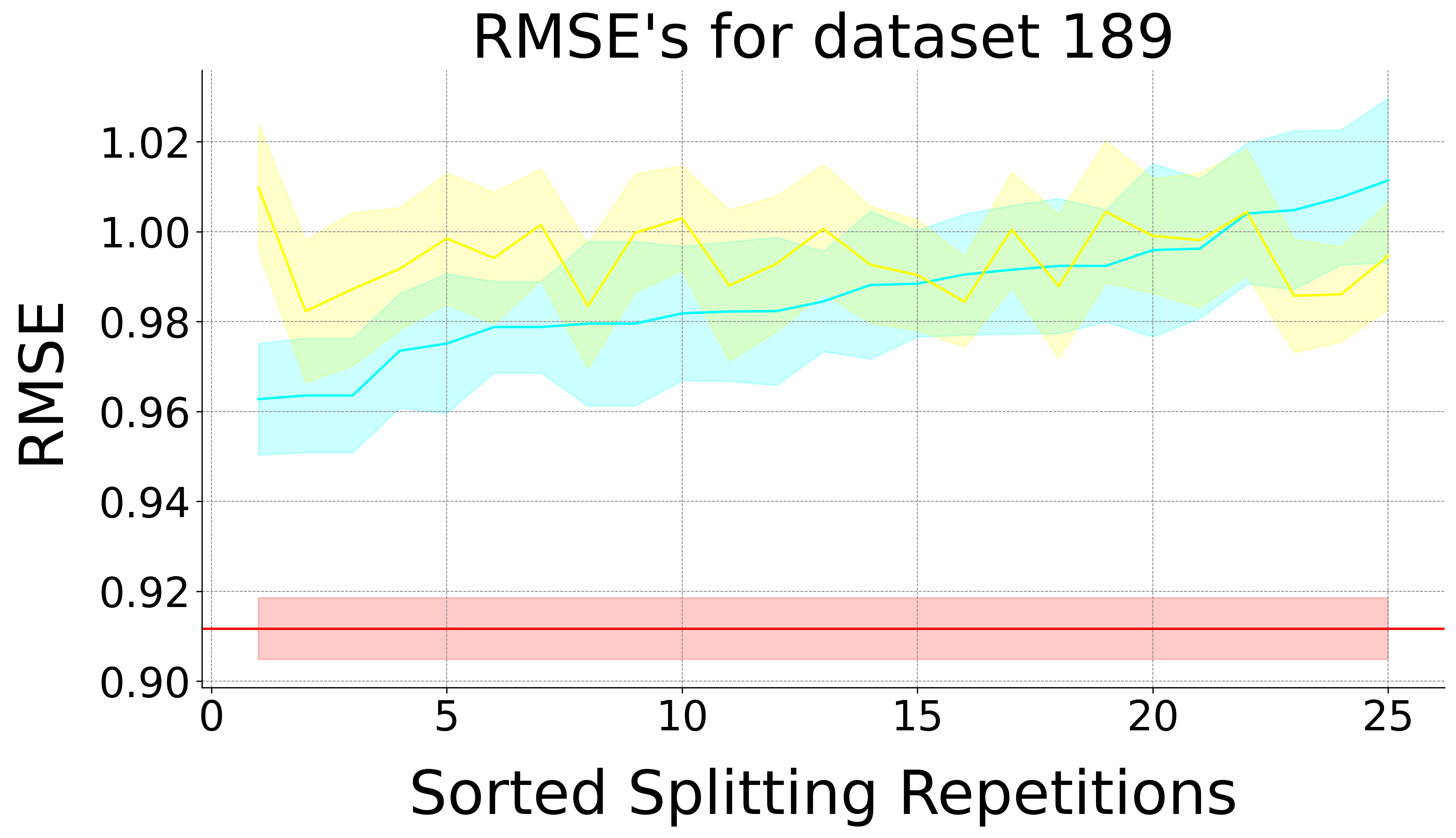}
        \includegraphics[width=0.32\textwidth]{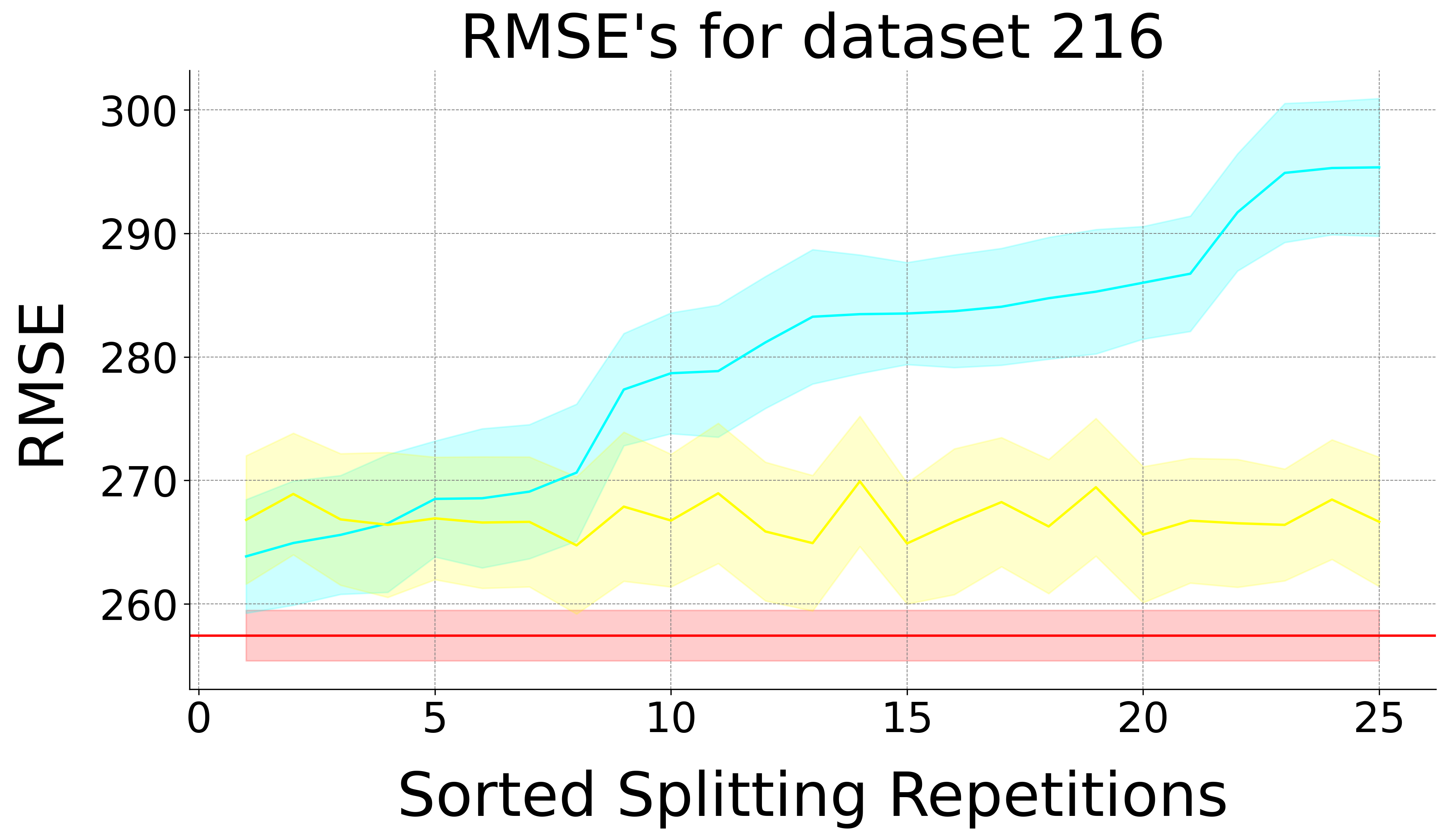}
        \includegraphics[width=0.32\textwidth]{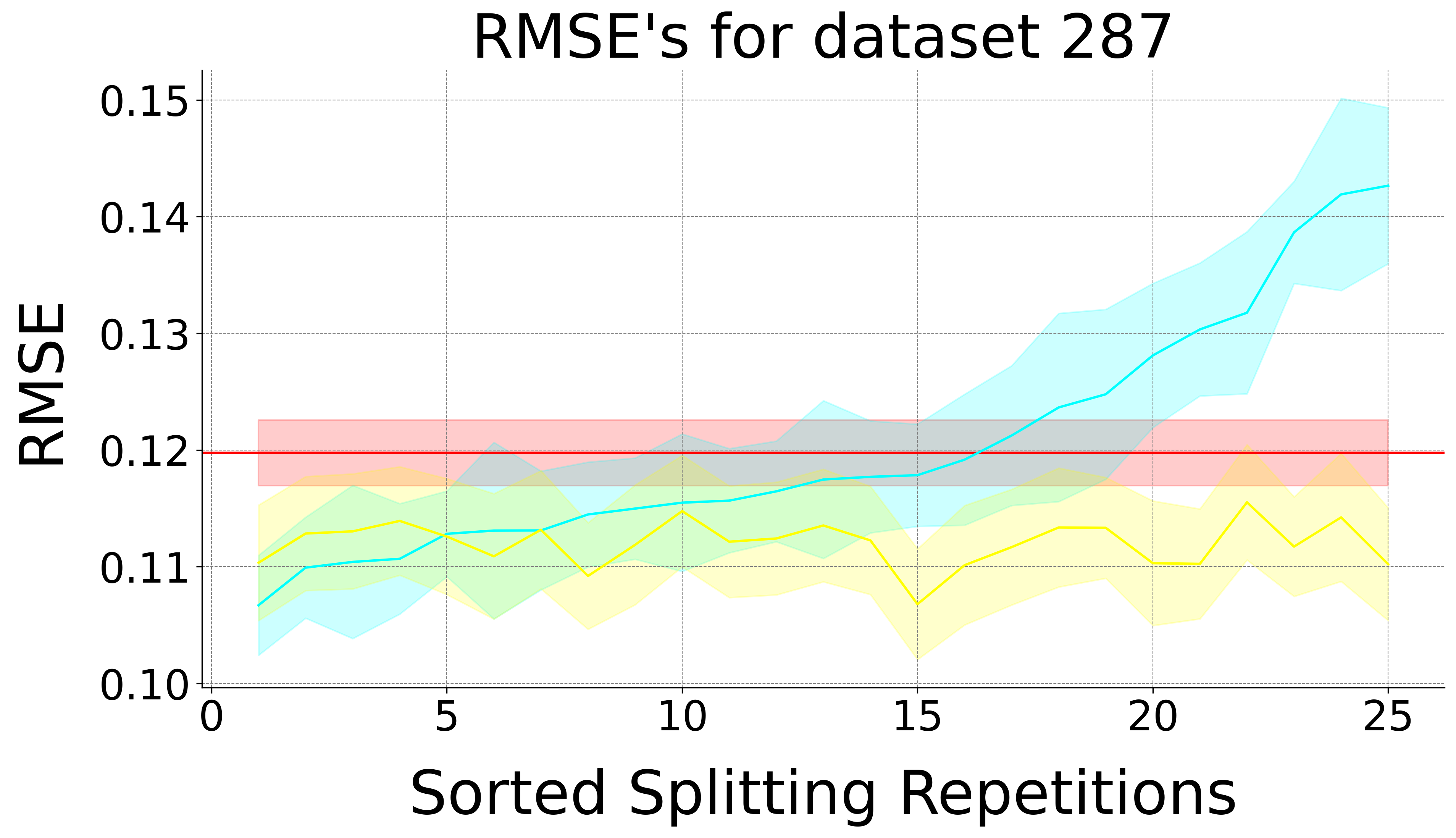}

        \caption{RMSE trends under MCAR missingness type.}
        \label{fig:plot_mcar}
    \end{subfigure}

    \vspace{0.5em}

    \begin{subfigure}[t]{\textwidth}
        \centering
        \includegraphics[width=0.32\textwidth]{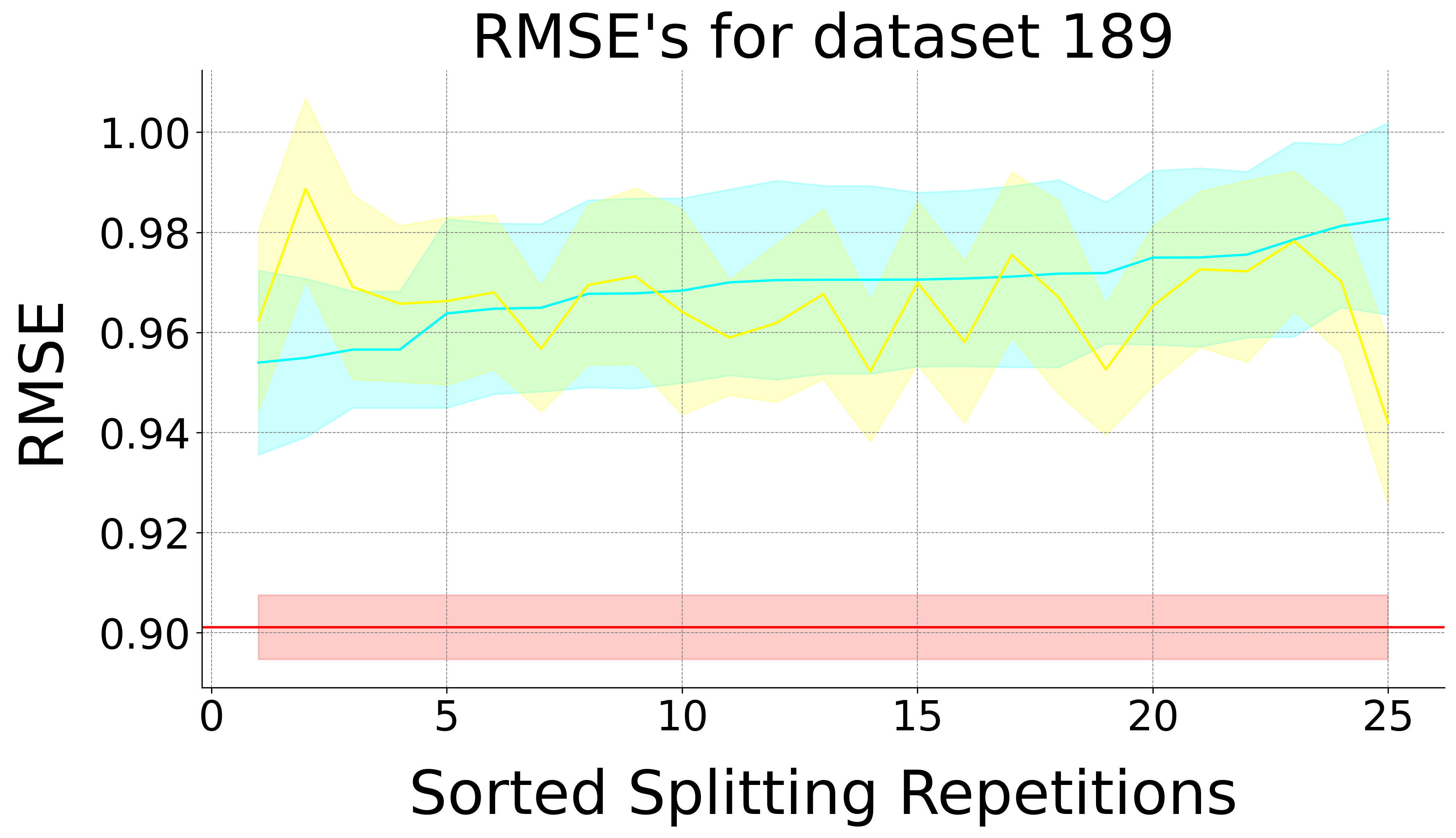}
        \includegraphics[width=0.32\textwidth]{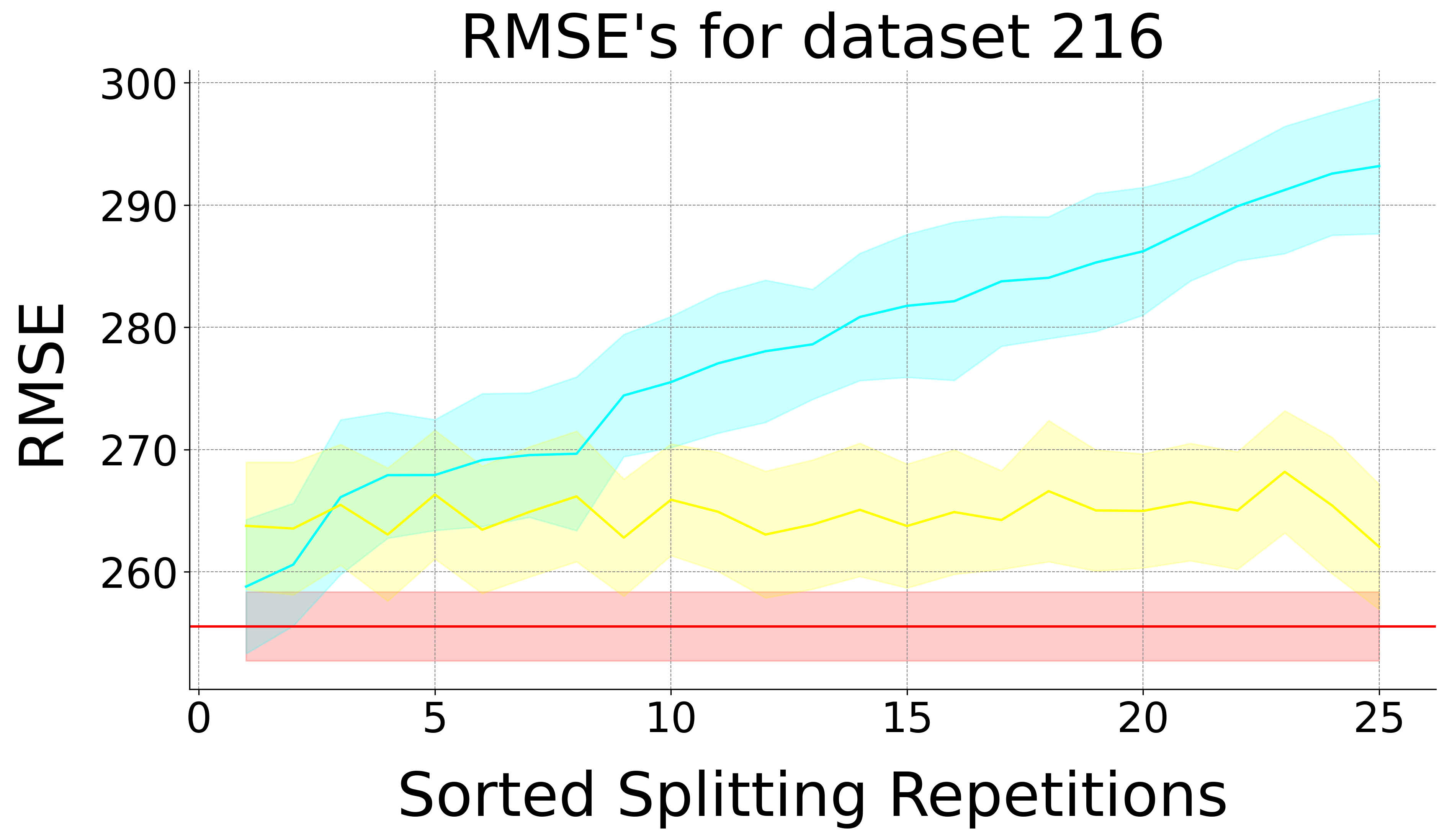}
        \includegraphics[width=0.32\textwidth]{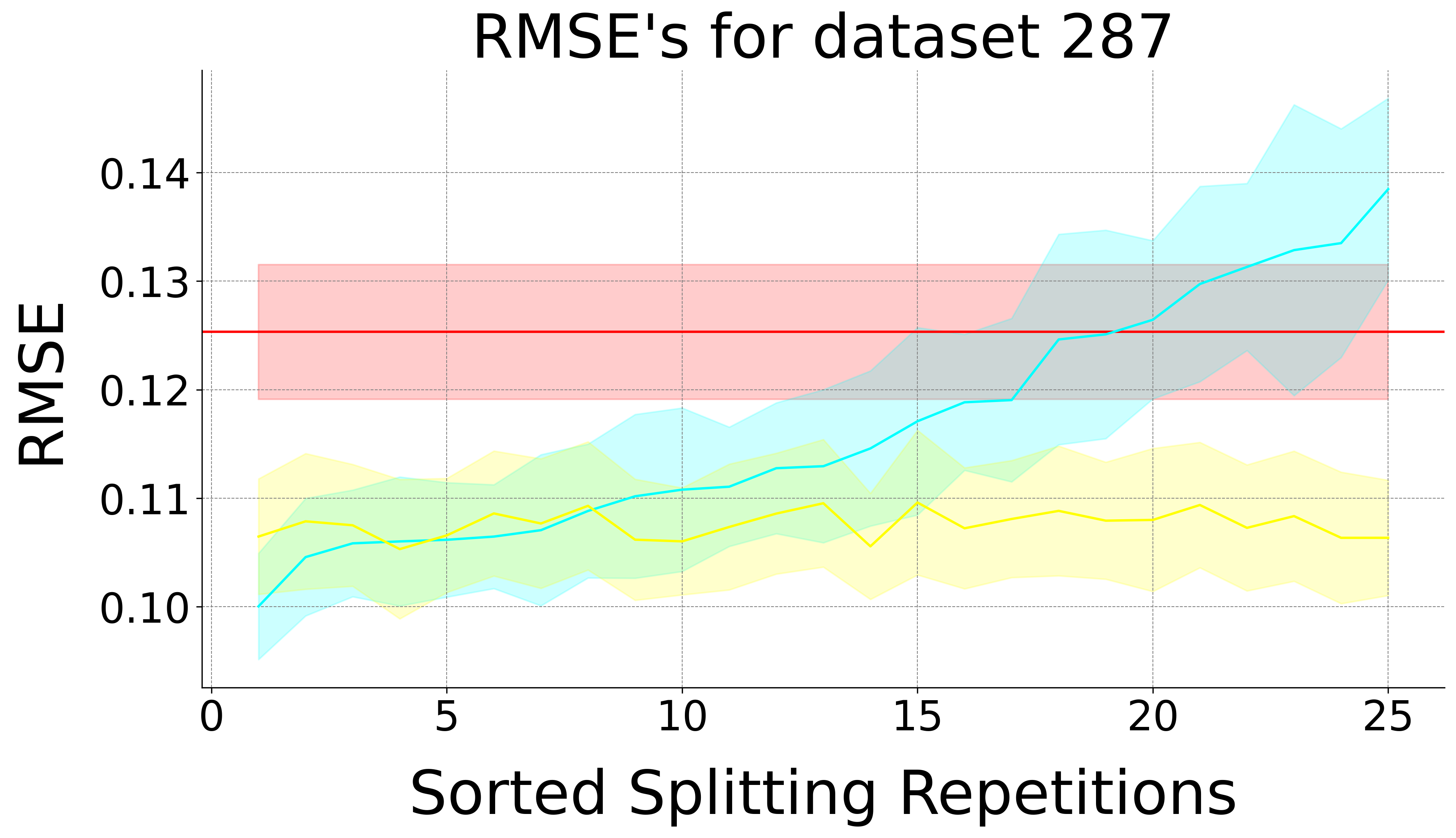}
        
        \caption{RMSE trends under MAR missingness type.}
        \label{fig:plot_mar}
    \end{subfigure}

    \vspace{0.5em}
    
    \begin{subfigure}[t]{\textwidth}
        \centering
        \includegraphics[width=0.32\textwidth]{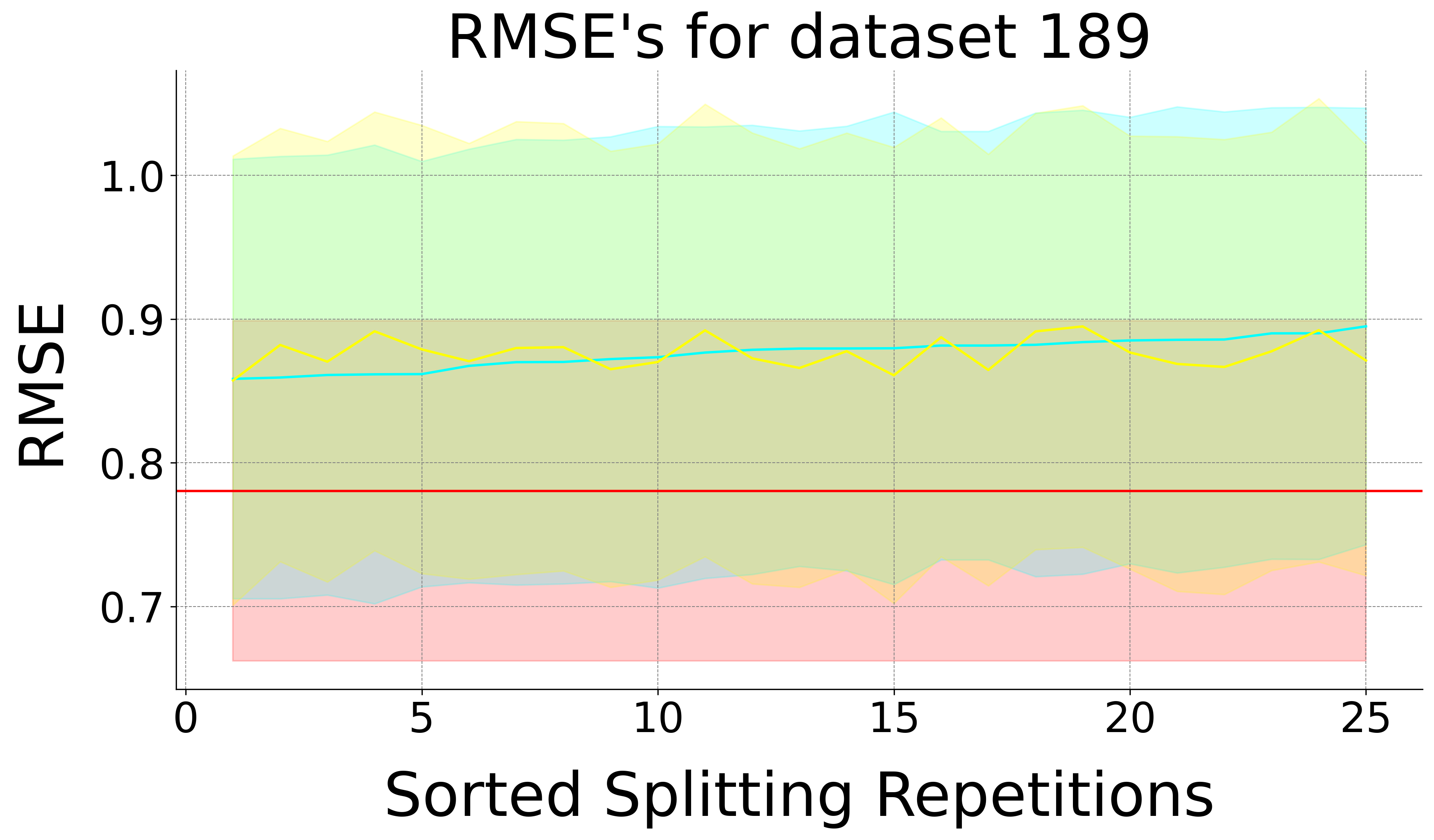}
        \includegraphics[width=0.32\textwidth]{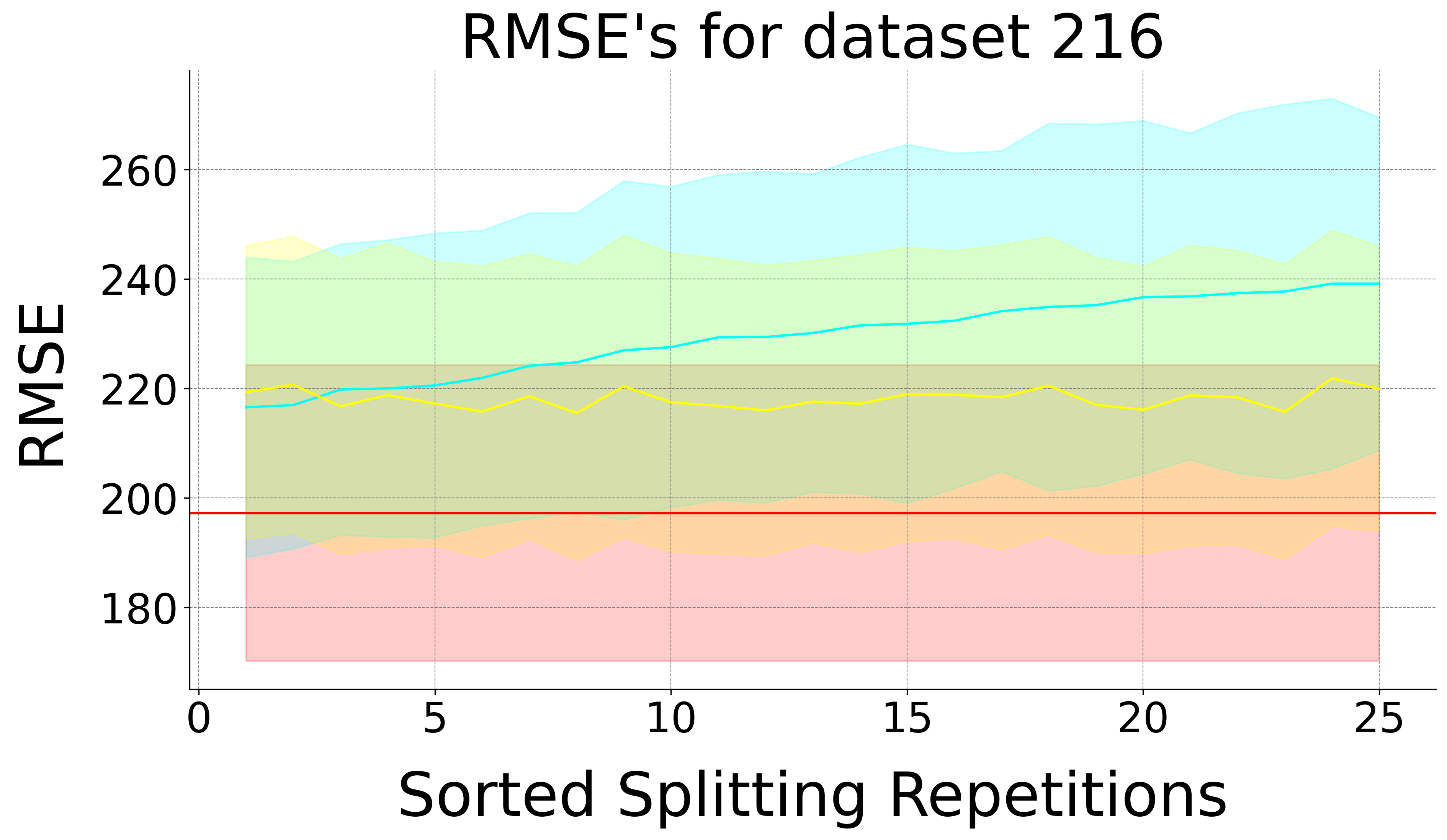}
        \includegraphics[width=0.32\textwidth]{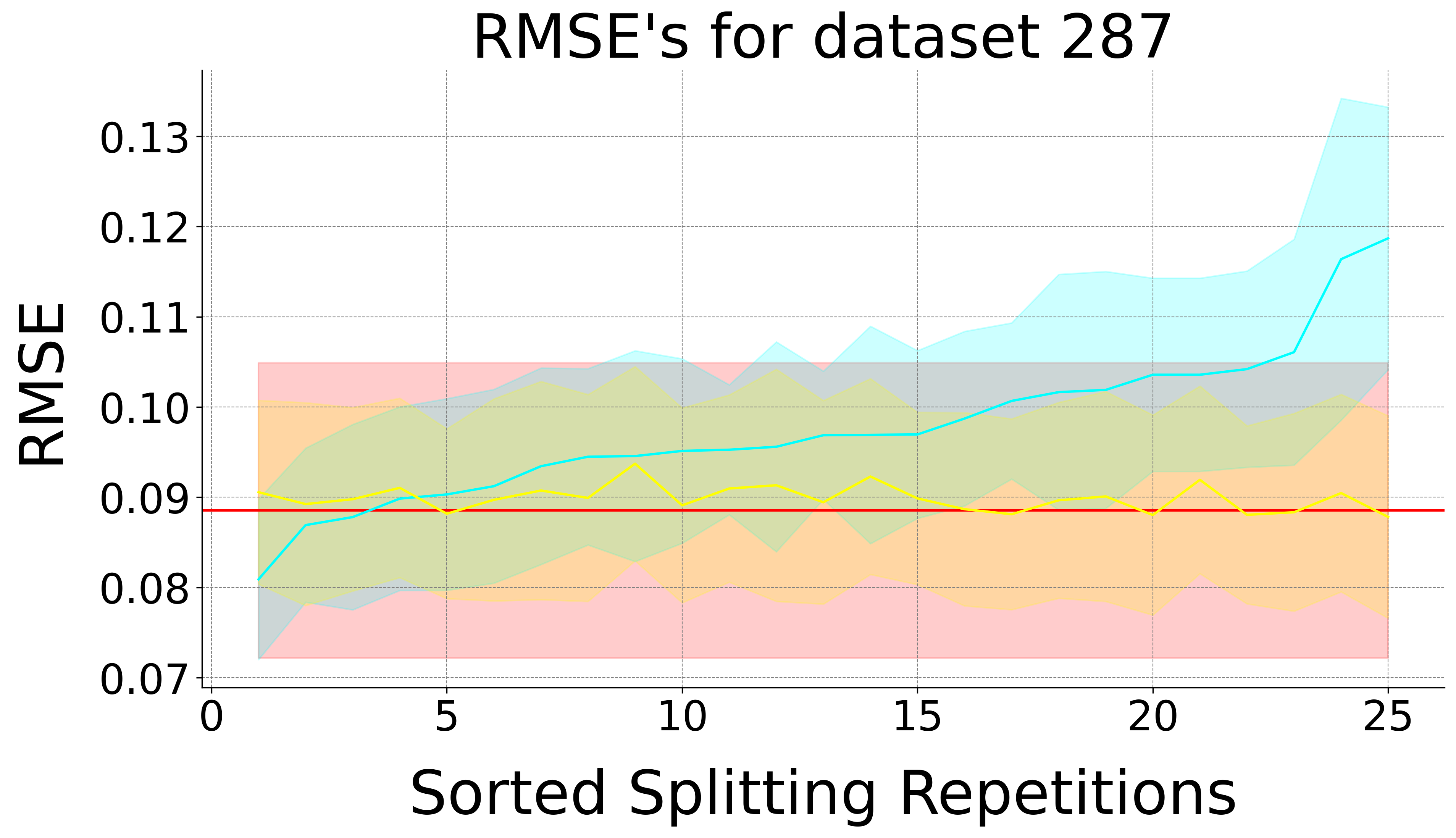}

        \caption{RMSE trends under MNAR missingness type.}
        \label{fig:plot_mnar}
    \end{subfigure}

    \caption{RMSE trends for datasets 189, 216, and 287 under MCAR, MAR, and MNAR missingness types. Splitting repetitions in each plot have been ordered by increasing Vertical $k$-NN RMSE. The width of the bands in each plot illustrates 95\% confidence intervals.}

    \label{fig:rmse_trends}
\end{figure*}

Our findings show that utilizing Alice's data in the computation in a privacy-preserving manner provides significant improvements in imputation accuracy over local imputation. This is due to the additional context and data points provided by Alice, which enables a more accurate imputation of the missing values. 
\fig{fig:rmse_trends} shows the results for three representative datasets, with a complete set of our results provided in \figs{fig:appendix_mcar}{fig:appendix_mnar} in Appendix~\ref{sec:extra-results}. Splitting repetitions in each plot have been ordered by increasing vertical $k$-NN RMSEs. The width of the bands in each plot illustrates the 95\% confidence intervals for the RMSE values.

For MCAR, as illustrated in \fig{fig:plot_mcar}, the RMSE results demonstrate the robustness of our protocols across different datasets. When comparing our imputation method -- denoted $r$-NN -- with the methods for vertically and horizontally partition data (50\%) using $k$-NN on a numerical column (mean computation), we find that we perform on average across all datasets and all splitting repetitions 20\% better than vertical $k$-NN, and 5\% better than horizontal $k$-NN. The scaled max for MCAR, shown in \tbl{tab:recovery_rates}, indicates that -- for a specific (worse) splitting -- our imputation method is up to \emph{5 times} more accurate in comparison with vertical $k$-NN and \emph{4.2 times} better than horizontal $k$-NN, highlighting the challenge of accurate imputation in this scenario.

For MAR, as shown in \fig{fig:plot_mar}, the RMSE trends closely mirror those observed under MCAR, but with a slight decrease in error margins due to the conditional dependencies influencing the missing data mechanism. On average, vertical $k$-NN is 10.6\% worse than our imputation method, while horizontal splitting performs marginally better, with an improvement of 0.7\%. However, \tbl{tab:recovery_rates} highlights that in the worst splitting scenario, our imputation method can be up to \emph{4.1 times} more accurate than vertical $k$-NN and about \emph{3 times} more accurate than horizontal $k$-NN. Thus, even though the MAR missigness type can introduce additional complexity, our protocol remains robust. 

The MNAR scenario, depicted in \fig{fig:plot_mnar}, presents the most significant challenge for accurate imputation where we experience a more pronounced decrease in RMSE due to its inherent bias in missingness. Our method on average outperforms vertical $k$-NN by 4\%, while horizontal $k$-NN exhibits an advantage of 7.6\% over our imputation approach. Despite this comparative disadvantage to horizontal $k$-NN in the overall average, the {worst splitting} scenario reveals that our method can be up to \emph{37.2 times} more accurate than vertical $k$-NN and \emph{3.4 times} more accurate than horizontal $k$-NN. Notably, this represents the largest performance gap achieved in any missingness type -- underscoring the substantial disparity posed by biased missingness and the randomness in the splitting.

    \begin{table}[t]
    \centering
    \caption{Scaled RMSE results across MCAR, MAR, and MNAR.}
    \label{tab:recovery_rates}
    \setlength{\tabcolsep}{0.5ex}
    \begin{tabular}{c | c c c c} \hline

    & \textbf{Metric} & \textbf{MCAR} & \textbf{MAR} & \textbf{MNAR} \\ 
    \hline\hline 

    \multirow{4}{*}{\shortstack[c]{ \textbf{Mean} \\ \textbf{(average)} \\\textbf{error} }}
       & $r$-NN (Full Data) & 100.00 & 100.00 & 100.00 \Tstrut\\
       & $k$-NN (Full Data) & 101.5  & 95.6   & 91.5   \\
       & Vertical $k$-NN    & 120.1  & 111.8  & 103.8  \\
       & Horizontal $k$-NN  & 104.9  & 99.3   & 92.5   \\
    \hline\hline

    \multirow{4}{*}{\shortstack[c]{ \textbf{Mean} \\ \textbf{(worst)} \\\textbf{error} }}
       & $r$-NN (Full Data) & 100.00 & 100.00 & 100.00 \\
       & $k$-NN (Full Data) & 117.06 & 96.88  & 306.75 \\
       & Vertical $k$-NN    & 499.2  & 410.7  & 3715.75 \\
       & Horizontal $k$-NN  & 426.1  & 299.9  & 341.71 \Bstrut \\
    \hline
    \end{tabular}
    \end{table}

The high-level summary of the error rates for vertical and horizontal $k$-NN relative to our method is provided in \tbl{tab:recovery_rates}. The table consolidates the average and maximum error rates, averaged across 25 runs with 25 splitting repetitions per run. These results underline the variability in performance across missingness types. 
Our approach's overage outperforms that of horizontal $k$-NN for typical missingness type MCAR and consistently outperforms vertical $k$-NN under all missingness types. Furthermore, the worst-case error for both vertical and horizontal $k$-NN is several times higher for all missingness types than that of our solution. 
This shows that our proposed $r$-NN-based method consistently achieves strong performance relative to the vertically and horizontally $k$-NN-based approaches. These findings highlight the practical resilience of our method, and emphasize the importance of selecting the appropriate splitting method based on the missingness type and dataset characteristics.

\subsection{Performance}
\label{sec:perf}

\begin{table*}[!t]
\caption{Performance evaluation of horizontal split protocols; the runtime is in seconds and communication in MB.}
\label{tab:Hsplit}
\setlength{\tabcolsep}{4.5pt}
\centering
\begin{tabular}{c c ||c| c c | c | c || c| c c | c| c } \hline
\multirow{3}{*}{Sampling} & \multirow{3}{*}{Network} & \multicolumn{5}{c||}{Medium Dataset} & \multicolumn{5}{c}{Large Dataset}\\ \cline{3-12}
 &  & \multirow{2}{*}{OPPRF} & \multicolumn{2}{c|}{ABY} & \multirow{2}{*}{Total} &\multirow{2}{*}{Comm.} & \multirow{2}{*}{OPPRF} & \multicolumn{2}{c|}{ABY} & \multirow{2}{*}{Total} & \multirow{2}{*}{Comm.} \\
 &  &  &  Build & Eval & & & & Build & Eval & \\ \hline\hline
\multirow{3}{*}{Plain, Mean} 
    & 10 Gbps  & $6.874$  & $0.001$ & $3.503$  & $10.378$ & \multirow{3}{*}{28.25} & $11.178$ & $0.002$ & $11.194$ & $22.374$ & \multirow{3}{*}{55.46}\Tstrut\\
 & 1 Gbps & $8.345$  & $0.001$ & $3.371$  & $11.718$ &  & $12.480$ & $0.002$ & $11.420$ & $23.902$ & \\
 & 100 Mbps & $12.601$ & $0.001$ & $4.114$  & $16.717$ &         & $19.869$ & $0.002$ & $12.211$ & $32.082$ & \Bstrut\\ \hline

\multirow{3}{*}{Plain, Random} 
    & 10 Gbps  & $6.919$  & $0.038$ & $3.397$  & $10.354$ & \multirow{3}{*}{31.37} & $11.060$ & $0.065$ & $11.568$ & $22.693$ & \multirow{3}{*}{62.35}\Tstrut\\
 & 1 Gbps & $8.312$  & $0.038$ & $3.868$  & $12.218$ &  & $12.693$ & $0.065$ & $12.697$ & $25.455$ & \\
 & 100 Mbps & $12.622$ & $0.038$ & $5.671$  & $18.331$ &         & $19.290$ & $0.065$ & $14.508$ & $33.863$ & \Bstrut\\ \hline
 
\multirow{3}{*}{Blind, Mean} 
 & 10 Gbps  & $6.903$ & $5.177$ & $8.6294$   & $20.709$ & \multirow{3}{*}{$561.93$} & $11.120$  & $10.216$ & $16.682$  & $38.018$ & \multirow{3}{*}{$1123.1$} \Tstrut\\
 & 1 Gbps & $8.037$ & $5.159$ & $10.546$   & $23.742$ &  & $12.976$  & $10.214$ & $19.673$  & $42.863$ & \\
 & 100 Mbps & $12.663$ & $5.166$ & $37.905$   & $55.734$ &         & $20.058$  & $10.259$ & $74.176$ & $104.493$ & \Bstrut\\ \hline

\multirow{3}{*}{Blind, Random} 
 & 10 Gbps & $3.393$ & $0.923$ & $1.697$   & $6.013$ & \multirow{3}{*}{$76.44$} & $3.859$  & $1.694$ & $2.854$  & $8.407$ & \multirow{3}{*}{$151.57$} \Tstrut\\
 & 1 Gbps   & $4.116$ & $0.947$ & $2.716$   & $7.779$ &  & $4.702$  & $1.718$ & $4.122$  & $10.542$ & \\
 & 100 Mbps & $6.268$ & $0.948$ & $8.174$   & $15.39$ &         & $7.668$  & $1.728$ & $13.236$ & $22.632$ & \Bstrut\\
\hline

\end{tabular}
\end{table*}

\begin{table*}[!t]
\caption{Performance evaluation of vertical split protocols; the runtime is in seconds and communication in MB.}
\label{tab:Vsplit}
\setlength{\tabcolsep}{4.5pt}
\centering
\begin{tabular}{c c ||c| c c | c | c || c| c c | c| c } \hline
\multirow{3}{*}{Sampling} & \multirow{3}{*}{Network} & \multicolumn{5}{c||}{Medium Dataset} & \multicolumn{5}{c}{Large Dataset}\\ \cline{3-12}
 &  & \multirow{2}{*}{PSI} & \multicolumn{2}{c|}{ABY} & \multirow{2}{*}{Total} &\multirow{2}{*}{Comm.} & \multirow{2}{*}{PSI} & \multicolumn{2}{c|}{ABY} & \multirow{2}{*}{Total} & \multirow{2}{*}{Comm.} \\
 &  &  &  Build & Eval & & & & Build & Eval & \\ \hline\hline
\multirow{3}{*}{Plain} 
    & 10 Gpbs & $0.532$  & \cellcolor{lightgray} & \cellcolor{lightgray}& $0.532$ & \multirow{3}{*}{$0.88$} & $0.570$  & \cellcolor{lightgray} & \cellcolor{lightgray} & $0.570$ & \multirow{3}{*}{$0.89$}\Tstrut\\
 & 1 Gbps   & $0.949$  & \cellcolor{lightgray} & \cellcolor{lightgray} & $0.949$ & & $1.156$ & \cellcolor{lightgray} & \cellcolor{lightgray} & $1.156$ & \\
 & 100 Mbps & $2.979$  & \cellcolor{lightgray} & \cellcolor{lightgray} & $2.979$ & & $2.984$ & \cellcolor{lightgray} & \cellcolor{lightgray} & $2.984$ & \Bstrut\\ 
 \hline

\multirow{3}{*}{Blind, Mean} 
 & 10 Gbps  & $1.116$ & $0.154$ & $0.613$ & $1.883$ & \multirow{3}{*}{$7.23$} & $1.198$ & $0.229$ & $0.814$ & $2.241$ & \multirow{3}{*}{$15.13$} \Tstrut\\
 & 1 Gbps   & $1.936$  & $0.164$ & $1.387$ & $3.487$ &  & $2.173$  & $0.290$ & $1.496$ & $3.959$ & \\
 & 100 Mbps & $5.209$  & $0.151$ & $2.839$ & $8.199$ &  & $5.116$  & $0.263$ & $3.687$ & $9.066$ & \Bstrut\\ 
 \hline

\multirow{3}{*}{Blind, Random} 
 & 10 Gbps  & $1.143$ & $0.189$ & $0.642$ & $1.974$ & \multirow{3}{*}{$8.35$} & $1.255$ & $0.291$ & $0.811$ & $2.357$ & \multirow{3}{*}{$17.47$} \Tstrut\\
 & 1 Gbps   & $1.951$ & $0.195$ & $1.113$ & $3.259$ &  & $2.174$  & $0.318$ & $1.527$ & $4.019$ & \\
 & 100 Mbps & $5.355$ & $0.170$ & $2.945$ & $8.471$ &  & $5.770$  & $0.328$ & $3.644$ & $9.742$ & \Bstrut\\
\hline
\end{tabular}
\end{table*}

We provide performance results of our protocols for horizontally and vertically split data, 
showing their efficiency under realistic settings. 

\medskip \noindent \textbf{Implementation and Experimental Setup.}
Our implementation \cite{AnonymizedRepo} is written in C++ utilizing ABY~\cite{aby} for secure two-party computation and VOLE-PSI~\cite{volepsi} for private set intersection. Experiments were conducted on a machine with an Intel Xeon E-2374G CPU @ 3.70GHz, 128 GB RAM, and Ubuntu 22.04.4 LTS. All times were averaged over 25 runs.

We used two synthetic datasets and varied (i) the dataset size to reflect typical real-world datasets 
and (ii) the percentage of neighbors of a tuple. The latter allows us to compute parameters $c$ and $d$ when performing blind random sampling on horizontally partitioned data and is also used as the number of local neighbors with vertically partitioned data.

Our medium dataset has 50,000 tuples with 10 attributes and $1\%$ probability of a tuple being a neighbor. For a desired probability of failure $\varepsilon \leq 2^{-40}$, we set $c = 100$ and $d = 28$. Our large dataset has 100,000 tuples with 10 attributes and $0.5\%$ neighbor probability. For the same $\varepsilon$, we obtain $c = 200$ and $d = 28$.


To emulate network settings for range of typical use cases, we used the Linux Netem and Traffic Control (TC) tools. These allowed us to precisely vary the bandwidth and latency as follows:
\begin{compactitem}
\item 10 Gbps with 2 ms round trip time (RTT),
\item 1 Gbps with 15 ms RTT,
\item 100 Mbps with 100 ms RTT.
\end{compactitem}
We note that the VOLE-OPPRF implementation we use from~\cite{volepsi} significantly influenced our performance evaluation setup. Currently, VOLE-OPPRF requires datasets to be sized as powers of 2 and restricts each set to a maximum of about two million blocks. To accommodate these limitations, we could not test our protocols on larger datasets and padded the input size in the horizontal partitioning solutions to the nearest power of 2.
For instance, in the plain and blind mean solutions, padding expanded Alice’s dataset from 500K to over 524K entries for the large dataset and from 250K to over 262K in the medium dataset (5\% increase for both). In the blind random sampling solutions, Alice's dataset grew from 56K to over 65K for the large data and from 28K to almost 33K for the medium dataset (17\% increase for both). This inflates both the computation and communication overhead of OPPRF evaluation.


\medskip \noindent \textbf{Horizontally Split Data.}
The performance of horizontally split data imputation methods is given in \tbl{tab:Hsplit}. We subdivide the runtime into OPPRF building and evaluation and ABY offline (Build) and online (Eval) time. The Build time is primarily from Yao's circuit garbling in ABY and is nearly absent when the majority of ABY computation uses arithmetic representation. Four protocols were evaluated: plain mean computation, plain random sampling, blind mean computation, and blind random sampling, with the last two being more attractive based on their security properties. 

Blind random sampling consistently delivers the fastest imputation times across all network settings for both datasets because of its lower complexity dependent on parameters $c$ and $d$ as opposed to the number of items in Alice's dataset for all other solutions. It is followed by plain imputation methods (which also have the lowest communication), but we do not consider those as providing high security guarantees. Overall, secure imputation (blind methods) can be achieved for our datasets on the order of a few to tens of seconds.


\medskip \noindent \textbf{Vertically Split Data.}
The performance of vertically split data imputation methods is shown in \tbl{tab:Vsplit}. Plain imputation (the same for mean and random sampling) achieves the lowest runtime and communication overhead across all settings, but of course is not as secure. Among the secure methods, blind mean computation slightly outperforms blind random sampling in both runtime and communication. Overall, these protocols are noticeably faster, performing on the order of seconds, which is due to their dependency only on the number of neighbors in Alice's and Bob's datasets. 

\section{Conclusions}
\label{sec:conclusions}

In this work, we study data imputation as a necessary prerequisite to data analysis in practical scenarios when certain values in a dataset are missing and need to be filled in using values in similar records. We consider a dataset with sensitive data partitioned among two different entities, Alice and Bob, and construct privacy-preserving imputation solutions for both horizontally and vertically split data. 

Our solution uses a variant of nearest neighbor computation that determines similar records, where we quantize the attributes in a novel approach that can be efficiently realized using cryptographic building blocks. Our construction of minimal leakage for horizontally partitioned data is a combination of OPPRF evaluation and optimized sampling from or aggregation of Alice's and Bob's neighbor records. Our construction of minimal leakage for vertically partitioned data relies on efficient private set intersection followed by private sampling from or aggregation of the result. 

Our empirical evaluations demonstrate that collaborative imputation, on average, improves the accuracy of the results over locally determined imputed values. On some data splits, it improves accuracy dramatically. This is despite quantization that we employ to boost efficiency of the algorithms. Furthermore, imputing an item in a dataset with 100,000 tuples and 10 attributes can be realized within seconds on a fast network and within 10--20 seconds on a slow network with high latency.

\ignore{

{\color{red}
TODO/Remaining (cross out when finished):
\begin{itemize}
\item Comparison with straightforward solutions. What about parameter tuning with secure kNN computation?
\end{itemize}
}

\section*{Open science}
In alignment with USENIX Security's open science policy, we commit to making the artifacts associated with our research publicly available to ensure availability, functionality, and reproducibility of our findings.
Upon acceptance of this paper, the source code, scripts, and relevant datasets will be hosted on a publicly accessible repository. Detailed instructions for setup and usage will also be provided to facilitate ease of replication of our results.

\section*{Ethics considerations}
Our paper considers data analyses and our primary stakeholders are data scientists as the user of our technology.
Our secondary stakeholders are subjects in the analyses performed by the data scientists.
We improve accuracy and privacy of the data analyses (at the expense of additional computational resources).
Improvements in accuracy and privacy have no negative ethical implications on either stakeholder.
We aim to keep the additional computational resources low, such that the environmental impact is justified.
We only use open data sources which have undergone ethical review.
}

\bibliographystyle{plain}
\bibliography{refs}

@book{goldreich2004,
  title={Foundations of Cryptography, Volume 2},
  author={Goldreich, Oded},
  year={2004},
  publisher={Cambridge university press Cambridge}
}

@inproceedings{gmw,
  author = "Goldreich, Oded and Micali, Silvio and Wigderson, Avi",
  title = "How to play any mental game or a completeness theorem for protocols with honest majority",
  booktitle = "Symposium on Theory of Computing (STOC)",
  pages = "218-–229",
  year = 1987
}

@inproceedings{garbled-circuits,
  author = "Yao, Andrew C.",
  title = "How to generate and exchange secrets",
  booktitle = "Foundations of Computer Science (FOCS)", 
  pages = "162–-167",
  year = 1986
}

@article{goldreich96,
  title={Software protection and simulation on oblivious {RAM}s},
  author={Goldreich, Oded and Ostrovsky, Rafail},
  journal={Journal of the ACM (JACM)},
  volume={43},
  number={3},
  pages={431--473},
  year={1996}
}

@inproceedings{chen20,
  author = {Chen, Hao and Chillotti, Ilaria and Dong, Yihe and Poburinnaya, Oxana and Razenshteyn, Ilya and Riazi, M. Sadegh},
  title = {{SANNS}: Scaling Up Secure Approximate k-Nearest Neighbors Search},
  booktitle = {USENIX Security Symposium (USENIX Security 20)},
  year = {2020},
  isbn = {978-1-939133-17-5},
  pages = {2111--2128},
  url = {https://www.usenix.org/conference/usenixsecurity20/presentation/chen-hao},
  publisher = {USENIX Association},
  month = aug
}

@article{pinkas18,
  author       = {Benny Pinkas and
                  Thomas Schneider and
                  Michael Zohner},
  title        = {Scalable Private Set Intersection Based on {OT} Extension},
  journal      = {{ACM} Transactions on Privacy and Security},
  volume       = {21},
  number       = {2},
  pages        = {7:1--7:35},
  year         = {2018}
}

@inproceedings{kerschbaum23,
  author       = {Florian Kerschbaum and
                  Erik{-}Oliver Blass and
                  Rasoul Akhavan Mahdavi},
  title        = {Faster Secure Comparisons with Offline Phase for Efficient Private
                  Set Intersection},
  booktitle    = {Network and Distributed System Security Symposium ({NDSS})},
  year         = {2023}
}

@inproceedings{volepsi,
  author       = {Peter Rindal and
                  Phillipp Schoppmann},
  title        = {{VOLE-PSI:} {F}ast {OPRF} and Circuit-PSI from Vector-OLE},
  booktitle    = {Advances in Cryptology -- {EUROCRYPT}},
  year         = {2021}
}

@inproceedings{fre04,
  title = "Efficient private matching and set intersection",
  author = "M. Freedman and K. Nissim and B. Pinkas",
  booktitle = "Advances in Cryptology -- EUROCRYPT",
  year = 2004,
  pages = "1--19"
}

@inproceedings{kis05,
  title = "Privacy-preserving set operations",
  author = "L. Kissner and D. Song",
  year = 2005,
  booktitle = "Advances in Cryptology -- CRYPTO",
  pages = "241--257"
}

@inproceedings{dec10,
  author = "E. {De Cristofaro} and G. Tsudik",
  title = "Practical Private Set Intersection Protocols with Linear
  Complexity",
  booktitle = "Financial Cryptography and Data Security (FC)",
  year = 2010,
  pages = "143--159"
}

@inproceedings{jar10,
  author = "S. Jarecki and X. Liu",
  title = "Fast Secure Computation of Set Intersection",
  booktitle = "International Conference on Security and Cryptography for
  Networks (SCN)",
  year = 2010,
  pages = "418--435"
}

@article{bla16,
  author = "Marina Blanton and Everaldo Aguiar", 
  title = "Private and Oblivious Set and Multiset Operations",
  journal = "International Journal of Information Security", 
  volume = 15, 
  number = 5, 
  pages = "493--518", 
  year = 2016
}

@inproceedings{aby,
  title={{ABY} -- {A} framework for efficient mixed-protocol secure two-party computation},
  author={Demmler, Daniel and Schneider, Thomas and Zohner, Michael},
  booktitle={Network and Distributed System Security Symposium (NDSS)},
  year={2015}
}

@misc{jen24,
  title={Privacy Preserving Data Imputation via Multi-party Computation for Medical Applications},
  author={Jentsch, Julia and {\"U}nal, Ali Burak and Ma{\u{g}}ara, {\c{S}}eyma Selcan and Akg{\"u}n, Mete},
  howpublished={arXiv Preprint Report 2405.18878},
  year={2024}
}

@article{ome17,
  title={Privacy-preserving of {SVM} over vertically partitioned with imputing missing data},
  author={Omer, Mohammed Z. and Gao, Hui and Mustafa, Nadir},
  journal={Distributed and Parallel Databases},
  volume={35},
  pages={363--382},
  year={2017}
}

@inproceedings{jag06,
  title={Privacy-preserving data imputation},
  author={Jagannathan, Geetha and Wright, Rebecca N.},
  booktitle={IEEE International Conference on Data Mining Workshops (ICDMW)},
  pages={535--540},
  year={2006}
}

@article{jag08,
  title = {Privacy-preserving imputation of missing data},
  journal = {Data \& Knowledge Engineering},
  volume = {65},
  number = {1},
  pages = {40-56},
  year = {2008},
  author = {Geetha Jagannathan and Rebecca N. Wright}
}

@article{cli22,
  title={Differentially private $k$-nearest neighbor missing data imputation},
  author={Clifton, Chris and Hanson, Eric J. and Merrill, Keith and Merrill, Shawn},
  journal={ACM Transactions on Privacy and Security},
  volume={25},
  number={3},
  pages={1--23},
  year={2022}
}

@inproceedings{bla23,
  title={Private collaborative data cleaning via non-equi {PSI}},
  author={Blass, Erik-Oliver and Kerschbaum, Florian},
  booktitle={IEEE Symposium on Security and Privacy (S\&P)},
  pages={1419--1434},
  year={2023}
}

@inproceedings{pinkas20,
  author       = {Benny Pinkas and
                  Mike Rosulek and
                  Ni Trieu and
                  Avishay Yanai},
  title        = {{PSI} from PaXoS: Fast, Malicious Private Set Intersection},
  booktitle    = {Advances in Cryptology -- {EUROCRYPT}},
  year         = {2020},
}

@ARTICLE{benchmark, 
AUTHOR={Jäger, Sebastian and Allhorn, Arndt and Bießmann, Felix},   
TITLE={A Benchmark for Data Imputation Methods},      
JOURNAL={Frontiers in Big Data},      
VOLUME={4},           
YEAR={2021}
}

@article{Schafer2002,
  title={Missing data: our view of the state of the art.},
  author={Joseph L. Schafer and John W. Graham},
  journal={Psychological methods},
  year={2002},
  volume={7},
  issue={2}
}

@inproceedings{Knott21,
  author       = {Brian Knott and
                  Shobha Venkataraman and
                  Awni Y. Hannun and
                  Shubho Sengupta and
                  Mark Ibrahim and
                  Laurens van der Maaten},
  title        = {CrypTen: Secure Multi-Party Computation Meets Machine Learning},
  booktitle    = {Conference on Neural Information Processing Systems (NeurIPS)},
  year         = {2021}
}

@article{Kairouz21,
  author       = {Peter Kairouz and
                  H. Brendan McMahan and
                  Brendan Avent and
                  Aur{\'{e}}lien Bellet and
                  Mehdi Bennis and
                  Arjun Nitin Bhagoji and
                  Kallista A. Bonawitz and
                  Zachary Charles and
                  Graham Cormode and
                  Rachel Cummings and
                  Rafael G. L. D'Oliveira and
                  Hubert Eichner and
                  Salim El Rouayheb and
                  David Evans and
                  Josh Gardner and
                  Zachary Garrett and
                  Adri{\`{a}} Gasc{\'{o}}n and
                  Badih Ghazi and
                  Phillip B. Gibbons and
                  Marco Gruteser and
                  Za{\"{\i}}d Harchaoui and
                  Chaoyang He and
                  Lie He and
                  Zhouyuan Huo and
                  Ben Hutchinson and
                  Justin Hsu and
                  Martin Jaggi and
                  Tara Javidi and
                  Gauri Joshi and
                  Mikhail Khodak and
                  Jakub Kone{\v{c}}n{\'y} and
                  Aleksandra Korolova and
                  Farinaz Koushanfar and
                  Sanmi Koyejo and
                  Tancr{\`{e}}de Lepoint and
                  Yang Liu and
                  Prateek Mittal and
                  Mehryar Mohri and
                  Richard Nock and
                  Ayfer {\"{O}}zg{\"{u}}r and
                  Rasmus Pagh and
                  Hang Qi and
                  Daniel Ramage and
                  Ramesh Raskar and
                  Mariana Raykova and
                  Dawn Song and
                  Weikang Song and
                  Sebastian U. Stich and
                  Ziteng Sun and
                  Ananda Theertha Suresh and
                  Florian Tram{\`{e}}r and
                  Praneeth Vepakomma and
                  Jianyu Wang and
                  Li Xiong and
                  Zheng Xu and
                  Qiang Yang and
                  Felix X. Yu and
                  Han Yu and
                  Sen Zhao},
  title        = {Advances and Open Problems in Federated Learning},
  journal      = {Foundations and Trends in Machine Learning},
  volume       = {14},
  number       = {1-2},
  year         = {2021},
}

@article{Zhou24,
  author       = {Ian Zhou and
                  Farzad Tofigh and
                  Massimo Piccardi and
                  Mehran Abolhasan and
                  Daniel Robert Franklin and
                  Justin Lipman},
  title        = {Secure Multi-Party Computation for Machine Learning: {A} Survey},
  journal      = {{IEEE} Access},
  volume       = {12},
  year         = {2024},
}

@book{ilyasbook,
author = {Ilyas, Ihab F. and Chu, Xu},
title = {Data Cleaning},
year = {2019},
publisher = {Association for Computing Machinery}
}

@techreport{google,
author = {Walker, Amanda and Patel, Sarvar and Yung, Moti},
title = {Helping organizations do more without collecting more data},
year = {2019},
institution = {Google Security Blog},
howpublished = {\url{https://security.googleblog.com/2019/06/helping-organizations-do-more-without-collecting-more-data.html}}
}

@inproceedings{jin19,
  author = "Jin, Haifeng and Song, Qingquan and Hu, Xia",
  year = 2019,
  title = "Auto-keras: An Efficient Neural Architecture Search System", 
  booktitle = "ACM SIGKDD International Conference on Knowledge Discovery \& Data Mining", 
  pages = "1946-–1956"
}

@inproceedings{kin14,
  author = "Kingma, Diederik P. and Welling, Max",
  year = 2014,
  title = "Auto-encoding Variational Bayes", 
  booktitle = "International Conference on Learning Representations (ICLR)", 
}

@inproceedings{yoo18,
  author = "Yoon, Jinsung and Jordon, James and van der Schaar, Mihaela",
  year = 2018,
  title = "GAIN: Missing Data Imputation Using Generative Adversarial Nets", 
  booktitle = "International Conference on Machine Learning (ICML)", 
  pages = "5675-–5684" 
}

@misc{AnonymizedRepo,
  title={Private Data Imputation: Anonymized Implementation},
  author={Anonymous},
  howpublished={\url{https://anonymous.4open.science/r/Private_Data_Imputation-FA42}},
  note={Anonymized for review},
  year={2025}
}

@inproceedings{biessmann,
author = {Biessmann, Felix and Salinas, David and Schelter, Sebastian and Schmidt, Philipp and Lange, Dustin},
title = {"Deep" Learning for Missing Value Imputationin Tables with Non-Numerical Data},
year = {2018},
booktitle = {ACM International Conference on Information and Knowledge Management},
pages = {2017–2025}
}

@inproceedings{schelter20,
author = {Schelter, Sebastian and Rukat, Tammo and Biessmann, Felix},
title = {Learning to Validate the Predictions of Black Box Classifiers on Unseen Data},
year = {2020},
booktitle = {ACM SIGMOD International Conference on Management of Data},
pages = {1289-–1299}
}

@Inproceedings{schelter21,
 author = {Sebastian Schelter and Tammo Rukat and Felix Biessmann},
 title = {JENGA: A framework to study the impact of data errors on the predictions of machine learning models},
 year = {2021},
 booktitle = {EDBT Industrial and Application Track},
}

@article{openml,
author = {Vanschoren, Joaquin and van Rijn, Jan N. and Bischl, Bernd and Torgo, Luis},
title = {OpenML: networked science in machine learning},
journal = {SIGKDD Explor. Newsl.},
volume = {15},
number = {2},
year = {2014},
pages = {49–60},
}

@article{lsh,
author = {Andoni, Alexandr and Indyk, Piotr},
title = {Near-optimal hashing algorithms for approximate nearest neighbor in high dimensions},
journal = {Association for Computing Machinery},
volume = {51},
number = {1},
year = {2008},
pages = {117–122},
}

@misc{chor1997private,
  title={Private information retrieval by keywords},
  author={Chor, Benny and Gilboa, Niv and Naor, Moni},
  year={1998},
  howpublished = "IACR Cryptology ePrint Report 1998/003"  
}

@article{chor1998private,
  title={Private information retrieval},
  author={Chor, Benny and Kushilevitz, Eyal and Goldreich, Oded and Sudan, Madhu},
  journal={Journal of the ACM (JACM)},
  volume={45},
  number={6},
  pages={965--981},
  year={1998},
  publisher={ACM New York, NY, USA}
}

@article{gasarch2004survey,
  title={A survey on private information retrieval},
  author={Gasarch, William},
  journal={Bulletin of the EATCS},
  volume={82},
  number={72-107},
  pages={113},
  year={2004},
  publisher={Citeseer}
}

\appendices

\section{Security Analysis}
\label{sec:security}

Recall that in the ideal case Bob learns only the imputed value $x_{\alpha,\beta}$ (which is necessarily a function of Alice's data) and no other information about Alice's dataset, while Alice learns nothing. The solutions proposed in this work, however, may disclose additional information. Thus, in this section we analyze the protection offered by our constructions. 

\medskip \noindent \textbf{Horizontally partitioned data.}
As mentioned earlier, the solution where Alice learns the set of $t_\alpha$'s neighboring tuples among her items, $I_A$, discloses a significant amount of information because it allows Alice to approximate Bob's tuple $t_\alpha$. For that reason, we consider the initial solutions with plain mean computation or sampling in Sections~\ref{sec:h-mean-plain} and~\ref{sec:h-random-plain} as too revealing and treat the final solutions in Sections \ref{sec:h-mean-blind} and~\ref{sec:h-random-blind} as our main construction. 

In our solutions, Alice learns the value of $\beta$ that corresponds to the missing attribute in one of Bob's tuples. This is a small amount of information to learn and this information may not be secret to Alice: given that Alice possesses tuples with the same attribute, she may already know based on her own dataset that some of the records have their $\beta$th attribute missing. However, hiding $\beta$ will inevitably increase the cost of the computation. 

%

We consider a standard simulation-based definition of security, where all a party's view can be simulated solely from their input and output. If such a simulated view is shown to be indistinguishable from their view during a real protocol execution, the protocol execution does not reveal any private information. A view is formed by all of the information a party possesses, including its inputs, messages received during the protocol execution, randomness used during the protocol, and the output obtained as a result of the protocol. We sketch security proofs for Alice and Bob for our construction.

Bob has his private input dataset and initially engages in OPPRF evaluations with Alice. OPPRF's properties are such that Bob does not learn any new information as part of that interaction. Next, Alice and Bob engage in private sampling or mean calculation by means of secure two-party computation. Bob inputs (among other values) the $\beta$ attribute from all of the neighbor records $I_B$ among his tuples and learns a value to use for $x_{\alpha,\beta}$. Because the secure two-party protocol has the desired properties (i.e., the computation can be simulated without the other party's input and the simulated view is indistinguishable from the real execution) and because Bob uses all of his neighbors, he is unable to learn any information beyond the final outcome. 

As far as Alice's view goes, she initially learns $\beta$ and the result of OPPRF evaluation on each attribute in her tuples. Due to the properties of an OPPRF, any evaluation on a new point is indistinguishable from a randomly generated string. We program the OPPRF differently for each tuple and for each attribute, which ensures that all values Alice obtains are independent of each other and indistinguishable from uniformly chosen random values.

Alice and Bob consequently engage in secure two-party computation. Alice enters the results of OPPRF evaluations for all of her tuples and receives no output as a result of that computation. Due to the properties of the secure two-party computation protocols, Alice's view during the computation is indistinguishable from a simulation that does not use any Bob's inputs and Alice does not learn any information, including which of her records were identified as neighbors of Bob's tuple $t_\alpha$.

\medskip \noindent \textbf{Vertically partitioned data.} In the case of vertically partitioned data, the computation involves PSI or circuit-PSI followed by secure two-party computation.

Consider Alice's view. In our solution Alice learns the index $\alpha$, i.e., that Bob is imputing a value in his tuple $t_\alpha$. This is a small amount of information that allows us to achieve performance with communication and the number of cryptographic operations being sub-linear in the size of Alice's dataset. Given that knowledge, Alice computes the set of neighbors of tuple $t_\alpha$ in her dataset and does not learn any other information throughout protocol execution. 

%

Specifically, when engaging in a PSI protocol, either Bob learns the resulting set intersection or the parties obtain secret shares of the indicator vector. In both of those cases, security of our solution relies on the security of the underlying PSI protocols and the view can be simulated using PSI's simulator.

With plain access in Sections \ref{sec:v-mean-plain} and~\ref{sec:v-random-plain}, no additional computation with private values is needed, but Bob learns the tuples in the intersection, i.e., global neighbors, which is undesirable. Thus, we next analyze enhanced solutions, with blind access in Sections \ref{sec:v-mean-blind} and~\ref{sec:v-random-blind}, that do not have this drawback. 

In that case, we combine circuit-PSI with secure two-party computation, neither of which provide any information to Alice. That is, both the PSI and secure two-party computation techniques come with simulators that will allow us to create Alice's view without Bob's data which will be indistinguishable from the protocol execution. Similarly, Bob's view can be simulated from his output alone by relying on the security of the building blocks we invoke. 

\tbl{tab:disclosure} summarizes information disclosure of different constructions.

\section{Extensions}
\label{sec:ext}

Recall that our previously described solutions have a small amount of information leakage beyond the function output. In particular, in the horizontally split case, Alice learns the attribute $\beta$ Bob wants to impute in his tuple; in the vertically split case, Alice learns the index $\alpha$ of the tuple in which Bob wants to impute a value. In this section, we comment on extensions to our protocols that eliminate that information disclosure. 

\subsection{Hiding $\beta$ with Horizontally Partitioned Data}

In Section \ref{sec:h}, we have assumed that both Alice and Bob know $\beta$, but it is actually not necessary for Bob to reveal $\beta$ to Alice.
Remedying this requires Alice and Bob to perform an imputation for each possible value of $\beta$ and then Bob privately selecting the correct one.
This is very inefficient, yet there is an opportunity for Alice and Bob to reduce the cost in case of multiple imputations.
Alice and Bob can first partition their data using a locality-sensitive hash function (LSH)~\cite{lsh} and then operate over each partition. LSHs are a family of
hash functions satisfying the following property: for any given pair of vectors in space, if the vectors are close, the probability of their hashes colliding is high, else the probability of their hashes colliding is low.
Note that Alice and Bob need to compute the same number of imputations for each bucket of the LSH to not reveal those that have missing values.
Hence, this method only provides a performance gain in case there are multiple imputations.



\subsection{Hiding $\alpha$ with Vertically Partitioned Data}

So far, we have assumed in Section \ref{sec:v} Alice computes her neighbors locally knowing $\alpha$.
However, we can also hide $\alpha$ from Alice.
For this, we use a technique called private information retrieval (PIR) and combine it with Diffie-Hellman based PSI.
In Diffie-Hellman PSI, for each tuple index $m_i \in I_A$, Alice computes and sends $H(m_i)^a$ to Bob. Similarly, for each index $m_j \in I_B$, Bob computes and sends $H(m_j)^{b}$ to Alice. Alice computes $(H(m_i)^b)^a$ in $\mathbb{G}$ and Bob computes $H((m_j)^a)^b$. It holds that
$$m_i = m_j \Leftrightarrow H(m_i)^{ab} = H(m_j)^{ba}.$$
PIR achieves privacy by ensuring that no single server or group of colluding servers can infer the query's target, thereby preserving the client's data access patterns.
In PIR, Alice has a database and Bob accesses an element at an index in the database without revealing which index it is, which naturally generalizes to retrieving buckets from a hash table \cite{chor1997private, chor1998private, gasarch2004survey}.

We construct the database as follows.
For each tuple $t_k$, Alice computes her set of neighbors $I_{A, k}$.
For each $m_i \in I_{A, k}$ Alice computes the first step in Diffie-Hellman based PSI $H(m_i)^a$.
The element at index $k$ in the database is set $\{ H(m_i)^a \ |\ m_i \in I_{A, k} \}$.

We continue the PSI protocol as follows.
Instead of Alice sending $H(m_i)^a$, Bob retrieves $H(m_i)^a$ for index $\alpha$ using PIR.
The protocol then proceeds as standard Diffie-Hellman based PSI (Bob computes and returns $H(m_i)^{ab}$) and if we stop after Alice's step (of exponentiating with $a^{-1}$) we can use it to build circuit-PSI.

\section{Additional Accuracy Evaluation Results}
\label{sec:extra-results}

\begin{figure*}
    \begin{minipage}[c][\textheight][c]{\textwidth}
        \centering
        \includegraphics[width=0.6\textwidth]{figures/legend.png}
        
        \includegraphics[width=0.32\textwidth]{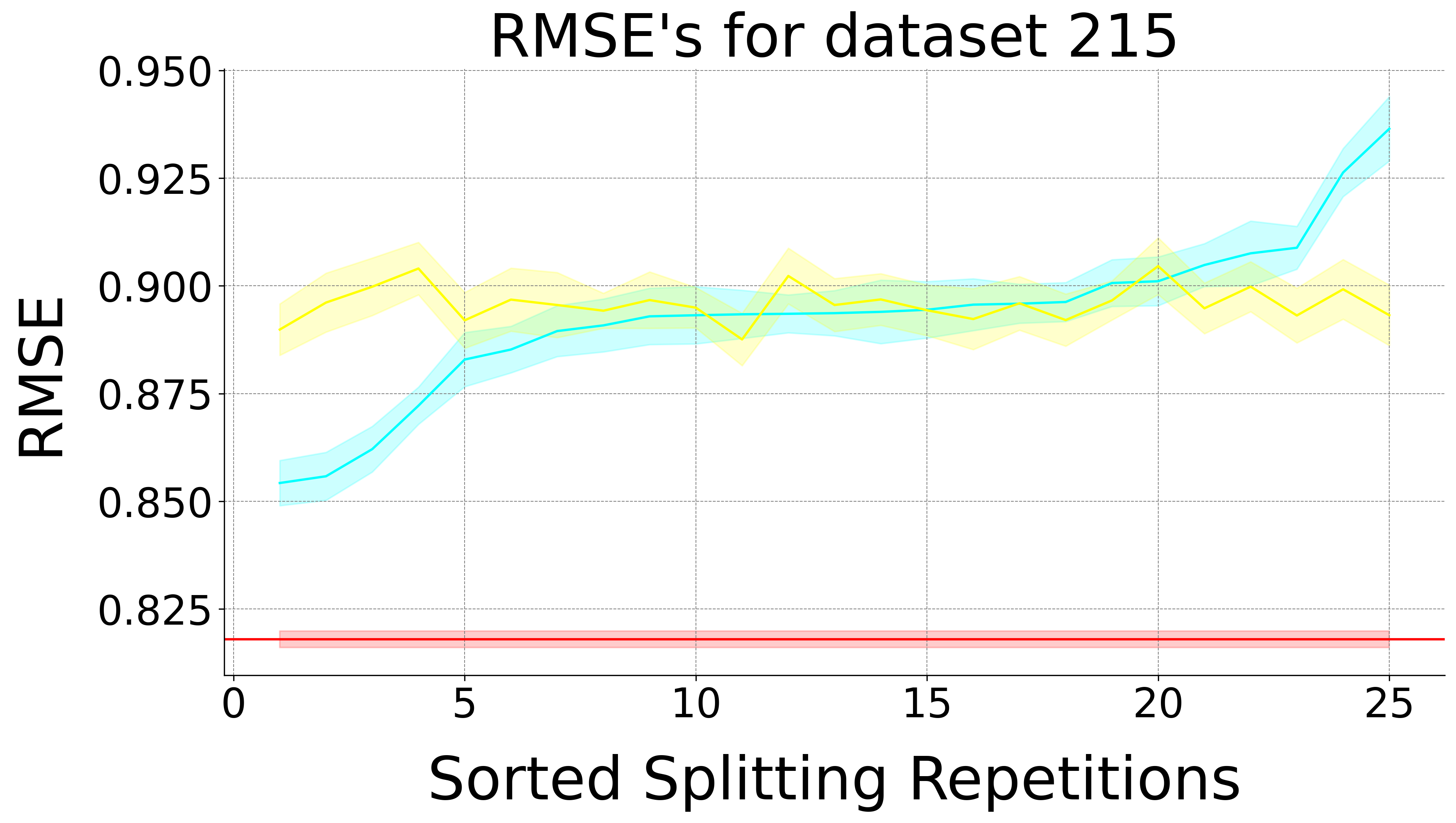}
        \includegraphics[width=0.32\textwidth]{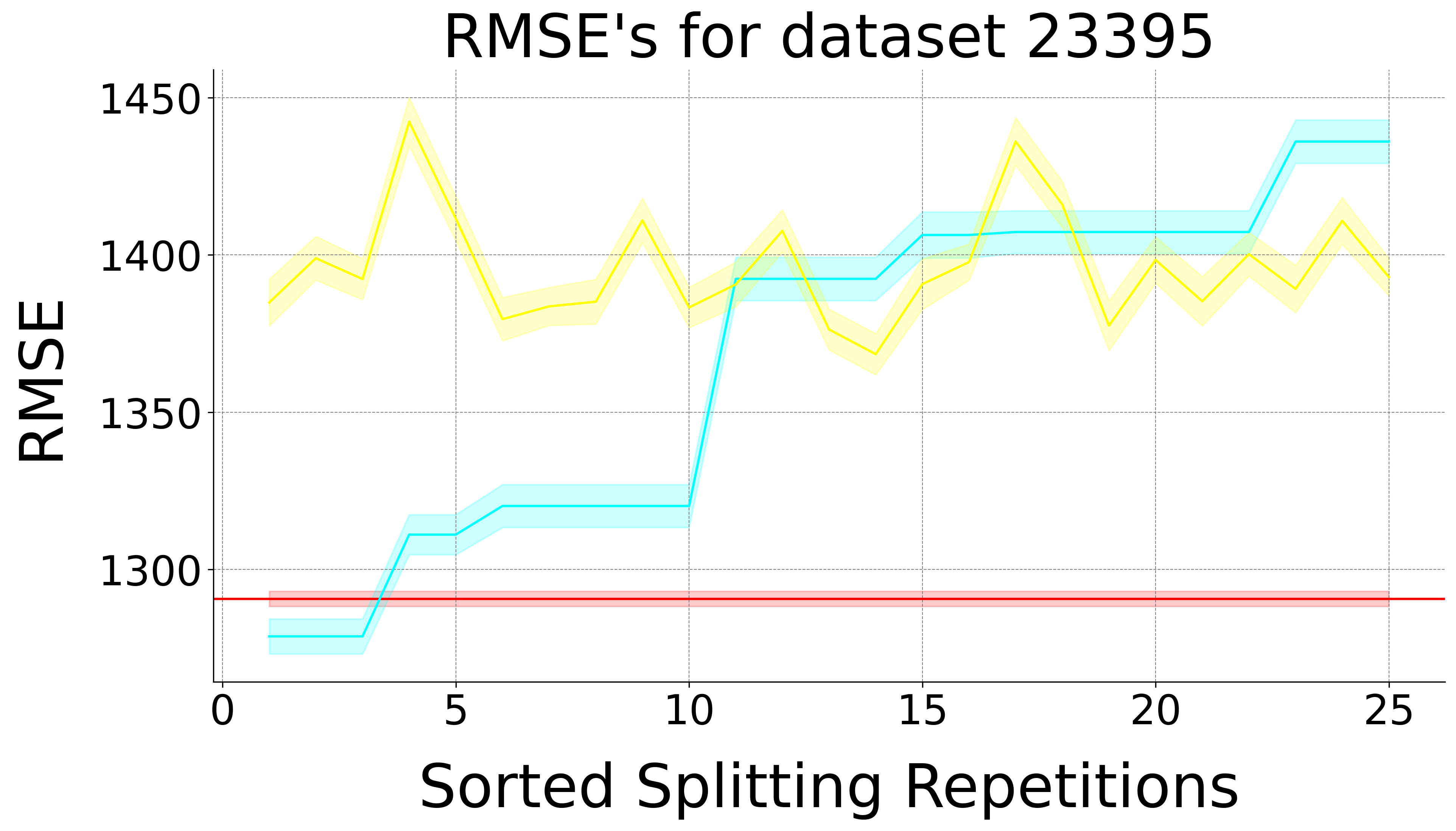}
        \includegraphics[width=0.32\textwidth]{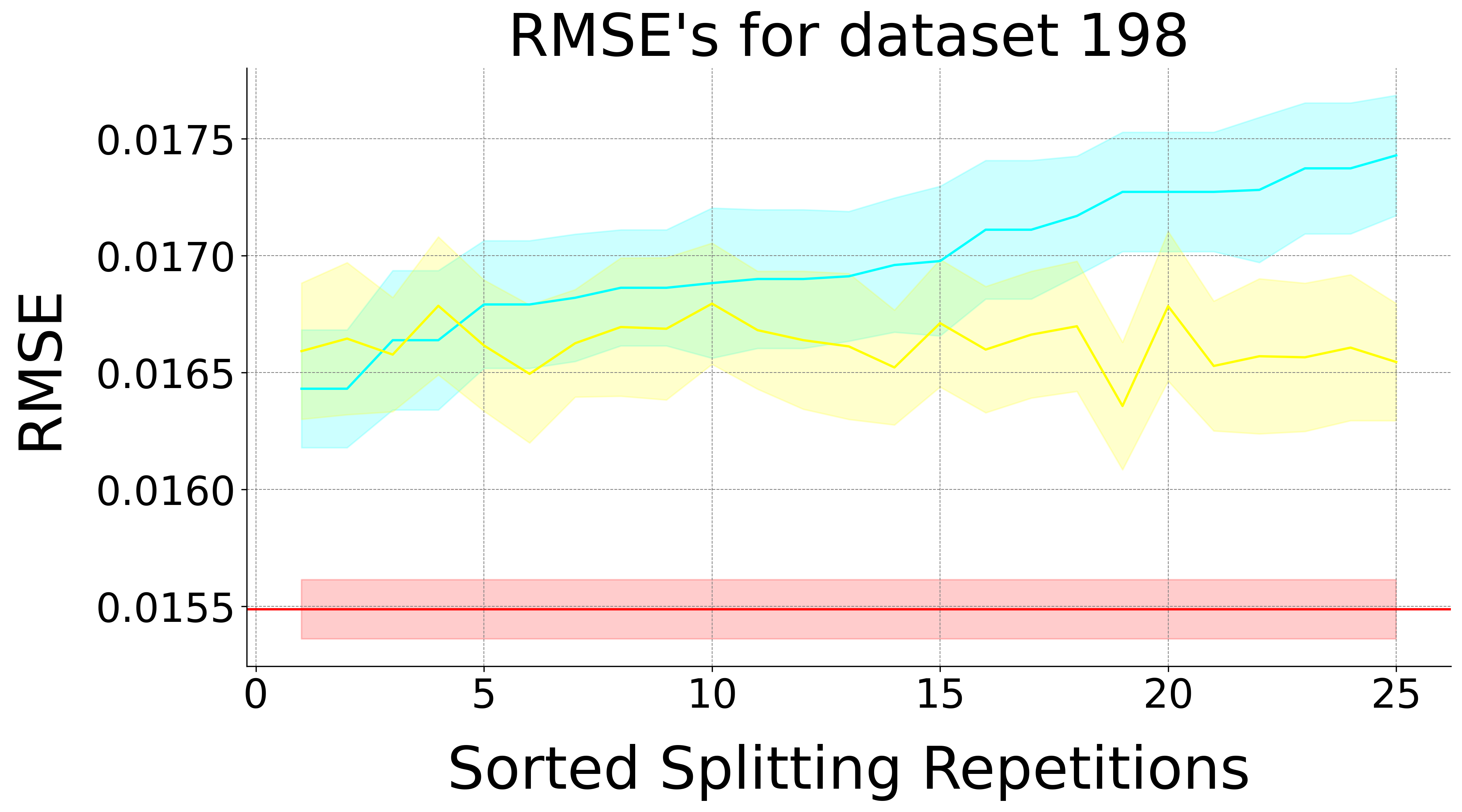}

        \vspace{2.5em}
        \includegraphics[width=0.32\textwidth]{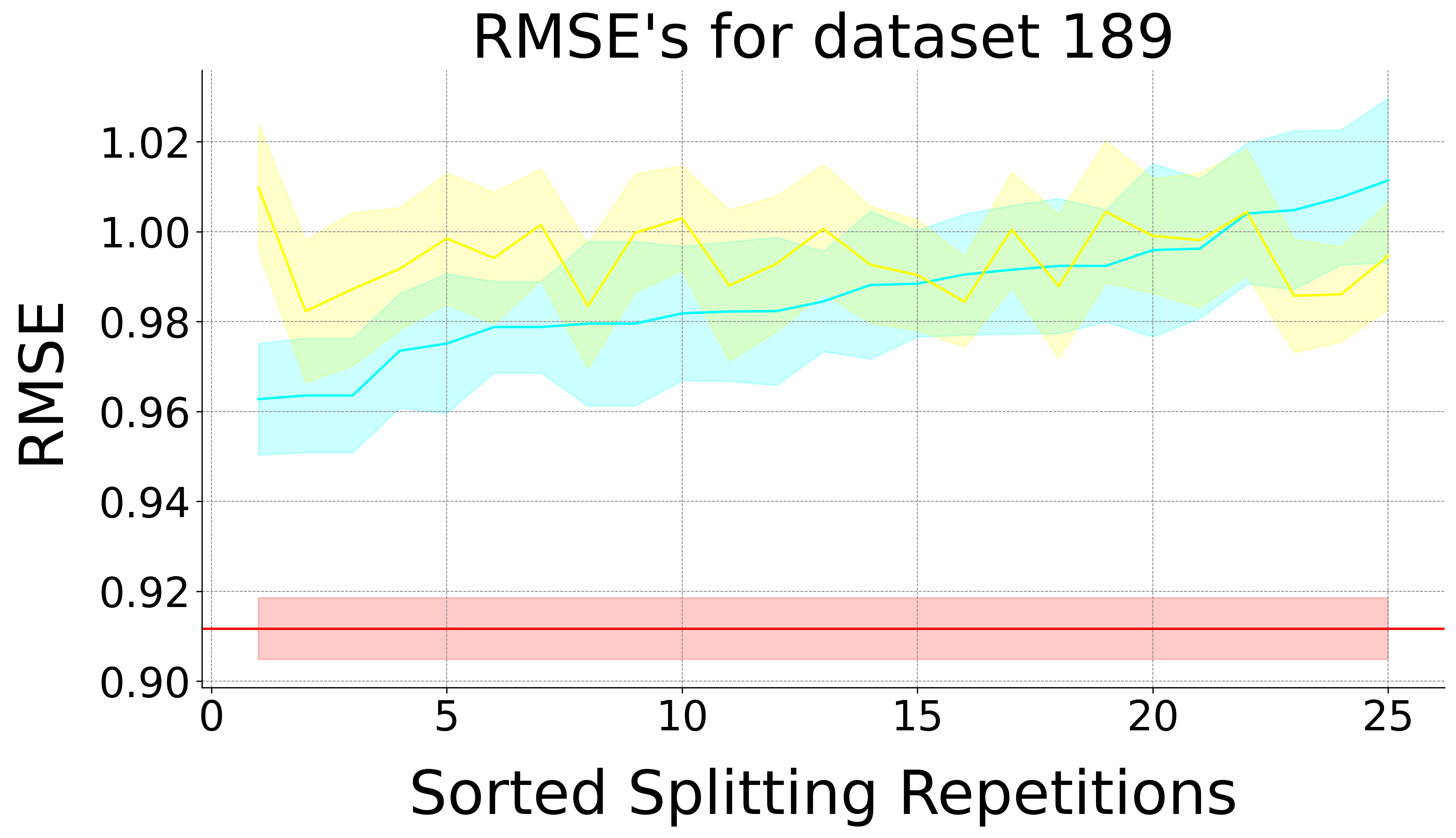}
        \includegraphics[width=0.32\textwidth]{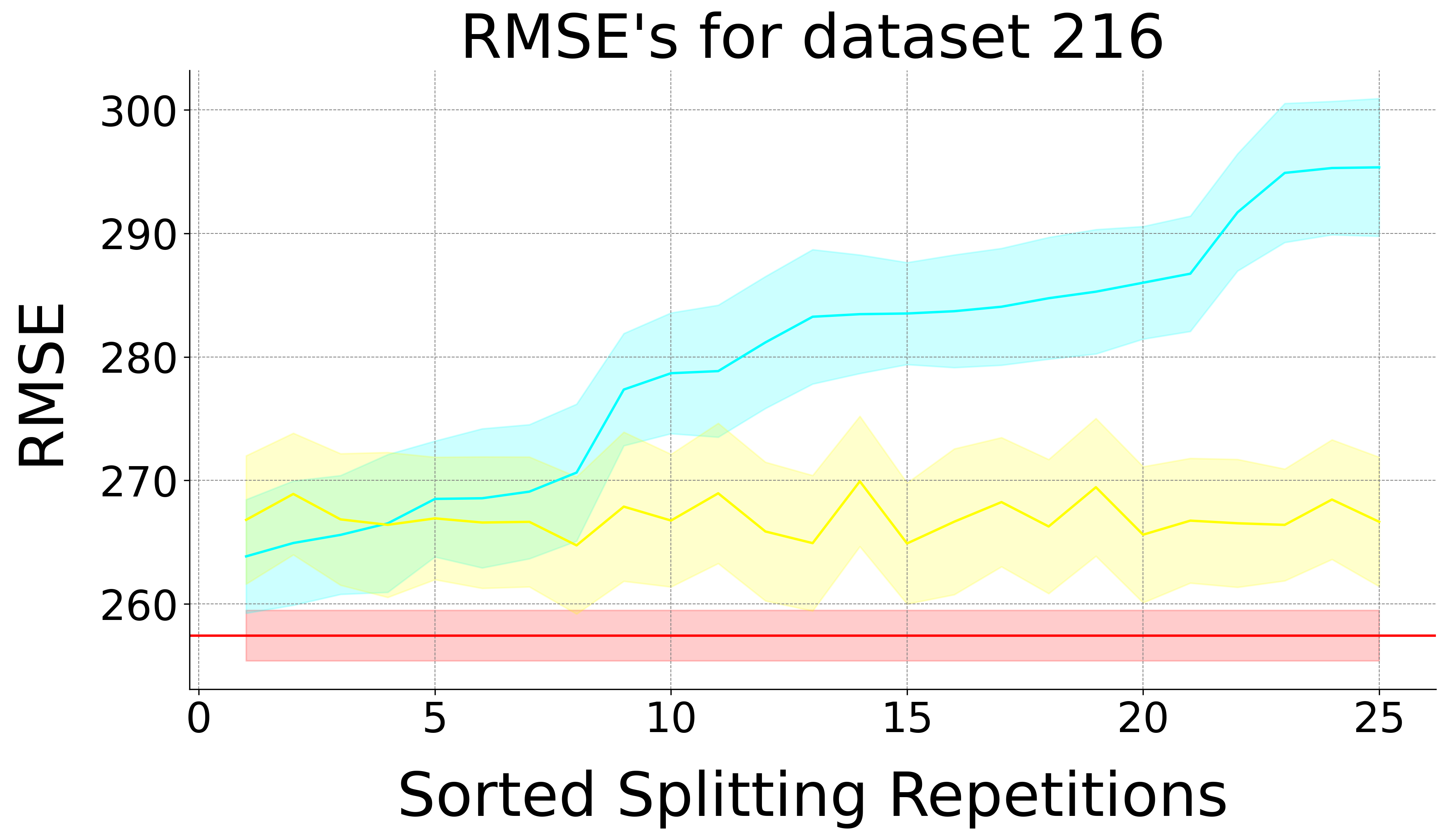}
        \includegraphics[width=0.32\textwidth]{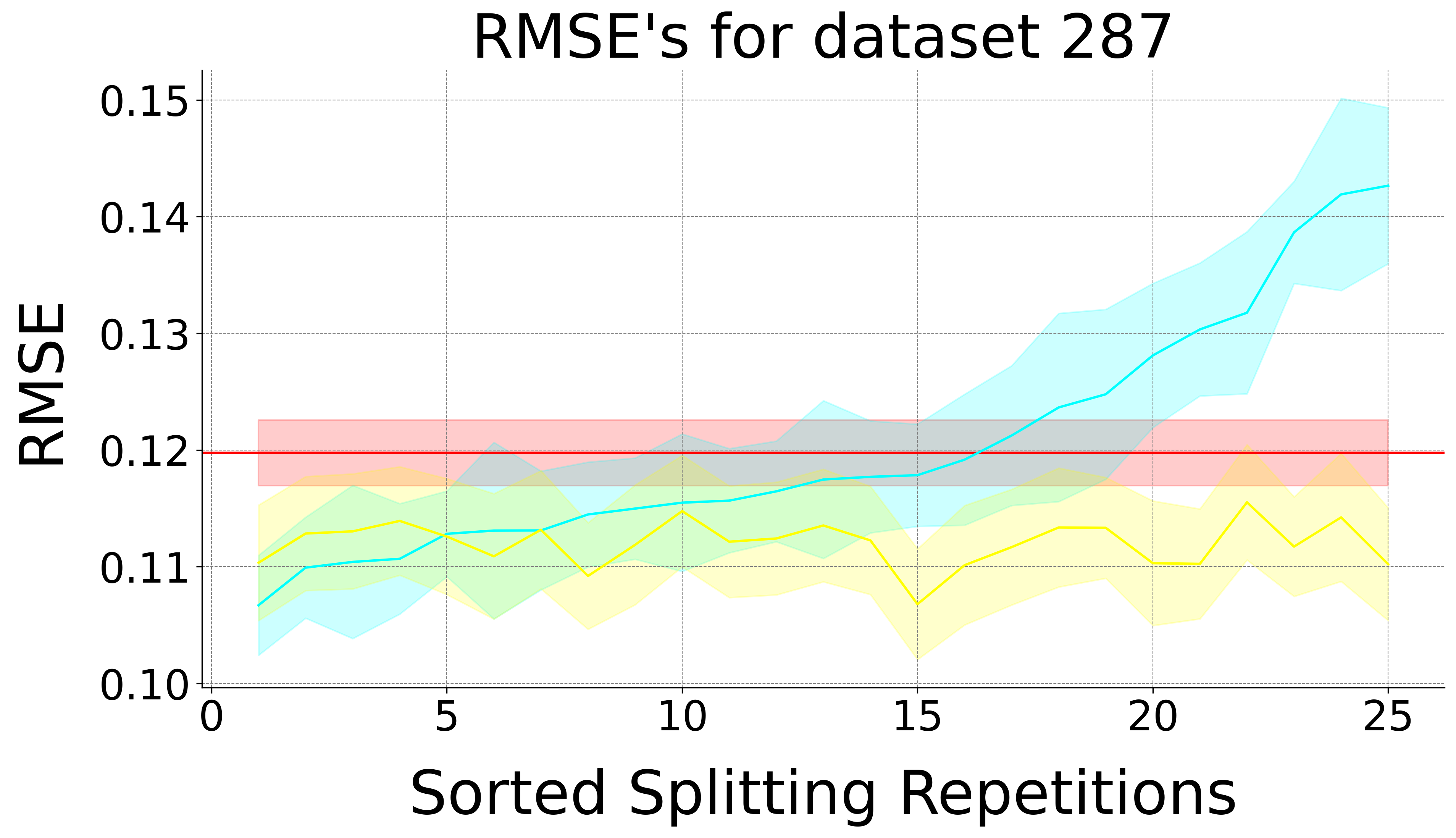}

        \vspace{2.5em}
        \includegraphics[width=0.32\textwidth]{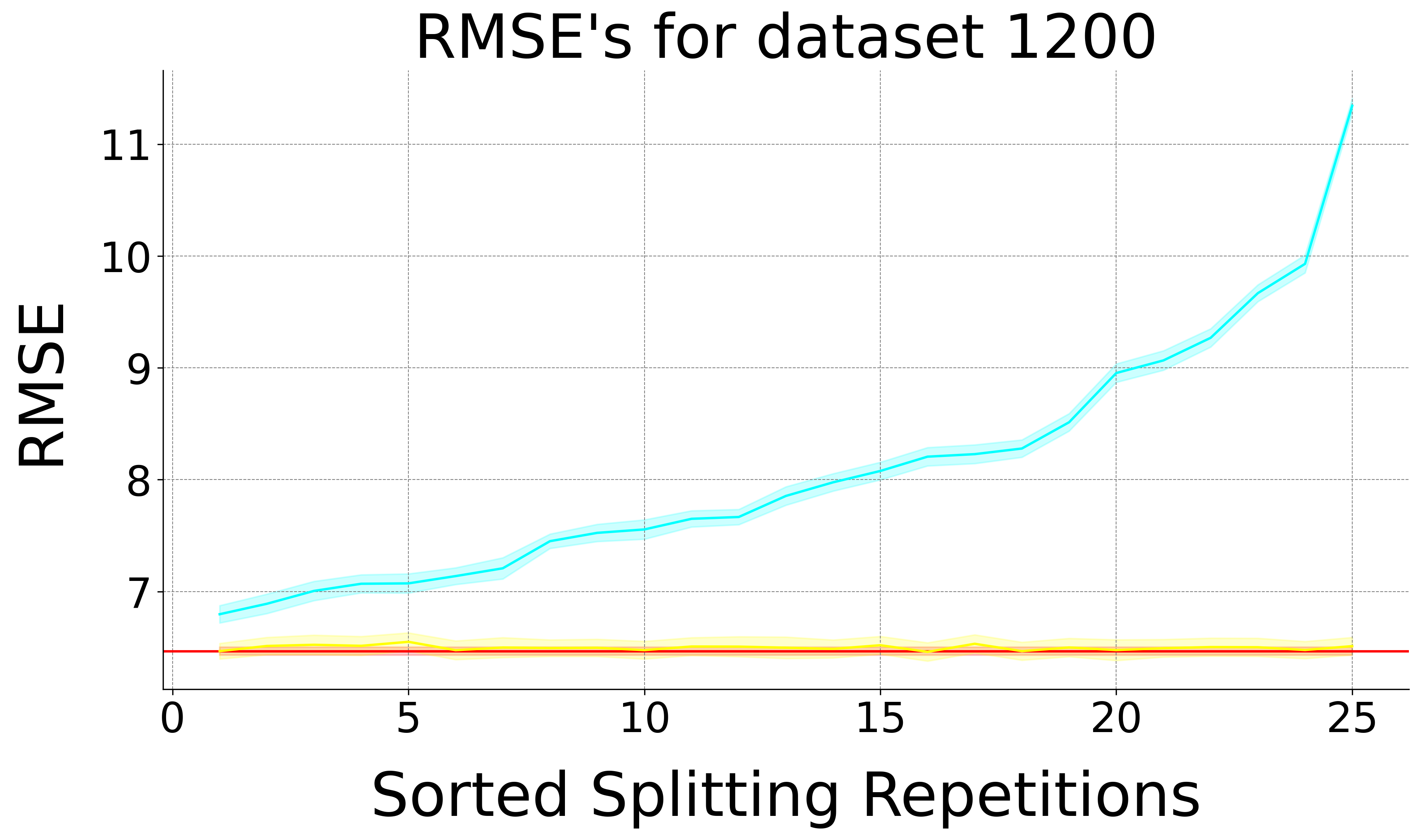}
        \includegraphics[width=0.32\textwidth]{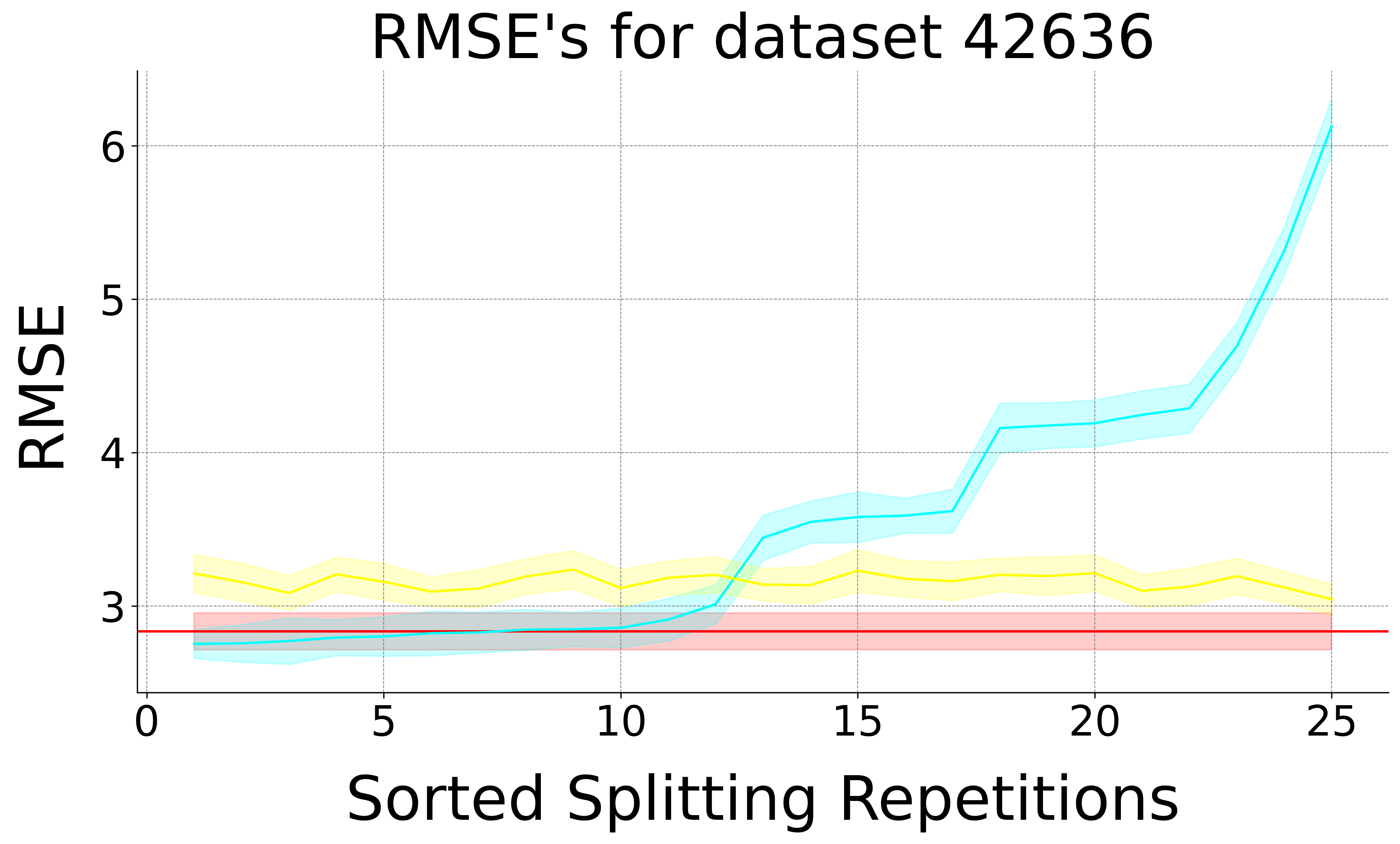}
        \includegraphics[width=0.32\textwidth]{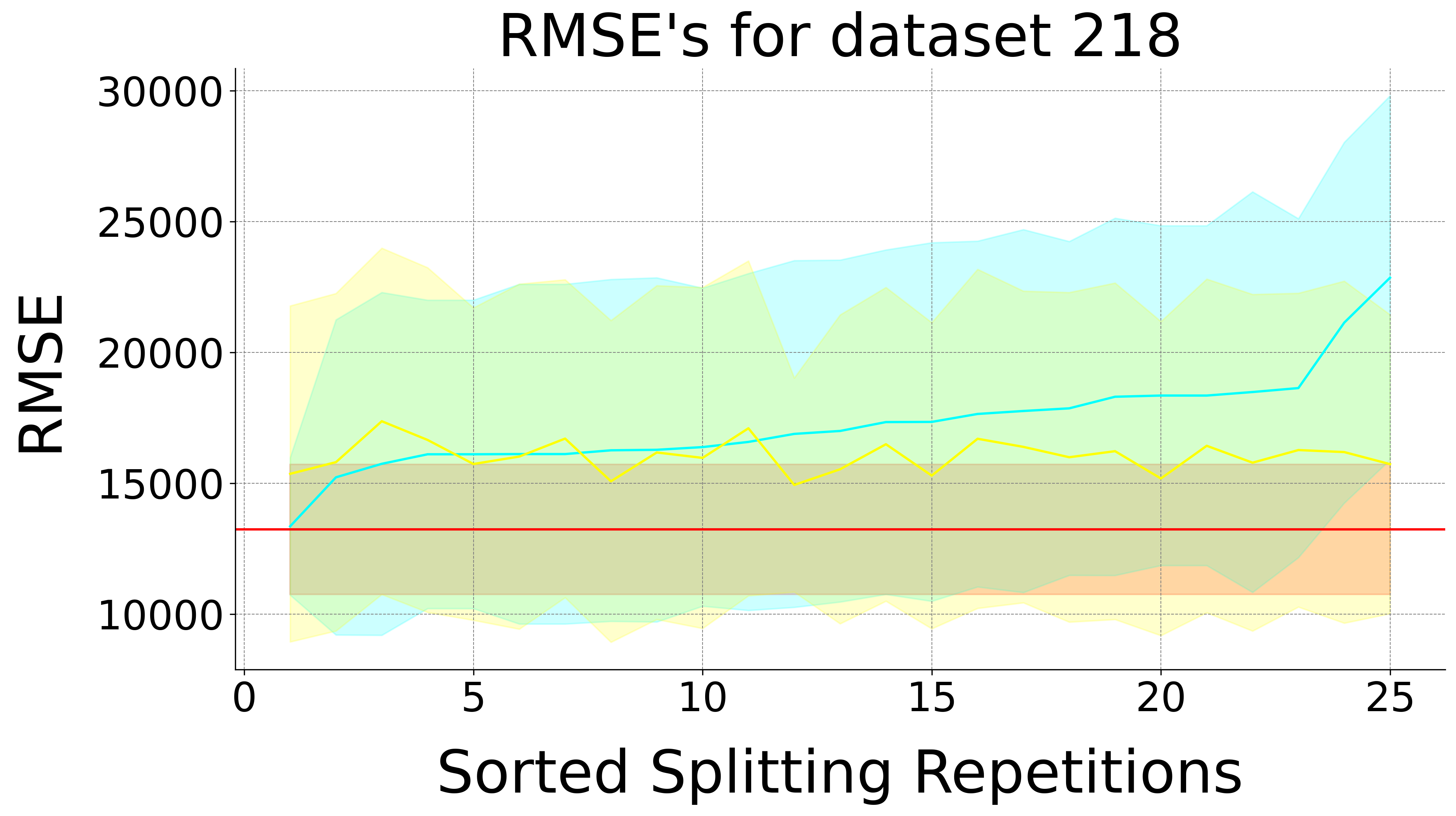}

        \vspace{2.5em}
        \includegraphics[width=0.32\textwidth]{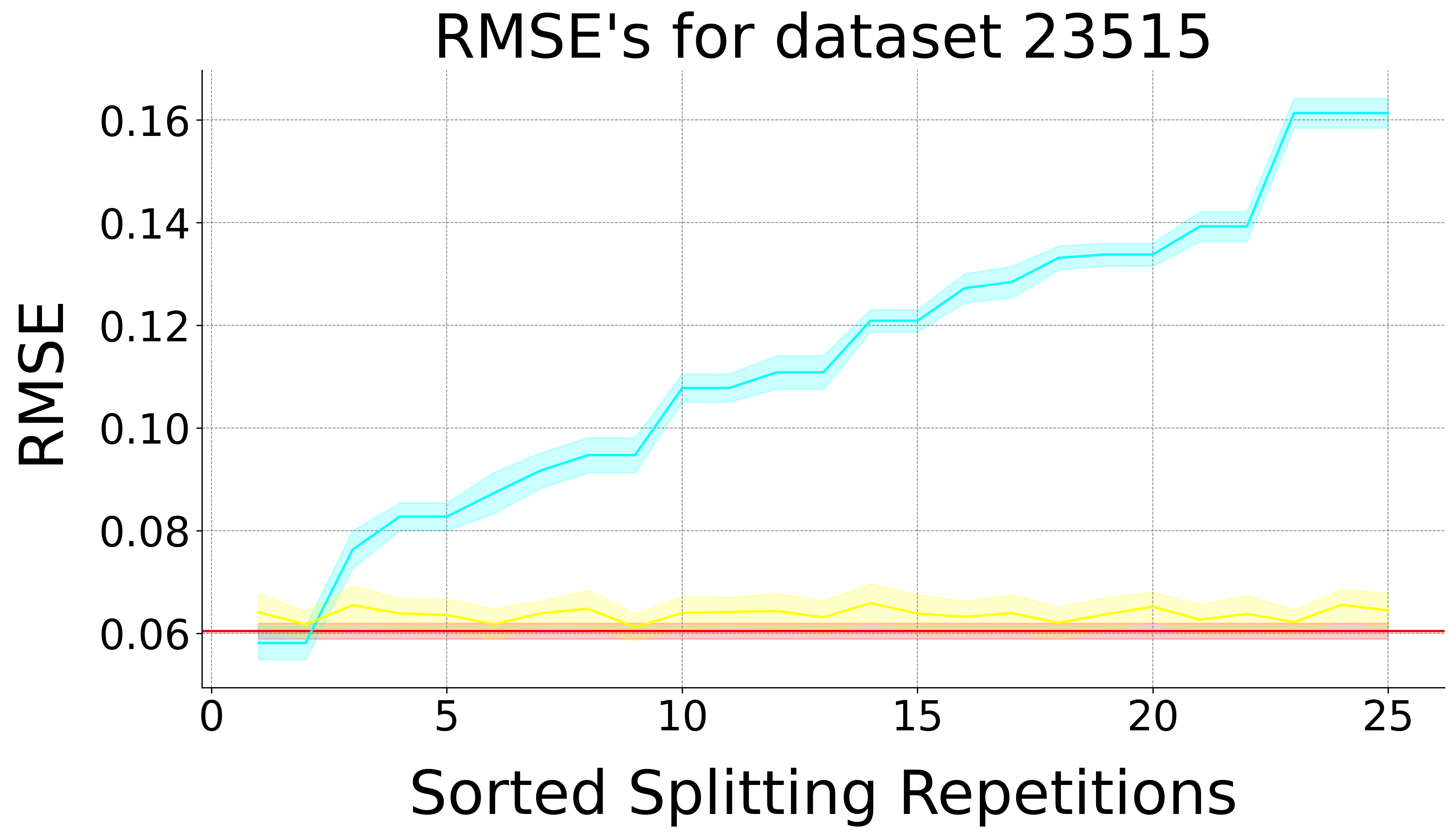}
        \includegraphics[width=0.32\textwidth]{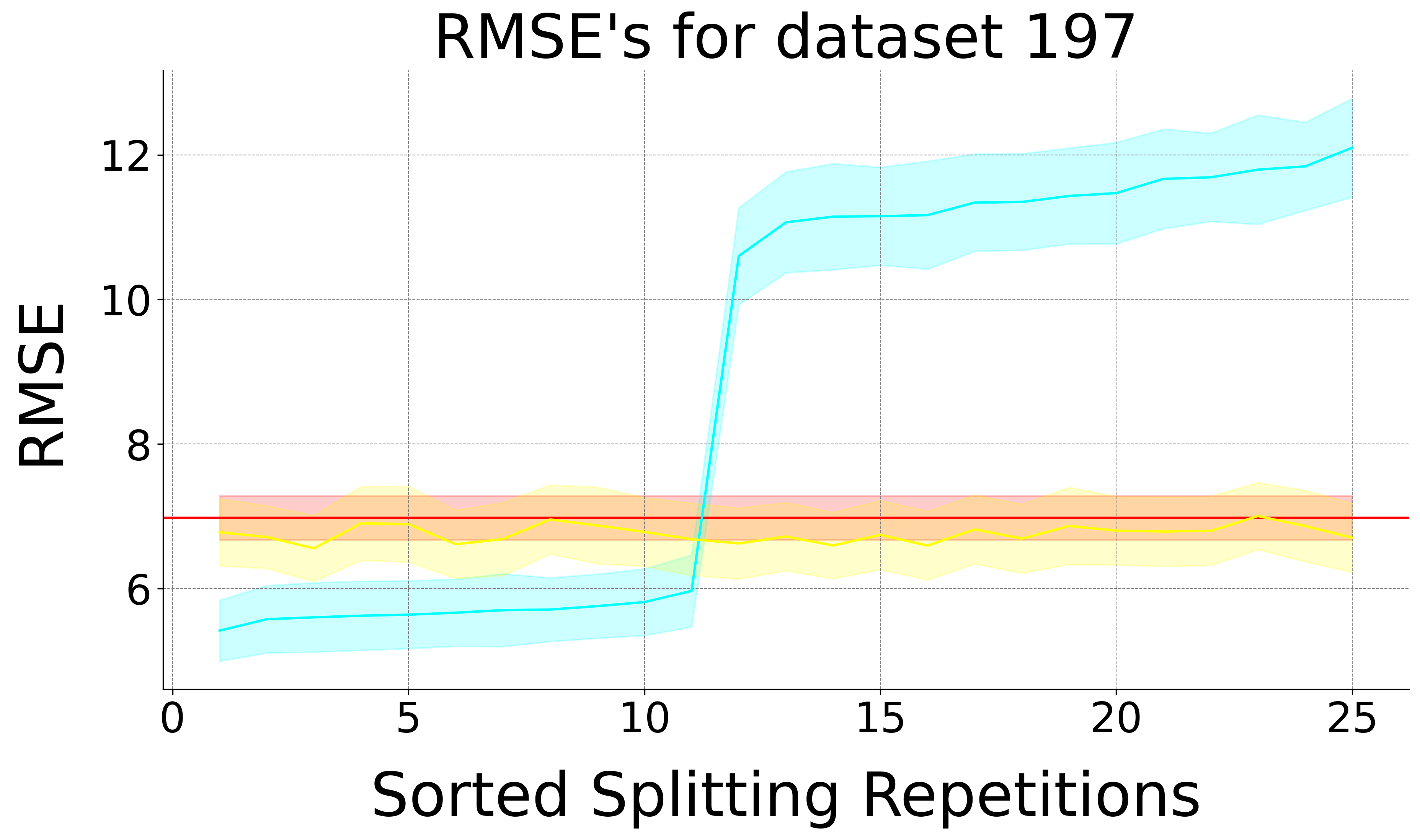} 
        \includegraphics[width=0.32\textwidth]{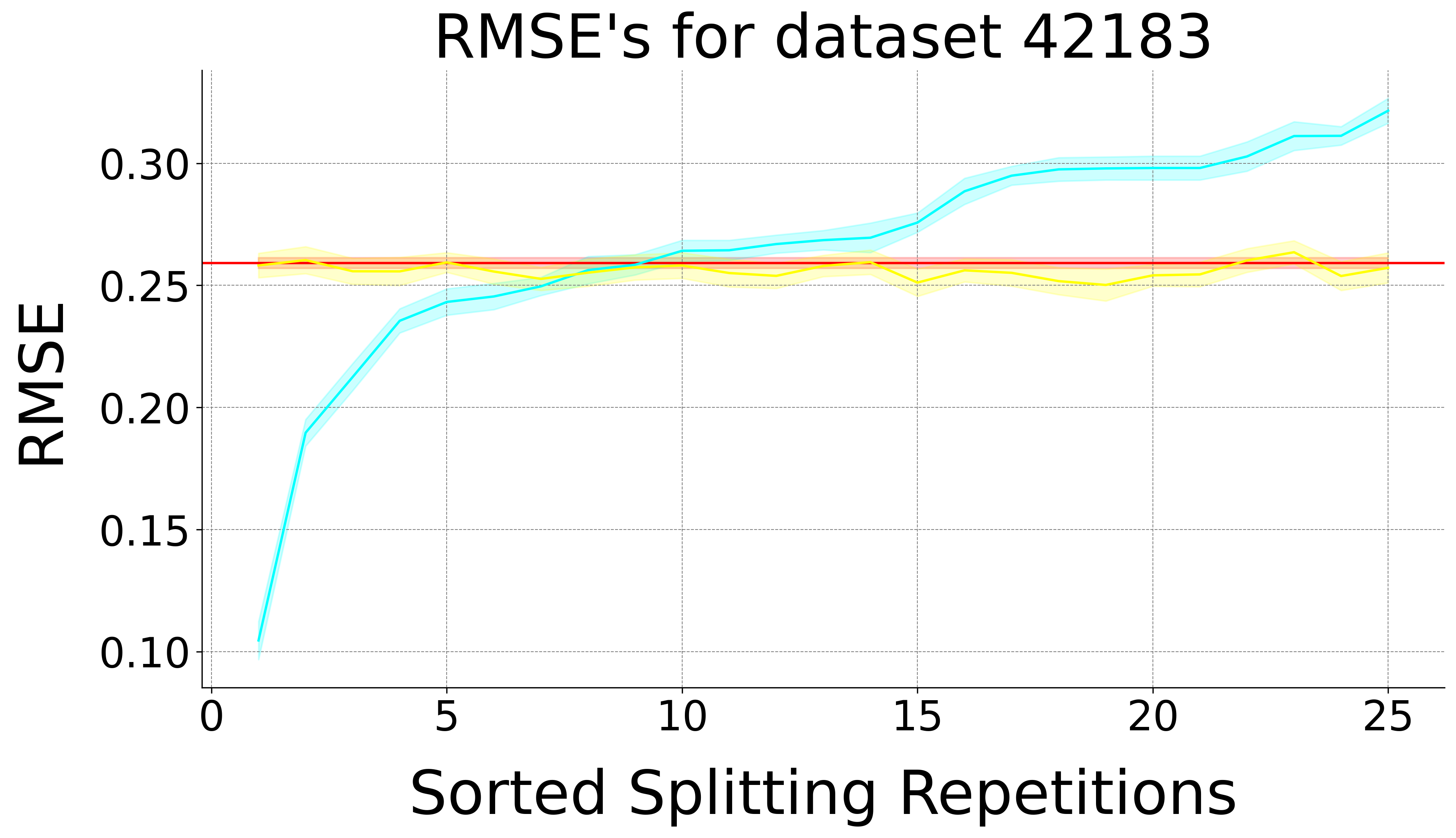}

        \caption{Detailed RMSE trends for all evaluated datasets under the MCAR missingness type.}
        \label{fig:appendix_mcar}
    \end{minipage}
\end{figure*}

We report comprehensive accuracy results for all evaluated datasets under MCAR, MAR, and MNAR missingness types in \figs{fig:appendix_mcar}{fig:appendix_mnar}.

Each figure represents the performance of the imputation for a specific dataset as detailed in \tbl{table:overview}. Splitting repetitions in each plot have been ordered by increasing Vertical $k$-NN RMSE. The width of the bands in each plot illustrates 95\% confidence intervals.

\begin{figure*}
    \begin{minipage}[c][\textheight][c]{\textwidth}
        \centering
        \includegraphics[width=0.6\textwidth]{figures/legend.png}
        
        \includegraphics[width=0.32\textwidth]{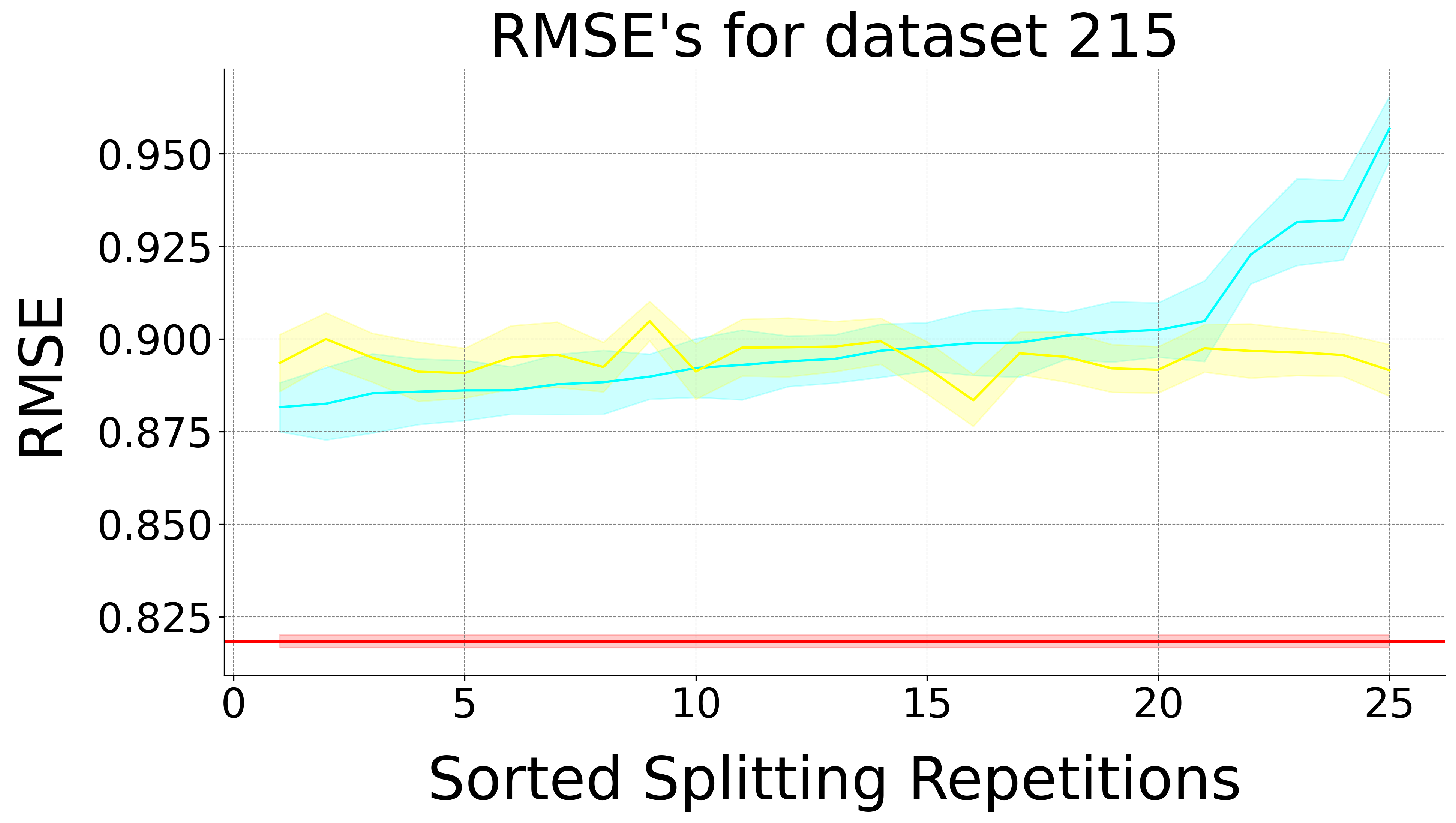}
        \includegraphics[width=0.32\textwidth]{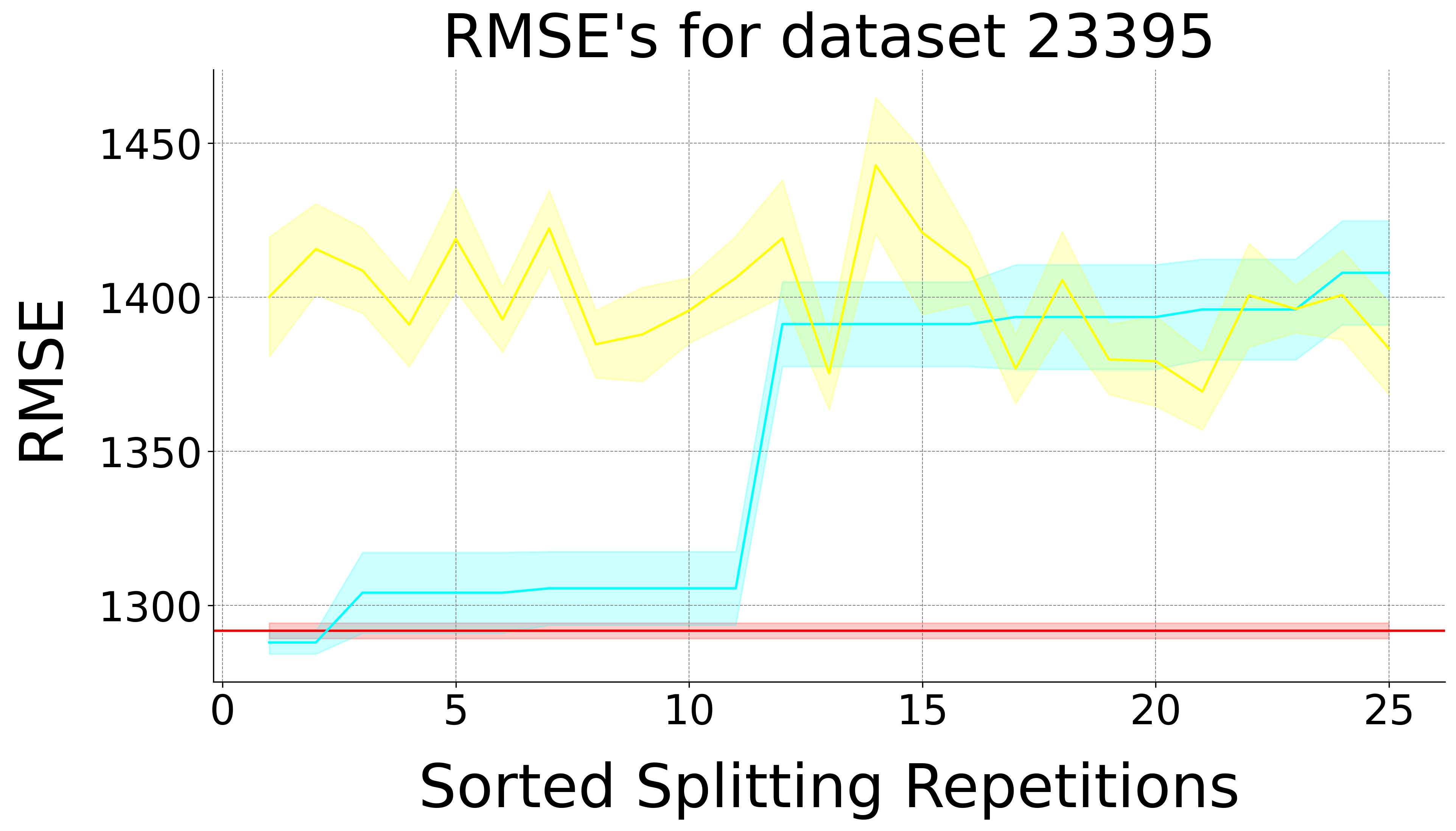}
        \includegraphics[width=0.32\textwidth]{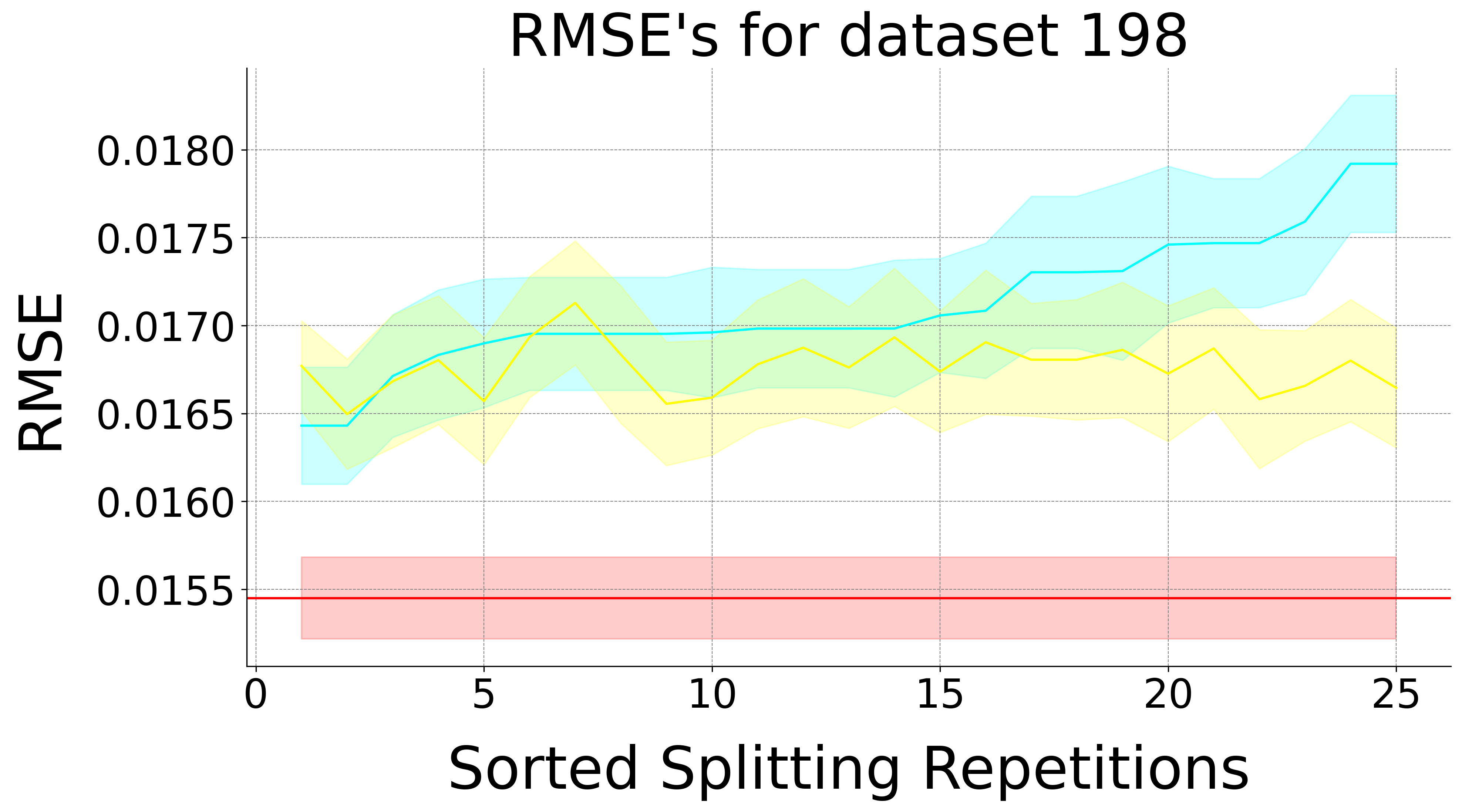}

        \vspace{2.5em}
        \includegraphics[width=0.32\textwidth]{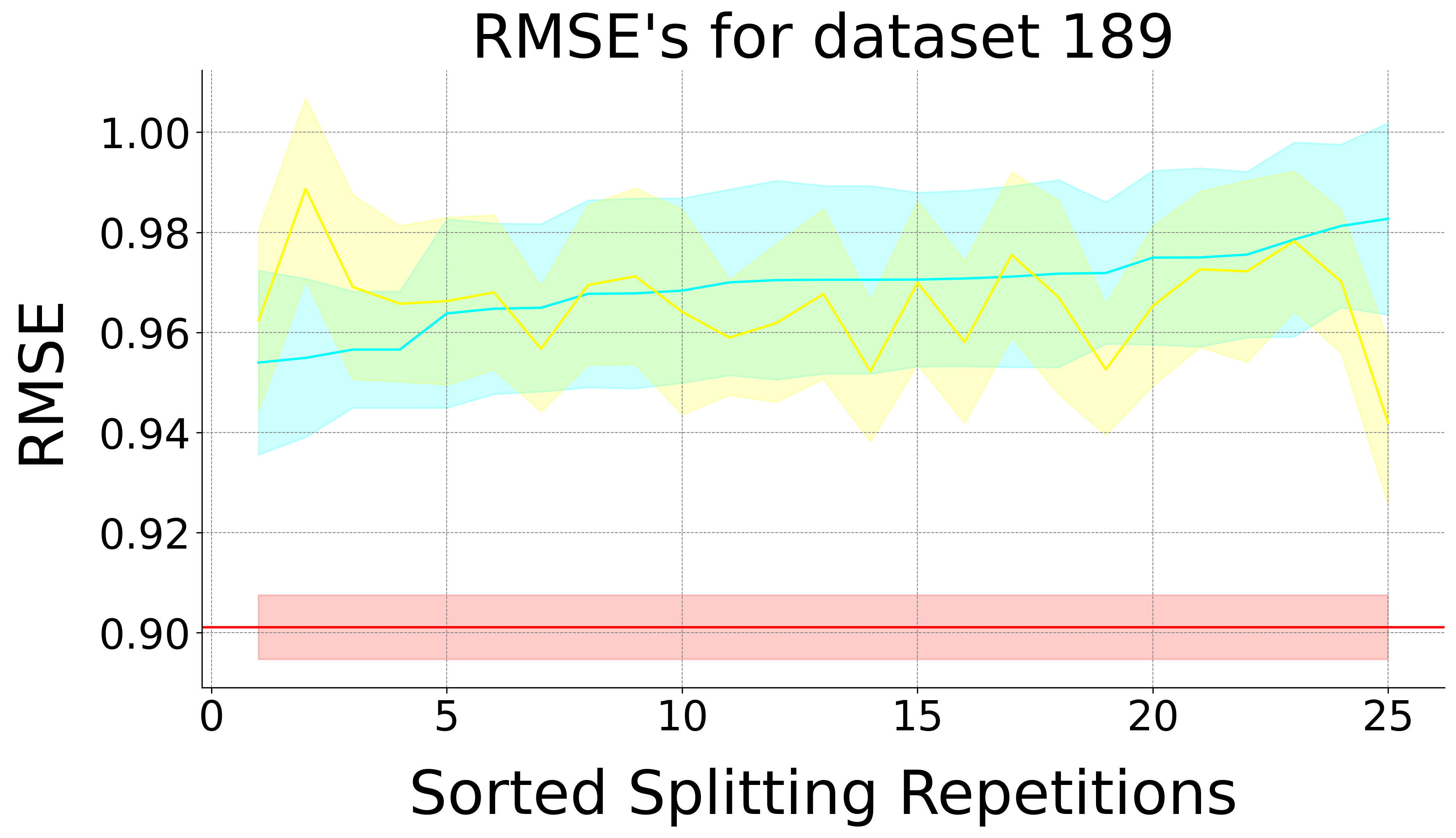}
        \includegraphics[width=0.32\textwidth]{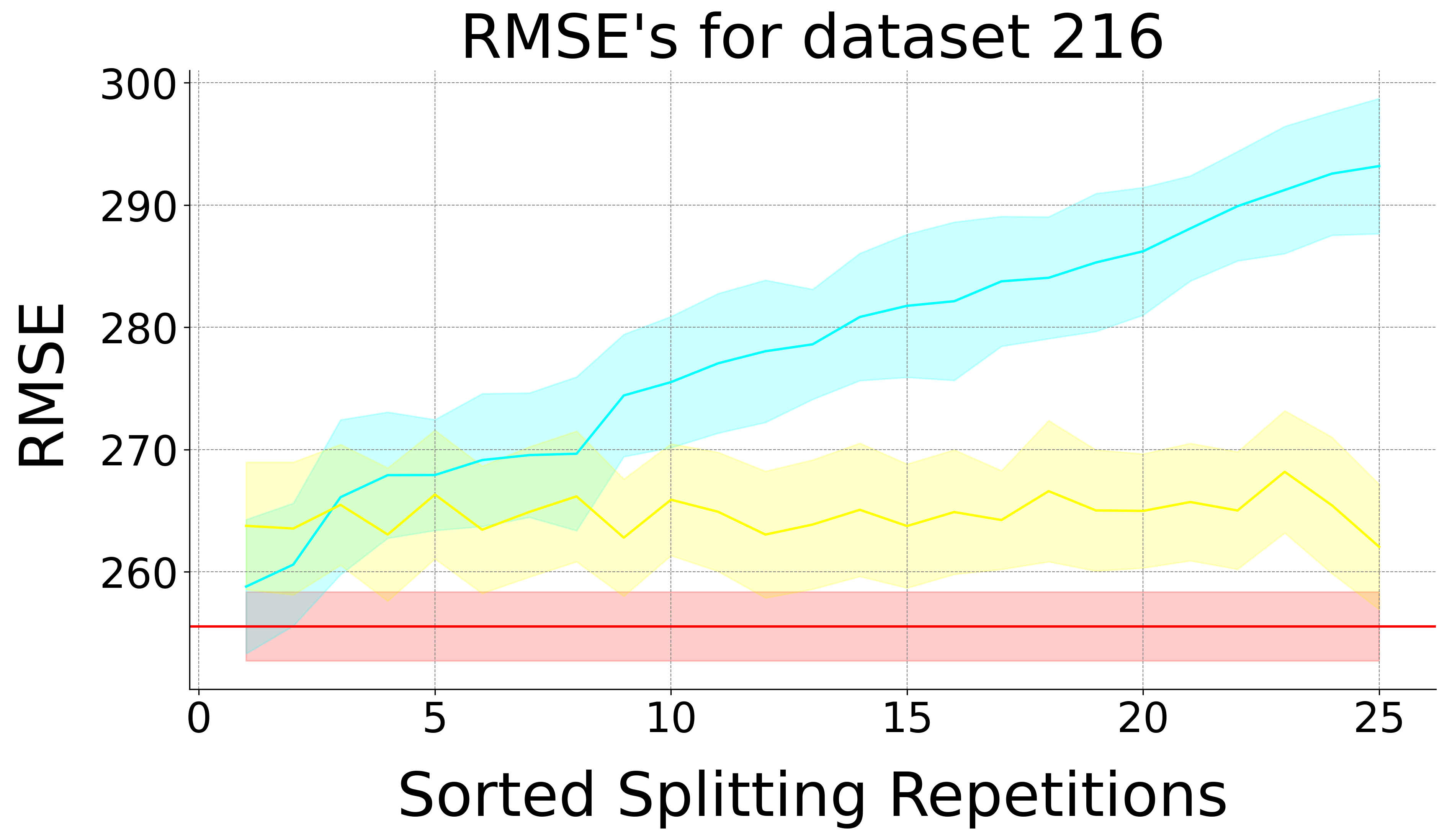}
        \includegraphics[width=0.32\textwidth]{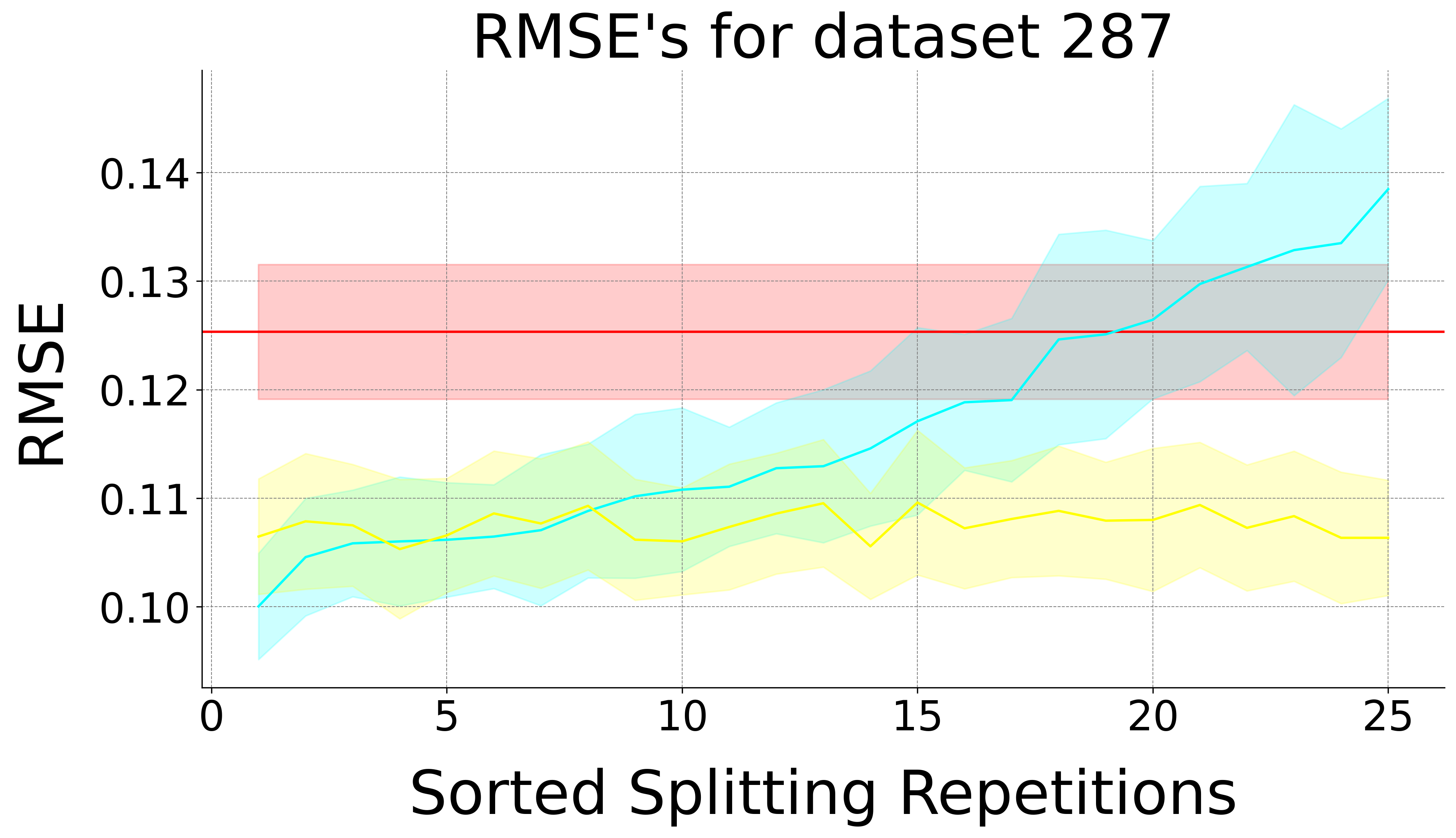}

        \vspace{2.5em}
        \includegraphics[width=0.32\textwidth]{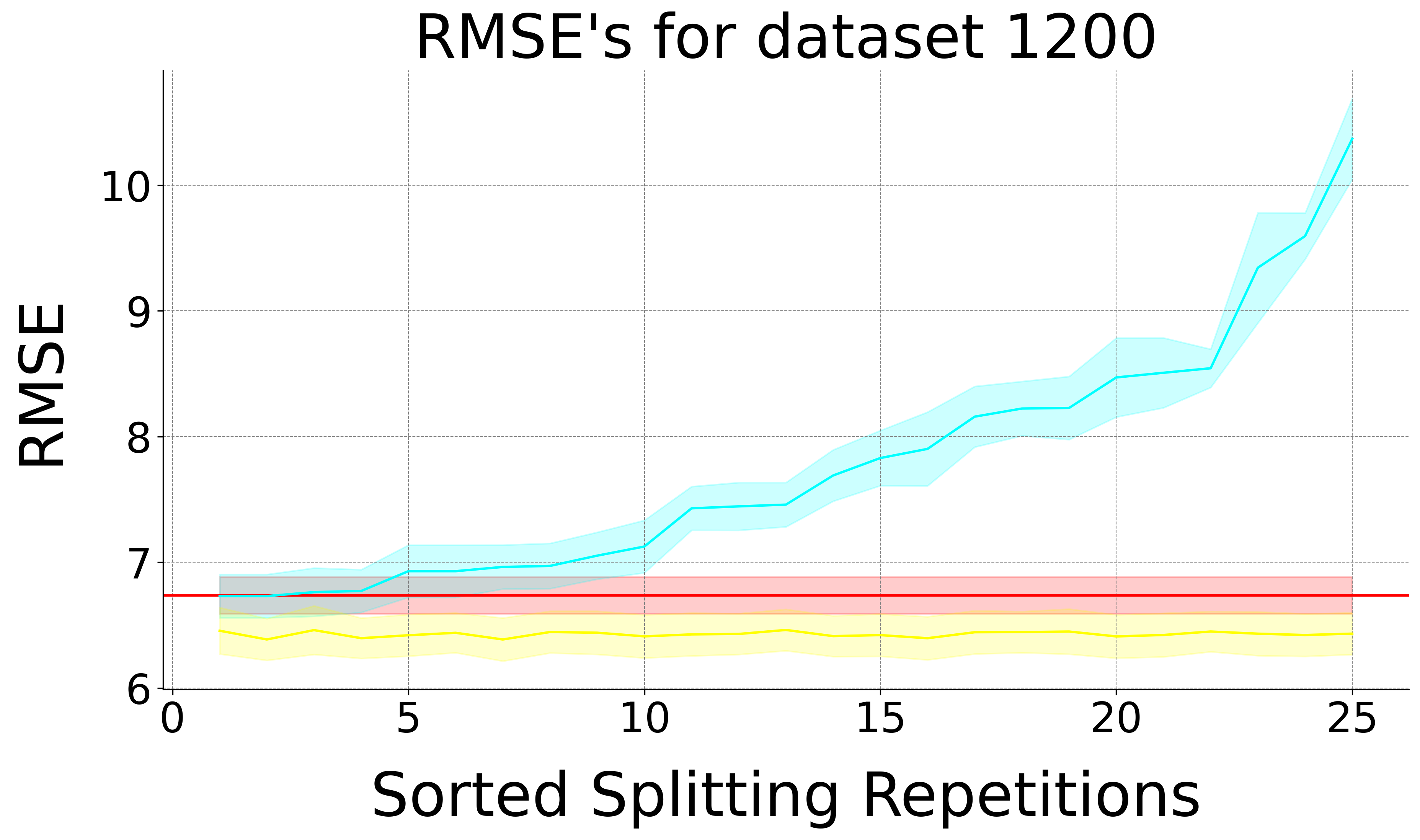}
        \includegraphics[width=0.32\textwidth]{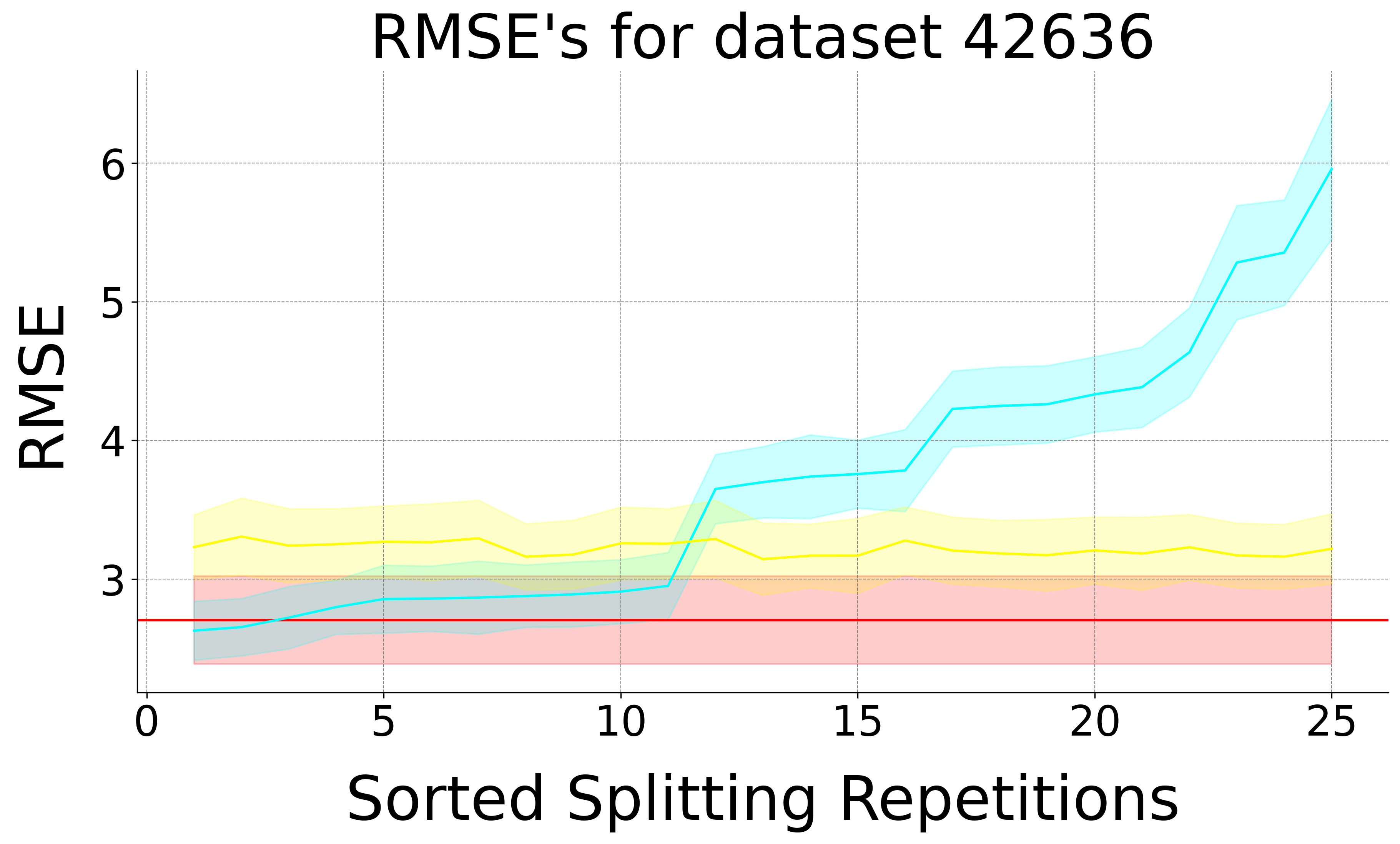}
        \includegraphics[width=0.32\textwidth]{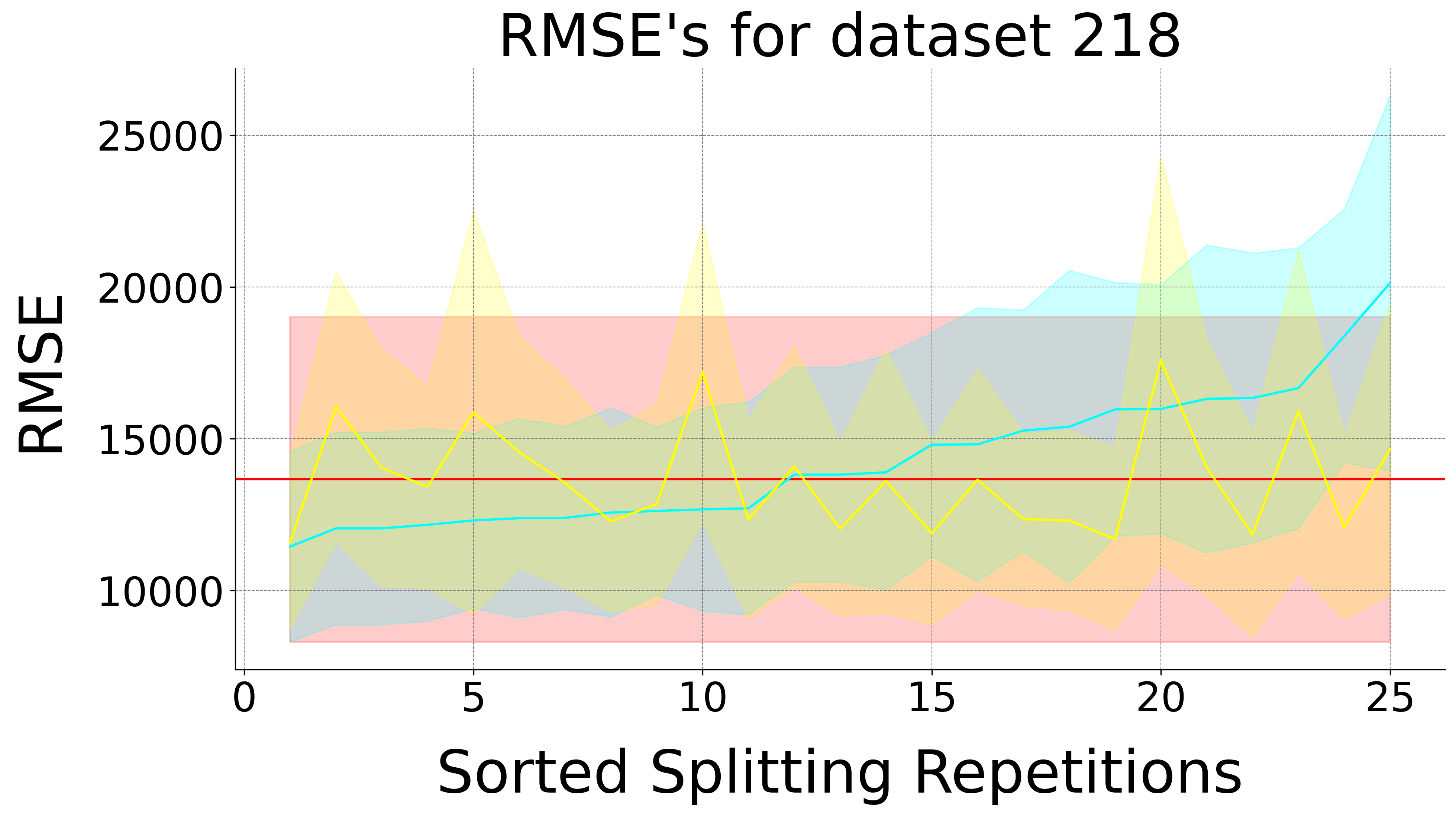}

        \vspace{2.5em}
        \includegraphics[width=0.32\textwidth]{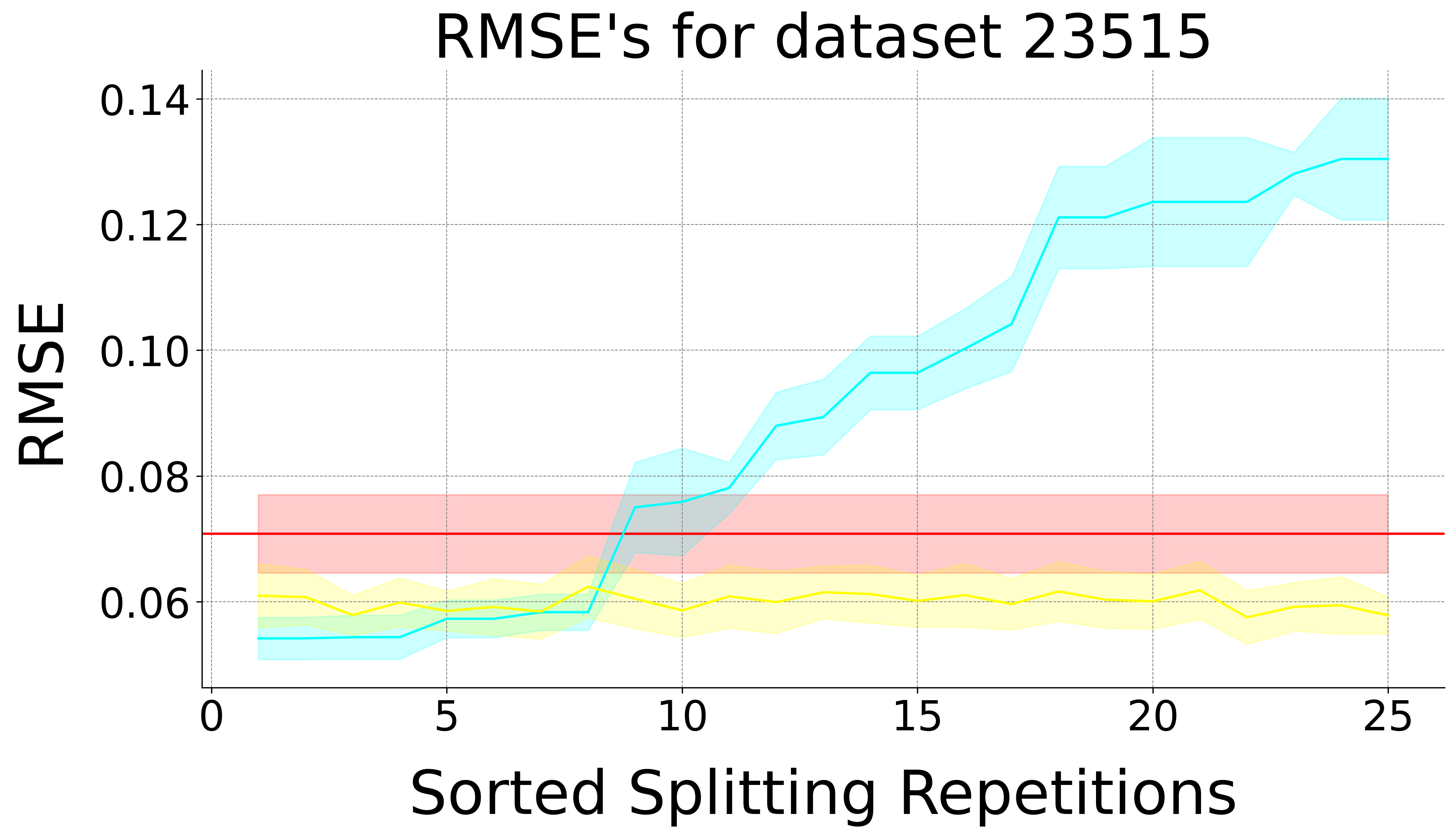}
        \includegraphics[width=0.32\textwidth]{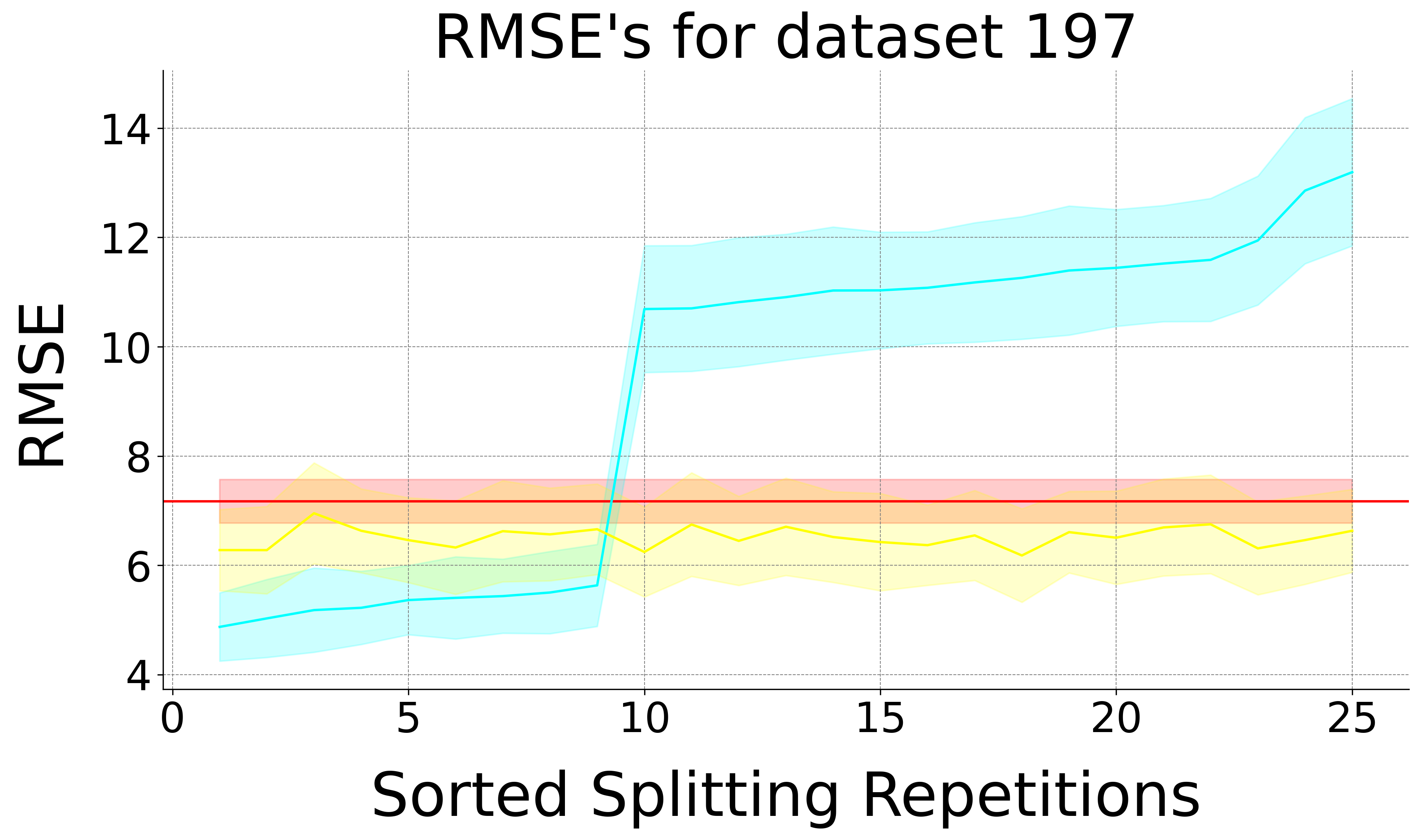}
        \includegraphics[width=0.32\textwidth]{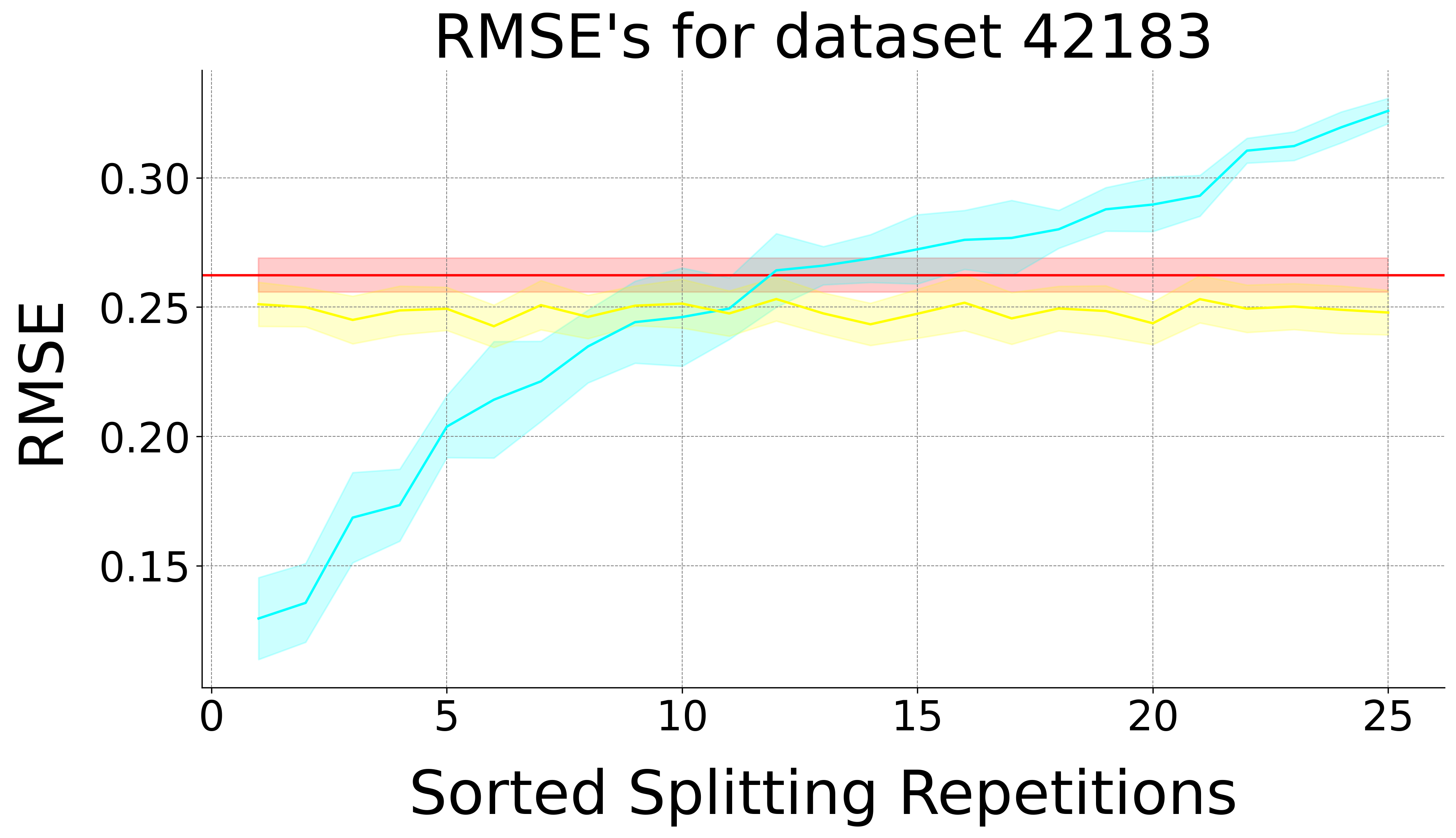}

        \caption{Detailed RMSE trends for all evaluated datasets under the MAR missingness type.} 
        \label{fig:appendix_mar}
    \end{minipage}
\end{figure*}

\begin{figure*}
    \begin{minipage}[c][\textheight][c]{\textwidth}
        \centering
        \includegraphics[width=0.6\textwidth]{figures/legend.png}
        
        \includegraphics[width=0.32\textwidth]{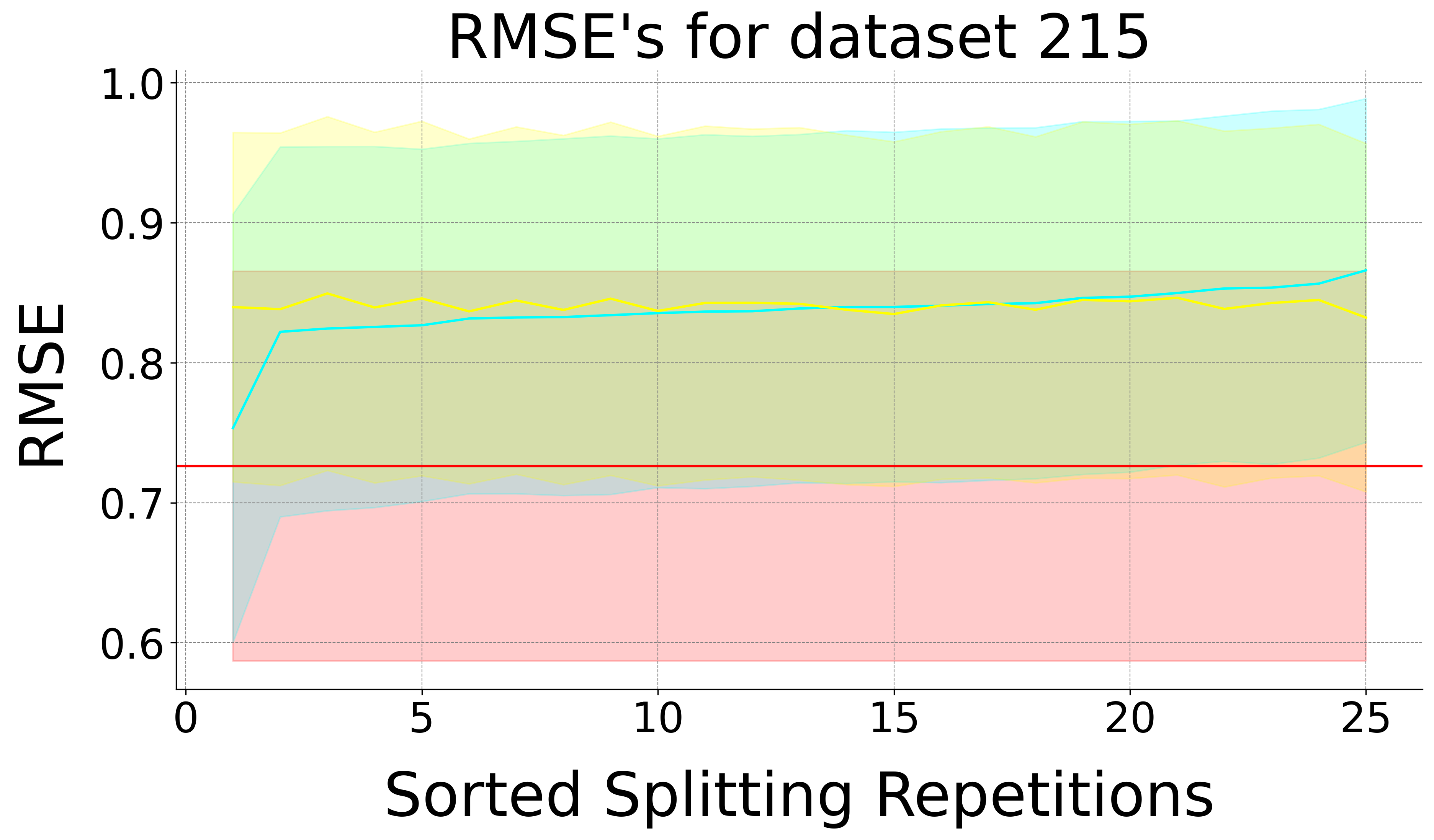}
        \includegraphics[width=0.32\textwidth]{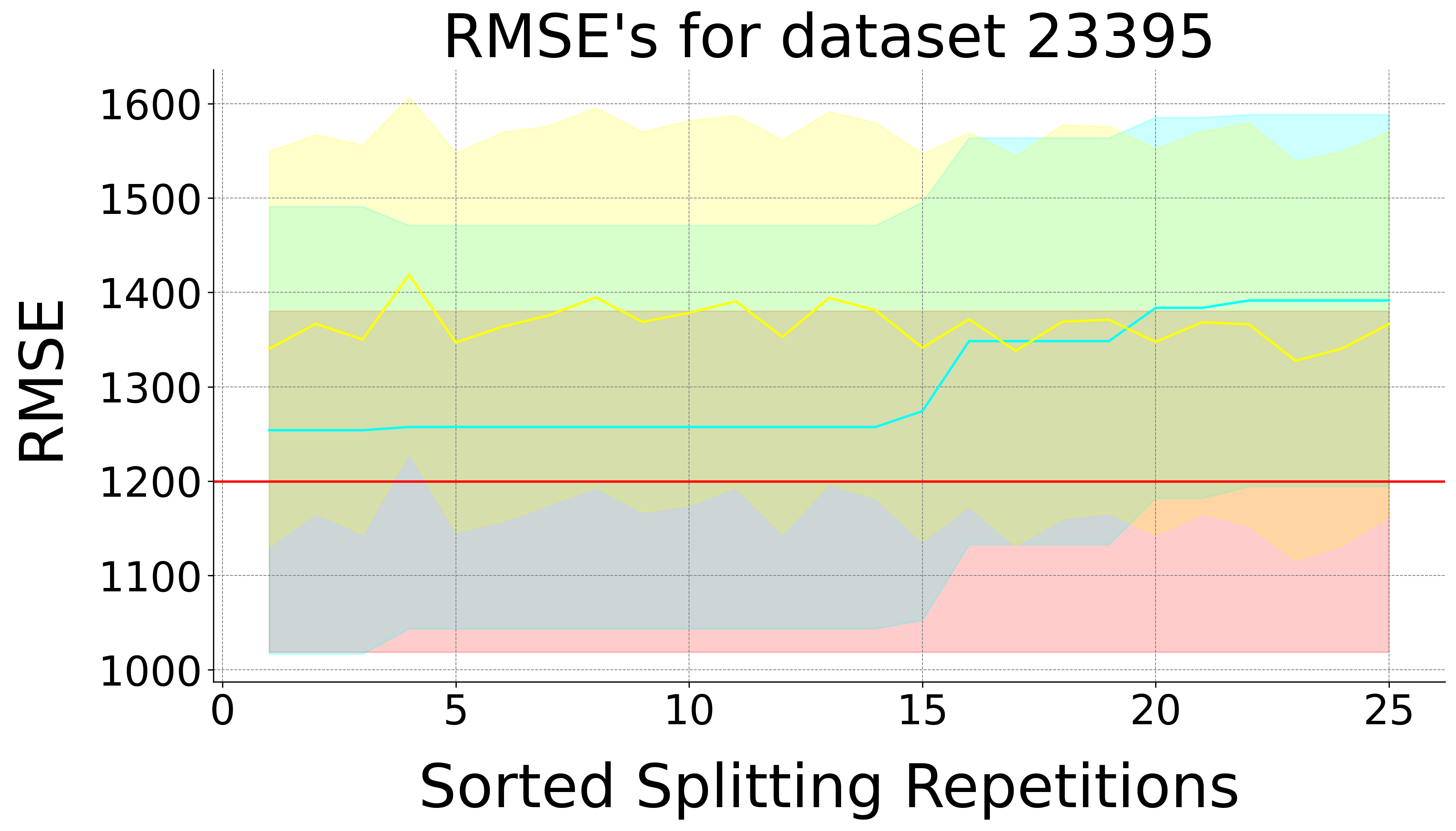}
        \includegraphics[width=0.32\textwidth]{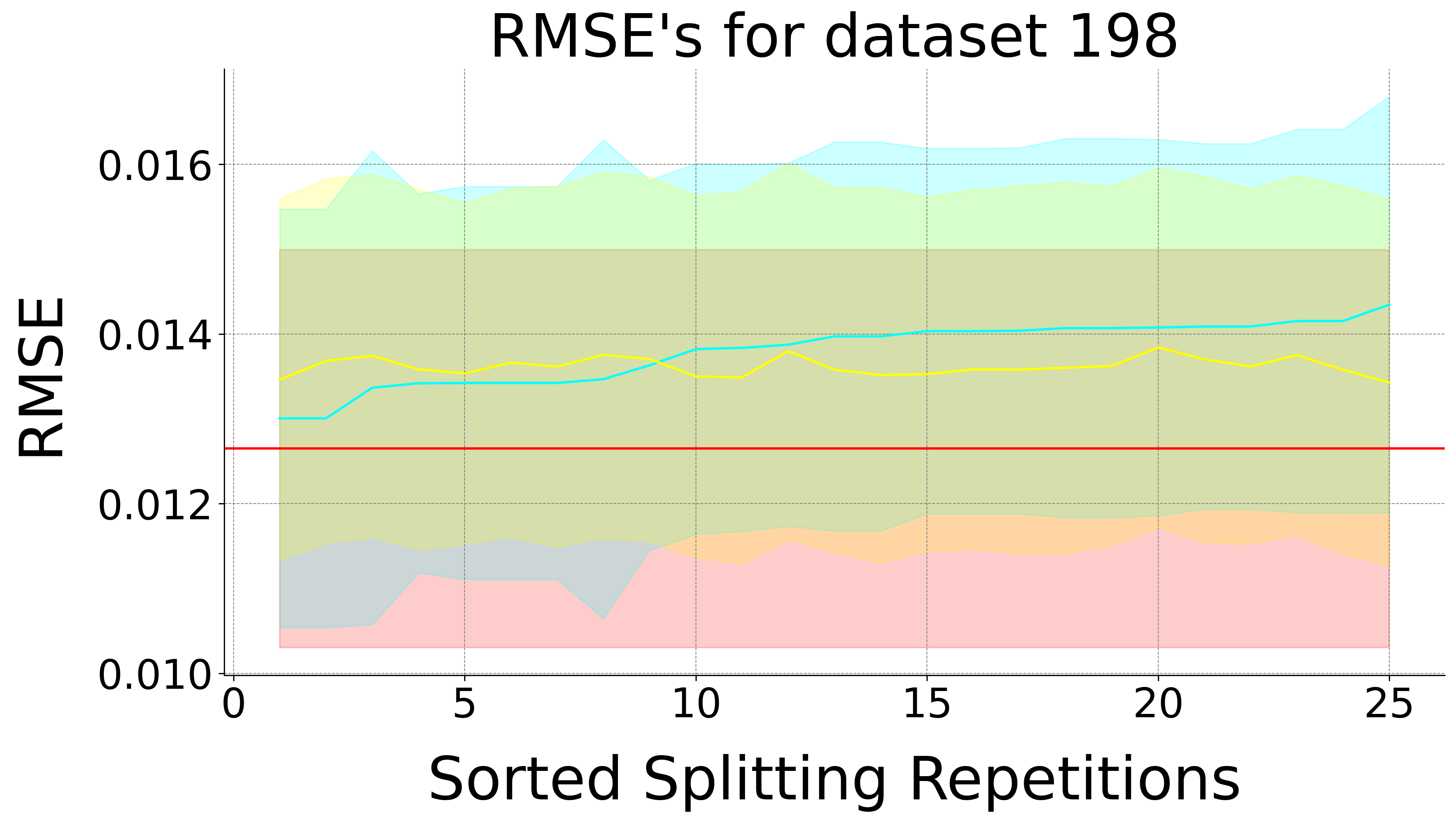}

        \vspace{2.5em}
        \includegraphics[width=0.32\textwidth]{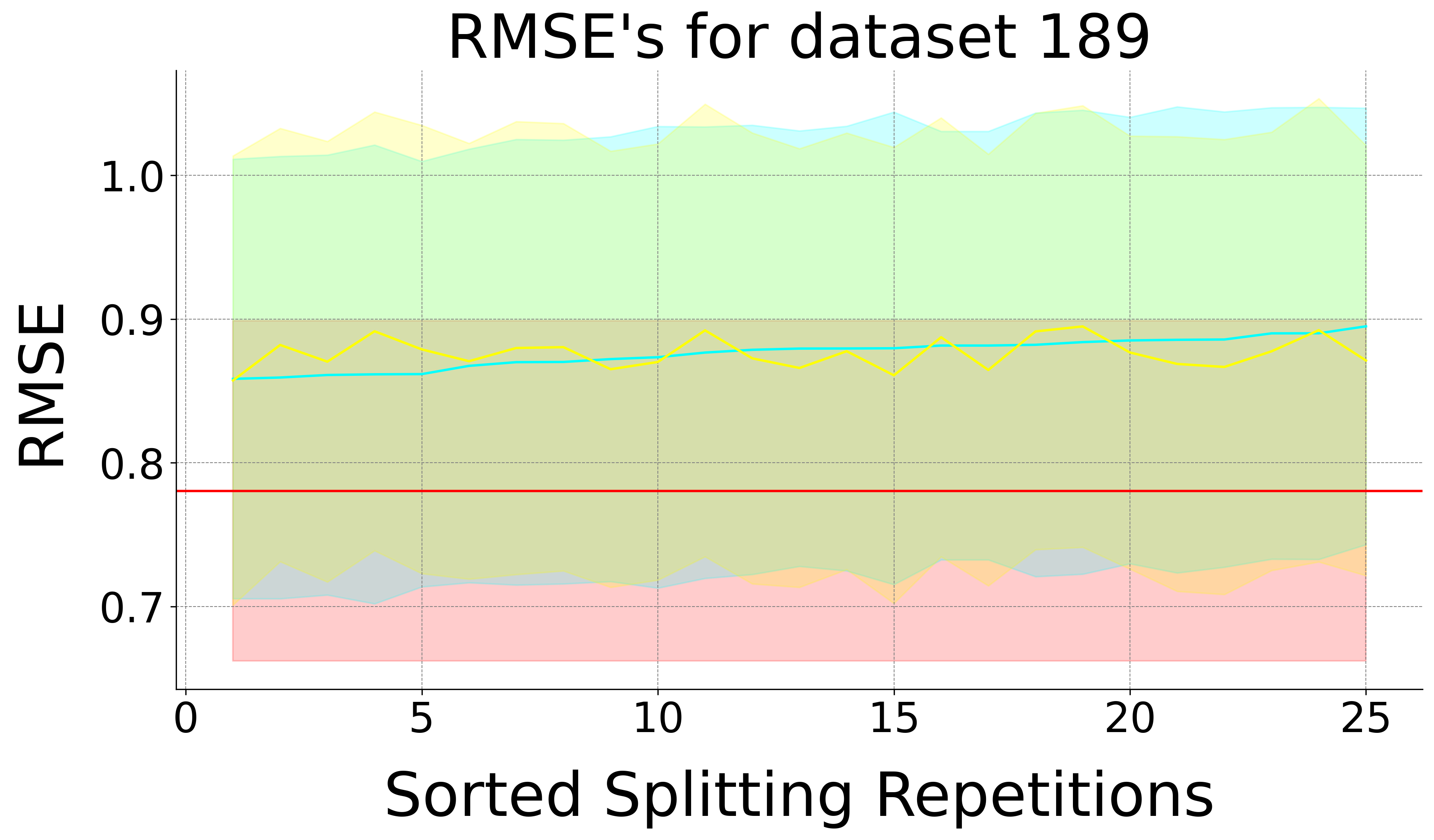}
        \includegraphics[width=0.32\textwidth]{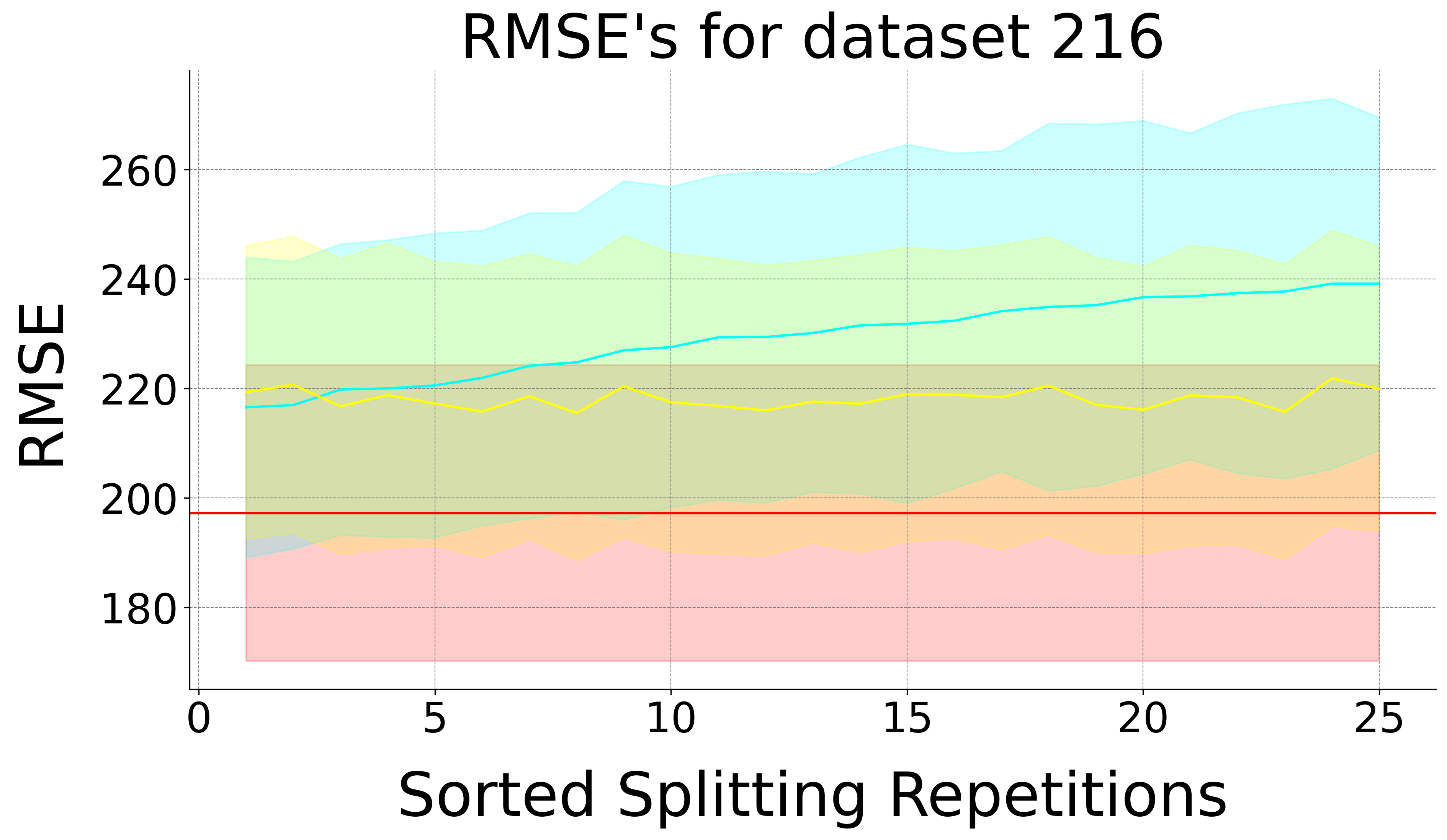}
        \includegraphics[width=0.32\textwidth]{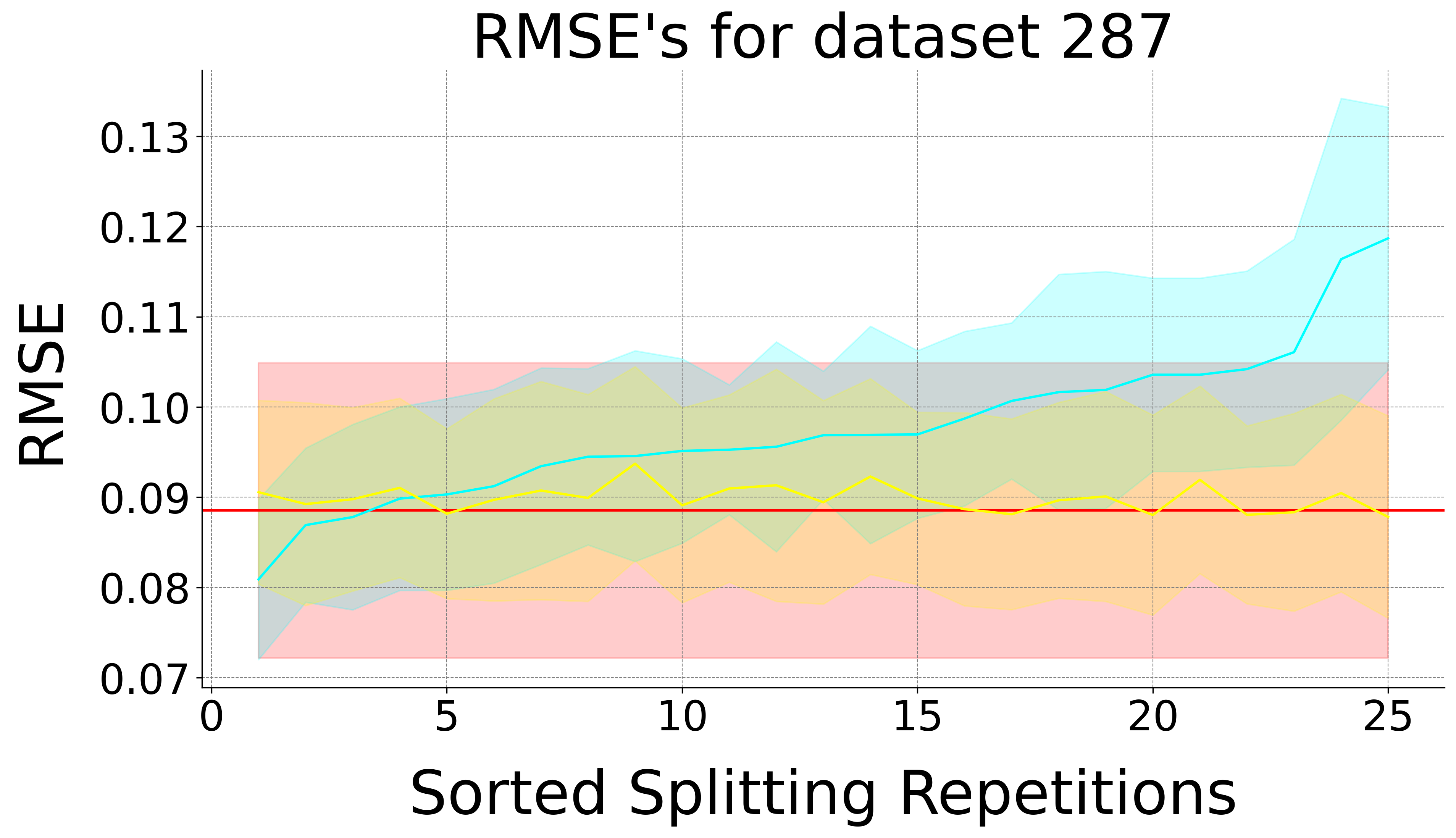}

        \vspace{2.5em}
        \includegraphics[width=0.32\textwidth]{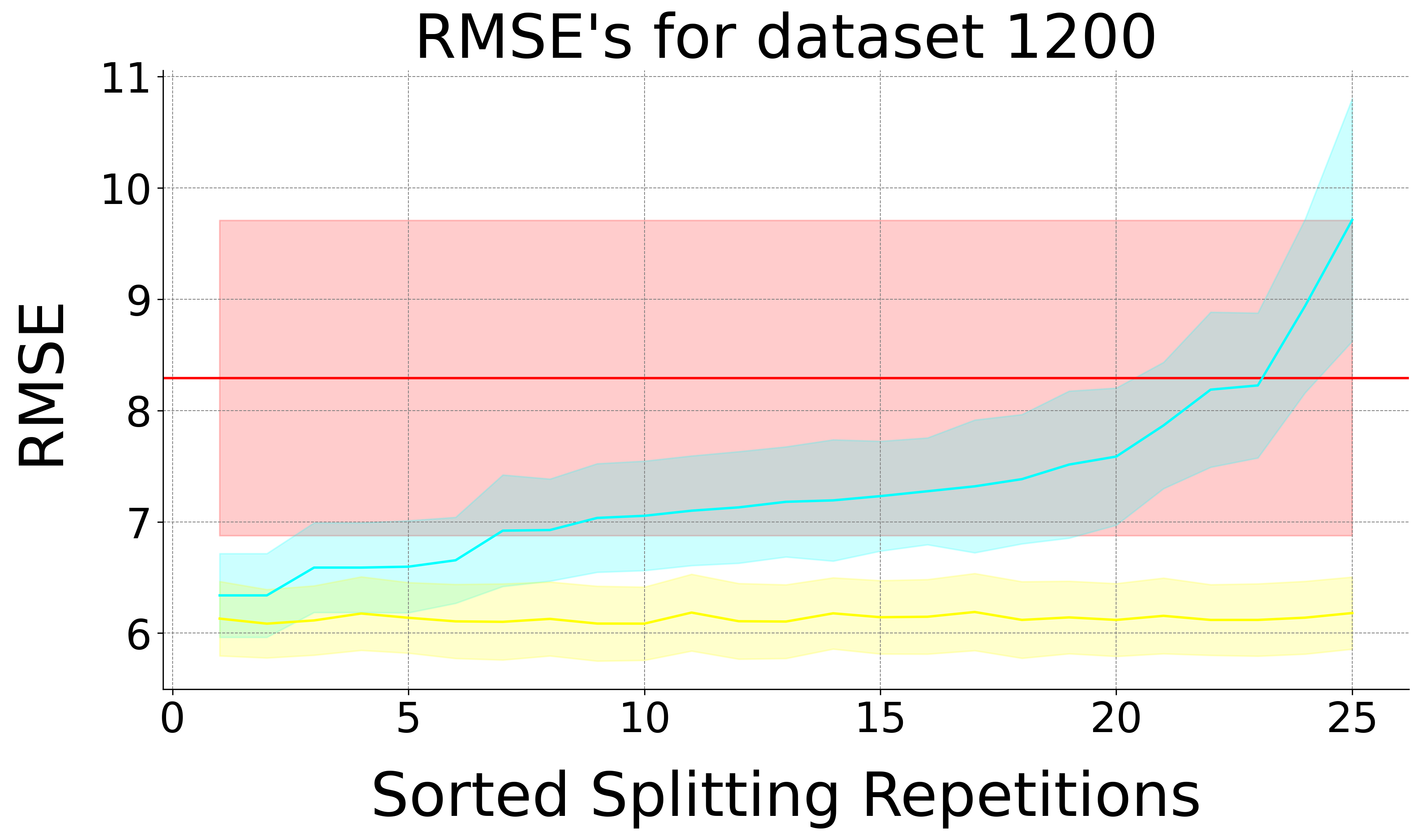}
        \includegraphics[width=0.32\textwidth]{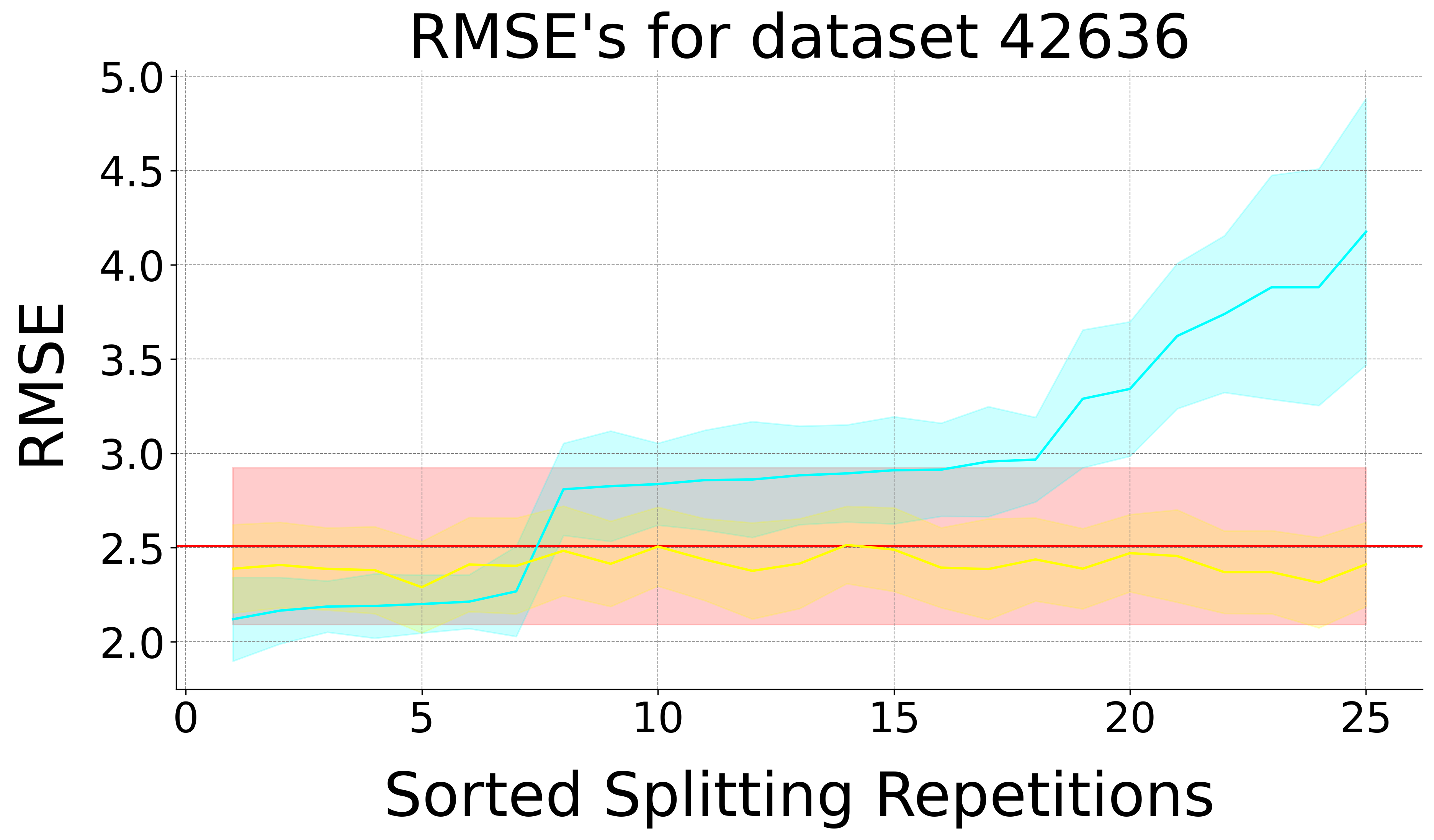}
        \includegraphics[width=0.32\textwidth]{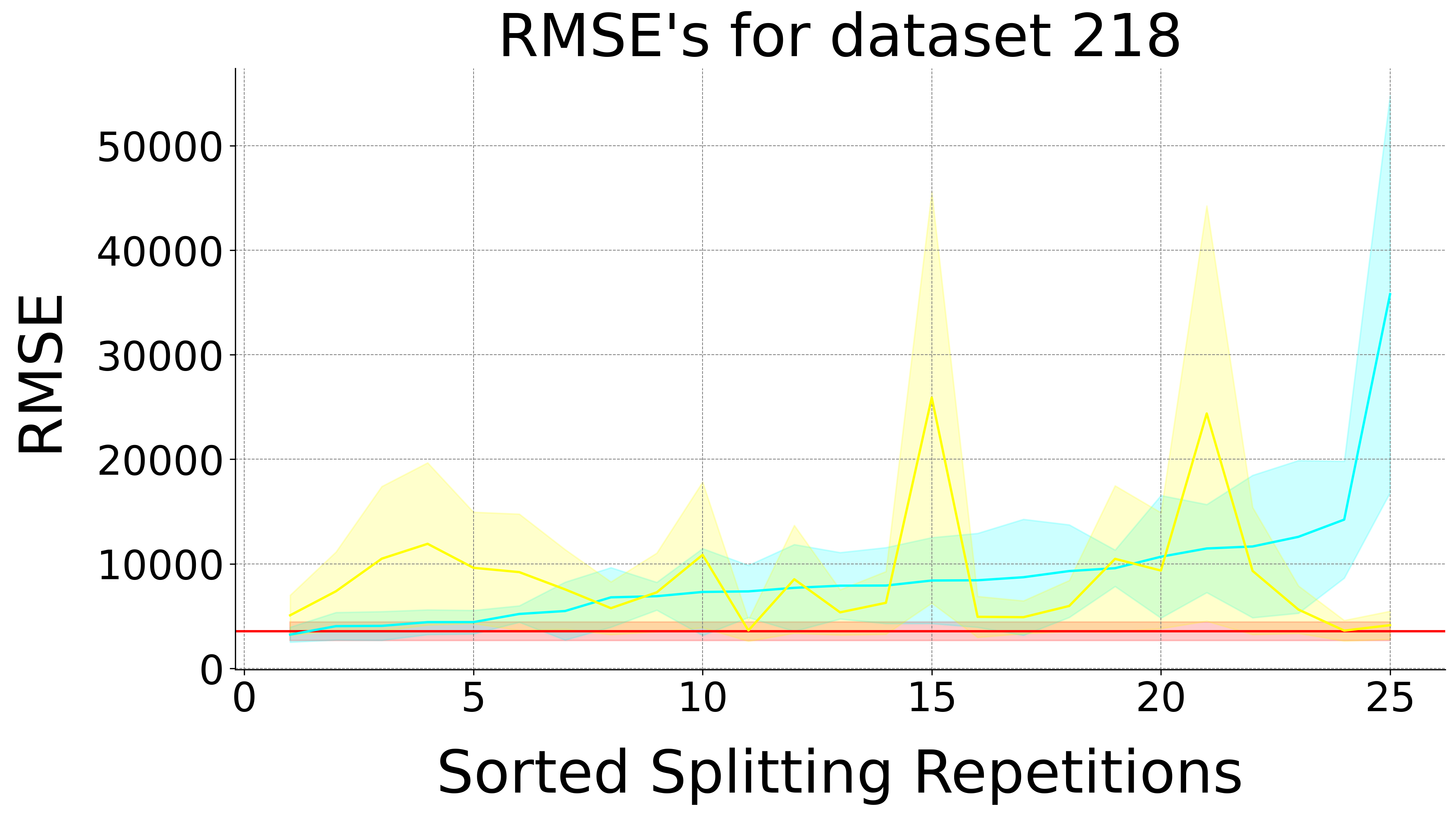}

        \vspace{2.5em}
        \includegraphics[width=0.32\textwidth]{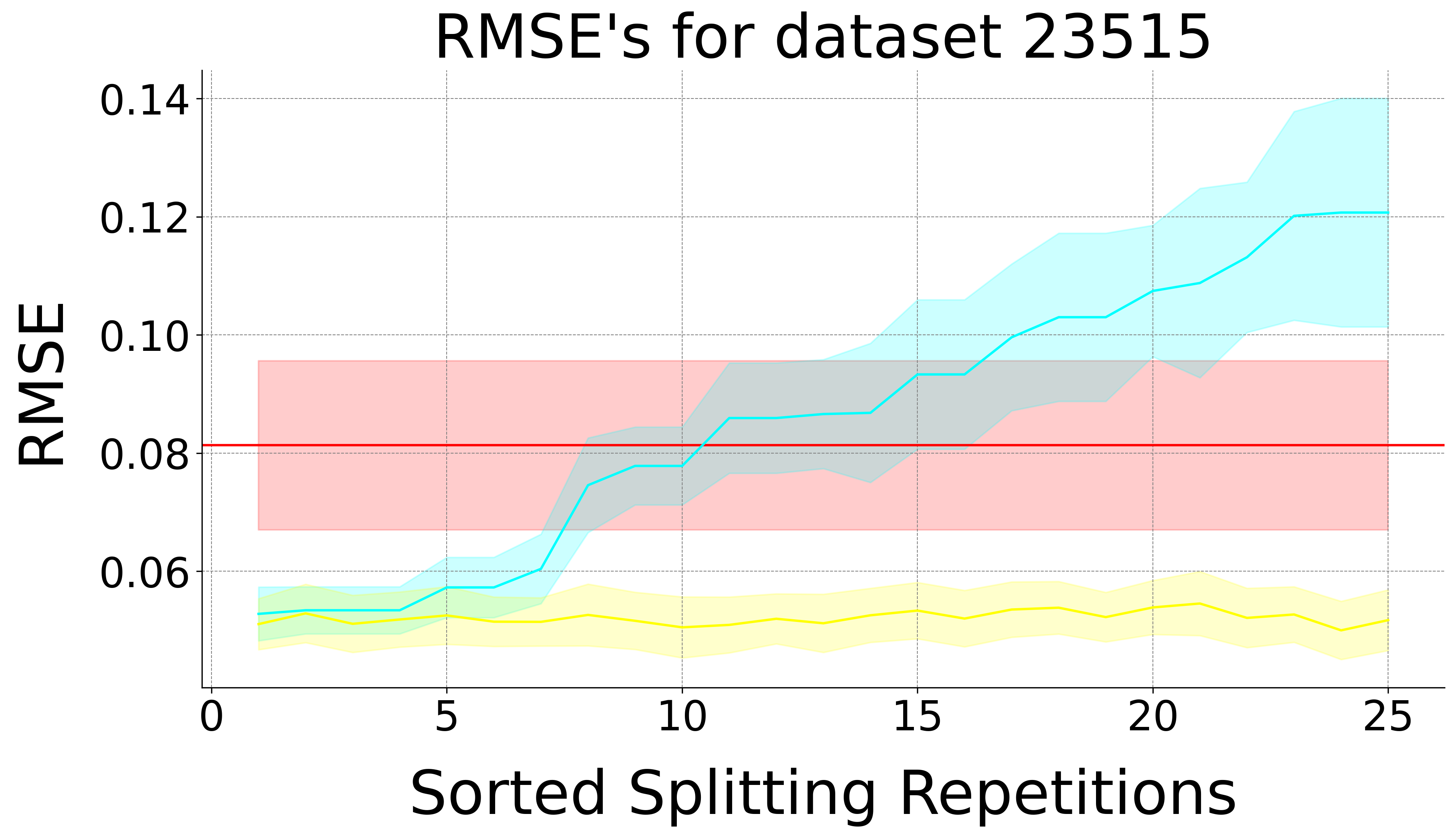}
        \includegraphics[width=0.32\textwidth]{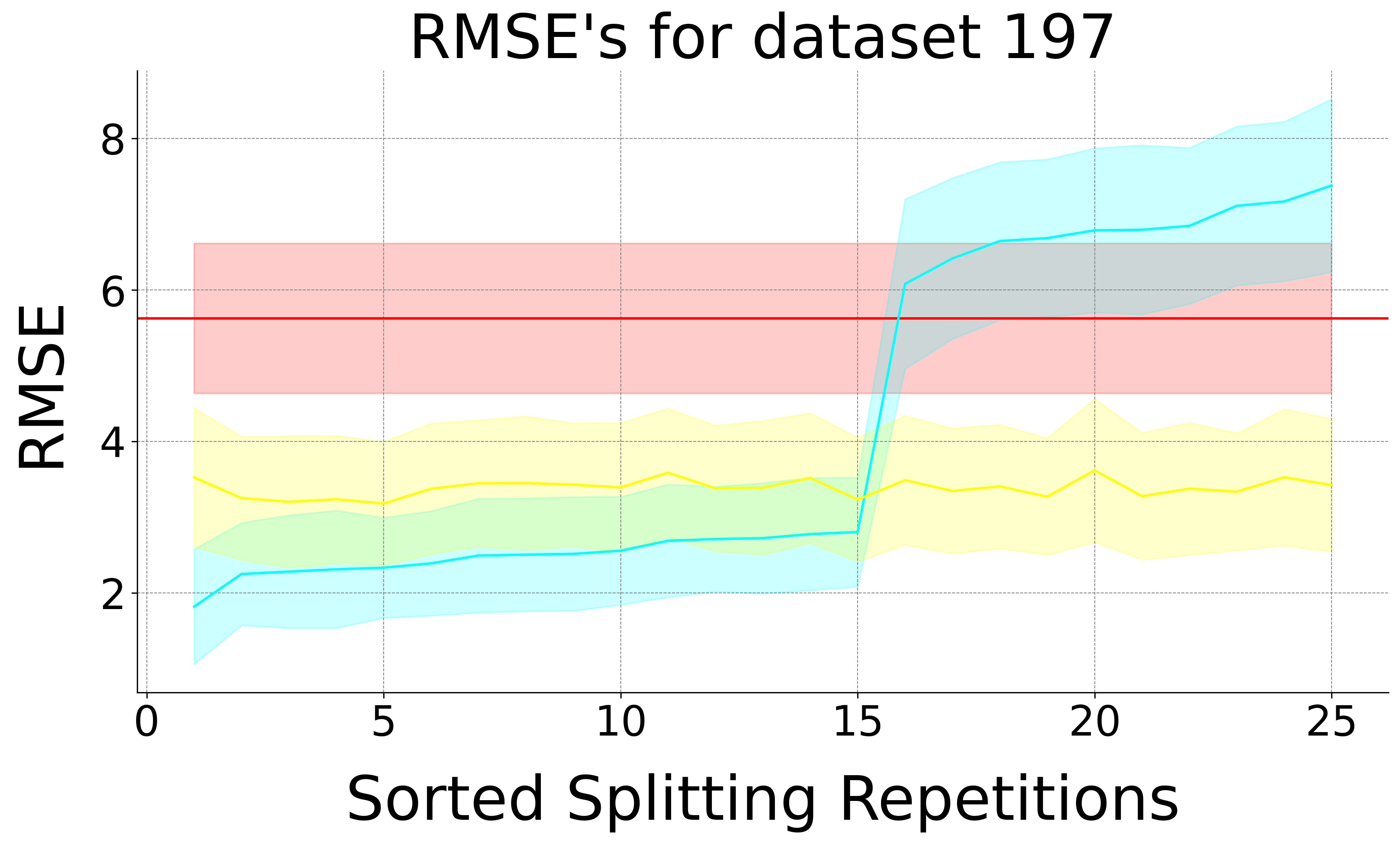}
        \includegraphics[width=0.32\textwidth]{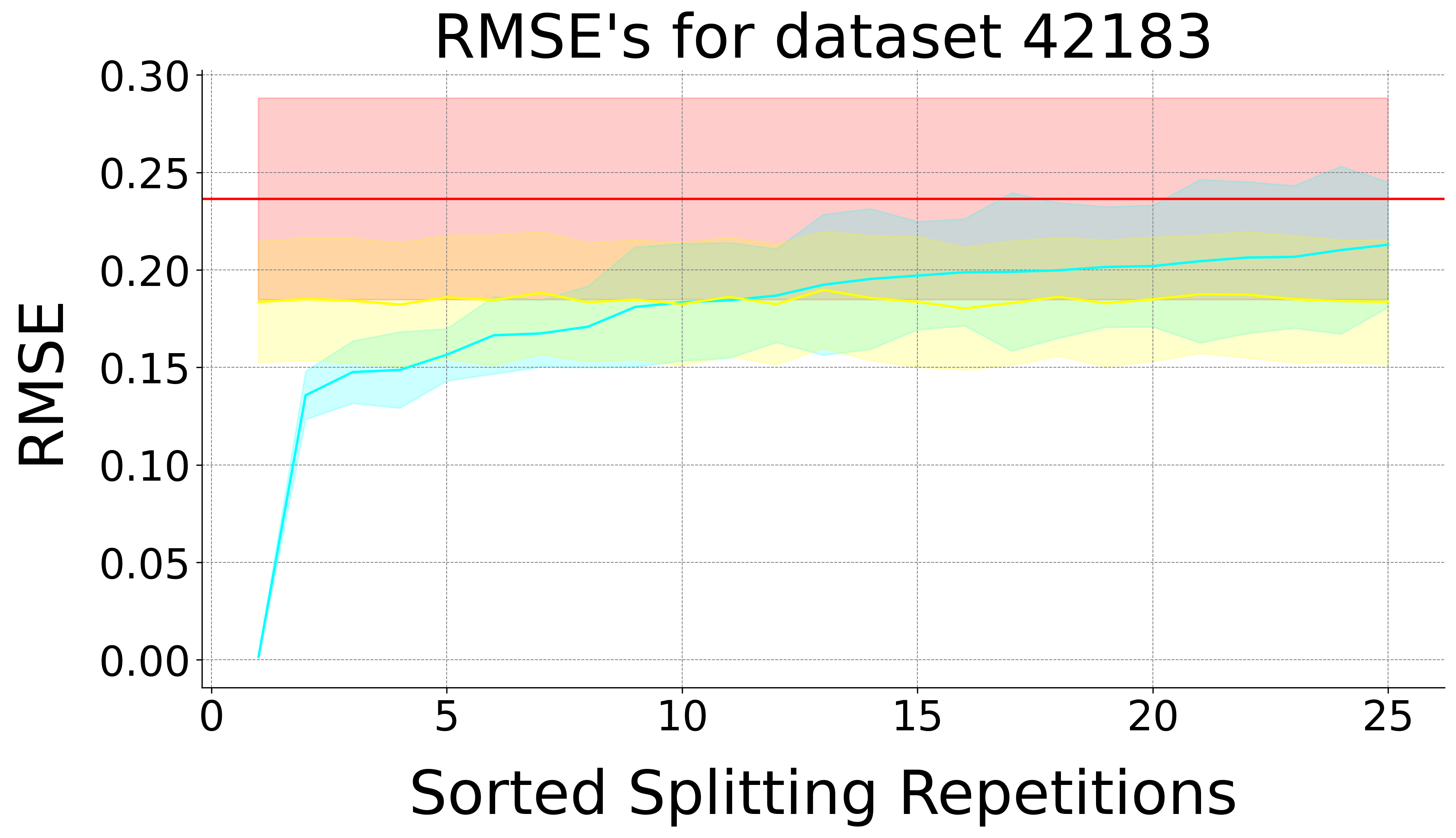}

        \caption{Detailed RMSE trends for all evaluated datasets under the MNAR missingness type.} 
        \label{fig:appendix_mnar}
    \end{minipage}
\end{figure*}

\end{document}